\documentclass[article,twocolumn,aps]{revtex4-1}

\usepackage{graphicx}
\usepackage{dcolumn}
\usepackage{amsmath,amssymb}
\usepackage{graphicx}
\usepackage{float}
\usepackage{xcolor}
\usepackage{mathtools}
\usepackage{hyperref}

\newcommand*{\affaddr}[1]{#1} 
\newcommand*{\affmark}[1][*]{\textsuperscript{#1}}

\begin{document}

\title{Critical properties of the Floquet time crystal within the Gaussian approximation}

\author{%
Muath Natsheh\affmark[1], Andrea Gambassi\affmark[2]\affmark[,3], and Aditi Mitra\affmark[1]\\
\affaddr{\affmark[1]\textit{Center for Quantum Phenomena, Department of Physics,\\
New York University, 726 Broadway, New York, New York, 10003, USA}}\\
\affaddr{\affmark[2]\textit{SISSA - International School for Advanced Studies, via Bonomea 265, 34136
Trieste, Italy}}\\
\affaddr{\affmark[3]\textit{INFN - Sezione di Trieste, via Bonomea 265, 34136, Trieste, Italy}}\\
}

\date{\today}

\begin{abstract}
The periodically driven \(O(N)\) model is studied near the critical line separating a 
disordered paramagnetic phase from
a period doubled phase, the latter being an example of a Floquet time crystal.
The time evolution of one-point and two-point correlation functions are obtained within the Gaussian approximation and perturbatively in the drive amplitude.
The correlations are found to show not only period doubling, but also power-law decays at large spatial distances. 
These features are compared with the undriven \(O(N)\) model, within the Gaussian approximation,
in the vicinity of the paramagnetic-ferromagnetic critical point. 
The algebraic decays in space are found to be qualitatively different in the driven and the undriven cases.
In particular, the spatio-temporal order of the Floquet time crystal leads to position-momentum and momentum-momentum correlation functions
which are more long-ranged
in the driven than in the undriven model.
The light-cone dynamics associated with the correlation functions is also qualitatively different as the critical line of the Floquet time crystal shows a light-cone with two  distinct velocities, with the ratio of these two velocities scaling as the square-root of the dimensionless drive amplitude.
The Floquet unitary, which describes the time evolution due to a complete cycle of the drive, is constructed for modes with small momenta compared to the drive frequency, but having a generic relationship with the square-root of the drive amplitude.
At intermediate momenta, which are large compared to the square-root of the drive amplitude, the Floquet unitary is found to simply rotate the modes. On the other hand, at momenta which are small compared to the square-root of the
drive amplitude, the Floquet unitary is found to primarily squeeze the modes, to an extent which increases upon increasing the wavelength of the modes, with a power-law dependence on it.
\end{abstract}

\maketitle


\section{Introduction}
A time crystal is defined as a many-body system showing spontaneous breaking of time-translation symmetry (TTS) in the
ground state \cite{Wilczek2012_Q,Wilczek2012_C,Li2012}.
There has been much controversy surrounding this definition, and no-go theorems have been proven to show that such a state is
impossible in thermal equilibrium \cite{Bruno2013_1,Bruno2013_2,Bruno2013_3,Oshikawa2015}. Supporting
arguments for a time crystal in thermal equilibrium
have also emerged, where it has been
argued that multicomponent superfluids \cite{Svistunov2020} and easy-plane magnets in a perpendicular
magnetic field \cite{Else2017,Else2019,Khemani2019} satisfy the definition of a time crystal.
To add to this list, time crystals 
in the ground state of Hamiltonians with long-range interactions and in interacting gauge theories have been recently proposed and debated, see 
Refs.~\cite{Kozin2019,Sondhi2020,Kozin2020} and \cite{Wright19,Sacha20,Wright20}, respectively.

It is more widely accepted that time crystals can be realized by relaxing the requirement of the system being in the ground state.
For example, time crystal phases --- referred to as
Floquet time crystals (FTC) --- appear in periodically driven systems, where the spontaneous symmetry breaking in the spatial average of an order parameter is accompanied by
broken TTS, because the order parameter oscillates at frequencies that are subharmonic to the drive frequency (see Refs.~\citep{Sacha2018,Else2019,Khemani2019} for reviews). Since with
Floquet driving, the Hamiltonian has discrete TTS, the Floquet time-crystal is
an example of a system that breaks discrete rather than continuous TTS, and thus it is often referred to as a discrete time crystal.

In the study of FTCs, there is a further dichotomy between phenomena that are purely
quantum, as studied in Refs.~\cite{Sacha15,Else2016,Else2016a,Chandran2016,Else2017,Khemani2016,Keyserlingk2016,Keyserlingk2016_2,Keyserlingk2016_3,Yao2017,Abanin_Ho17,Moessner2017,Russomanno2017,Zeng2017,Huang2018,Gong2018,Kosior2018,Wang2018} and phenomena that emerge in classical driven-dissipative systems \citep{Yao2018,Heugel2019,Garrahan19}. 
In addition, FTCs have been
further characterized on the basis of their stability upon adding perturbations or thermalizing processes~\cite{Else2019,Khemani2019}.
Despite the controversies and the various naming conventions, the field has
remained very active and now includes many experimental examples \cite{Zhang2017,Choi2017,Barrett2018,Volovik2018,Sohan18,Smits18}.

An open and largely unexplored question is the nature of the transition between the ``trivial'' phase and the FTC phase, defined as specified below.
This is clearly a nonequilibrium phase transition which can be realized, for example, by tuning a microscopic parameter of the time-periodic Hamiltonian.
Motivated by the analogy with the behavior in equilibrium, we define the trivial phase of the Floquet system as the one in which the expectation value of an order parameter (e.g., the magnetization) in generic eigenstates of the time-evolution operator \(U\) over one drive cycle vanishes, and the two-point correlation functions of the order parameter are short-ranged in space. 
In addition, we require that the stroboscopic dynamics, i.e., the dynamics observed at integer multiples of the period of the drive, is synchronized with the drive frequency.

For the FTC phase, instead, one requires the existence of a sector of degenerate many-body eigenstates of \(U\). 
For a system with \({\mathbb Z}_2\) symmetry, this degeneracy is at least two-fold
as it corresponds to the two eigenstates of \({\mathbb Z}_2\). 
Strictly speaking, the energy-splitting between these pairs of eigenstates is exponentially small upon increasing the system size, but here we assume the system size to be infinite.
In the FTC phase, the dynamics induced by \(U\) spontaneously breaks \({\mathbb Z}_2\) symmetry
by selecting, for example, a positive value of the magnetization. Accordingly, the state is characterized by long-range spatial order. 
In addition, in order to qualify as a FTC, the dynamics of this state should have the feature that under the time evolution with \(U\), the order parameter 
oscillates with twice the period of the drive. 
This long-range spatio-temporal order, where the spatial average of the order-parameter is 
non-zero and its stroboscopic dynamics occurs at half the drive frequency,
is an example of a period-doubled FTC phase.
For a system with an underlying \({\mathbb Z}_{n>2}\) discrete symmetry, more complex FTC phases can be realized (see Ref.~\cite{Sacha18, Khemani2019,Fazio19}
and references therein).

It is natural to ask whether any universality or scaling is associated with the nonequilibrium phase transition between the trivial and the FTC phase, and if so, what the critical exponents are. 
This issue, which we address here for quantum systems, is even more intriguing in view of the existing discussion on the nature of the nonequilibrium phase transition for classical FTCs~\cite{Yao2018}.

In an attempt to answer the question above, we consider the periodically driven \(O(N)\) model which, in thermal equilibrium, captures, inter alia, the Ising and superfluid critical points \cite{Eyal1996,Moshe2003} depending on the value of \(N\).
Recently, a number of studies \cite{Sotiriadis2009,Sotiriadis2010,Sciolla2011,Sciolla2013,Chandran2013,Schmalian2014,Schmalian2015,Tavora2015,Maraga2015, Smacchia2015,Maraga2016,Chiocchetta2016,Lemonik2016,Chiocchetta2017}  
focused on the nonequilibrium dynamics of the isolated \(O(N)\) model due to a sudden change (global quantum quench) in its Hamiltonian and an emerging universality in the transient regime was identified \cite{Schmalian2014,Schmalian2015,Tavora2015,Maraga2015,Maraga2016,Chiocchetta2016,Lemonik2016,Chiocchetta2017}.
In the limit $N\to\infty$, this model also provides one of the few available examples of exactly solvable nonequilibrium dynamics in generic spatial dimension~\cite{Smacchia2015,Maraga2015,Lemonik2016}.

The periodically driven \(O(N)\) model was studied in Ref.~\cite{Chandran2016}. While it is expected that generic, isolated, periodically driven systems will eventually heat to infinite temperature~\cite{Rigol14,Lazarides14,Ponte15} and will therefore not support any non-trivial phase, Ref.~\cite{Chandran2016} showed that in the limit \(N\rightarrow \infty\), interactions can suppress heating and stabilize a FTC phase.  For finite \(N\), instead, the \(O(N)\) model supports a prethermal FTC, the temporal duration of which increases upon increasing \(N\). 
Within this prethermal regime, the existence of a trivial phase and a period-doubled FTC
phase can be identified. However, while these phases are known, the nature of the phase transition between them is largely unexplored. 

Our goal here is to explore this transition starting from its Gaussian approximation, which, as it is known from the theory of critical phenomena, is well-defined and exactly solvable  in spite of the fact that the very same existence of these phenomena hinges on the presence of interactions. The Gaussian approximation is key to establishing the emergence of scaling, if at all, and it is a stepping stone for exploring the role of interactions, which will be reported elsewhere \cite{Natsheh2020b}.

Since the FTC phase is not a phase in thermal equilibrium, its realization is not guaranteed, and it may depend in important ways on the initial conditions~\cite{Chandran2016}. 
Here we study how the FTC phase is approached after a quench \cite{calabrese2007,Mitra2018}, where the initial state of the system is the thermal equilibrium state of one Hamiltonian,  while the time evolution is determined by another.
We choose an initial state characterized by the absence of order and with spatial correlations
decaying over short distances.
We follow the time-evolution of this state under periodic driving and we identify the
parameters which allow this state to reach the FTC phase. 
We then determine the expressions of the correlation functions at or near criticality, 
within the Gaussian approximation. 

The paper is organized as follows. The model is introduced in Section \ref{model}, where we also review its phase diagram and explain the quench dynamics.
In Section \ref{FBth}, the Floquet-Bloch theory is used to determine the quasi-modes and quasienergies within the Gaussian approximation. Section \ref{corr} presents the expressions of the various relevant unequal-position and unequal-time correlation functions along the critical line, while in Section \ref{microm} we determine and discuss the Floquet unitary of the model.
Section \ref{conclu} presents our conclusions, while details of the various calculations are
outlined in several appendices.

\section{The 
model, the quench protocol, and the phase diagram}
\label{model}

In this section we present the model, outline the quench protocol, and discuss the phase diagram.

\subsection{The Model}
The periodically driven \(O(N)\) model in \(d\) spatial dimensions is defined by the Hamiltonian
\begin{multline}
H=\sum_{i=1}^{N} \int d^d x \frac{1}{2}\left[(r-r_1\cos{(\omega t)})\phi_i^2(\textbf{x}) \right. \\
\left.+(\vec{\nabla}\phi_i)^2 +\Pi_i^2(\textbf{x})\right] + V,\label{H}
\end{multline}
where \(\phi_i\) and \(\Pi_i\) are \(N\)-component bosonic fields which obey the canonical commutation relation
\begin{equation}
[\phi_j(\textbf{x}),\Pi_l(\textbf{y})]=i\delta_{jl}\delta^{(d)}(\textbf{x}-\textbf{y}).
\end{equation}
\(V\) is the interaction term
\begin{align}
V = \frac{u}{4! N}\int d^d x \biggl(\sum_{i=1}^N \phi_i^2\biggr)^2,
\end{align}
while \(r\) is the detuning parameter which, if assuming negative values, causes an instability
in the free, undriven model with \(V=r_1=0\), towards forming a ferromagnet.
The presence of interactions is actually necessary for stabilizing such a ferromagnetic phase.
In Eq.~\eqref{H}, \(r_1\) and \(\omega\) are  the amplitude and angular frequency, respectively, of the
periodic driving of the detuning parameter. Accordingly, \(H\) is periodic in time with period \(T = 2\pi/\omega\), i.e., \(H(t+T) = H(t)\).

In the limit \( N \rightarrow \infty\), the Hartree approximation for \(V\) becomes exact not only for the equilibrium properties \cite{Moshe2003} but also for the non-equilibrium dynamics (see, e.g., Refs.~\cite{Sotiriadis2010,Berges02,Gasenzer07} 
for undriven models),
and gives a more complex phase diagram than the undriven model \cite{Chandran2016}. 
We will discuss the phase diagram in detail below.
Corrections of order \(1/N\) and beyond, on the other hand, lead to heating effects,
making any possible non-trivial phases ultimately unstable at longer times.
Accordingly, the case we are studying is, strictly speaking, that of a prethermal FTC
the lifetime of which increases upon increasing \(N\).

Our goal is to understand the possible emergence of scaling behavior and critical exponents in the dynamics of this model. To this end, we will present predictions for the dynamics of the order parameter, defined as the expectation value \(\langle \phi_j(x,t)\rangle\). We will also
discuss the unequal-time and unequal-position correlation function \(\langle \phi_j(x,t) \phi_k(x',t')\rangle\) and its time derivatives. The latter correspond to correlations of the type \(\langle \phi_j(x,t) \Pi_k(x',t')\rangle, \langle \Pi_j(x,t) \Pi_k(x',t')\rangle\), i.e., position-momentum and momentum-momentum correlations, respectively.
We will derive these predictions within the Gaussian approximation and for the initial condition discussed below. We will also highlight the differences with the undriven model.

As we focus below on the Gaussian model corresponding to having \(V=0\) in Eq.~\eqref{H},
it is convenient to introduce the representation of the various fields in momentum space,
\begin{align}
\phi_i(\textbf{x})=\int^{\Lambda} \frac{d^d k}{(2\pi)^d} e^{i\textbf{k}.\textbf{x}}\phi_{i,\textbf{k}},
\label{phik}
\end{align}
with an analogous definition for the Fourier transform $\Pi_{i,\textbf{k}}$ of  $\Pi_i(\textbf{x})$. 
In terms of these fields, the resulting Hamiltonian can be written as
\begin{multline}
H=\sum_{i=1}^{N} \int^{\Lambda} \frac{d^d k}{(2\pi)^d}\ \frac{1}{2}\left[(r+k^2-r_1\cos{\omega t})|\phi_{i,\textbf{k}}|^2 \right. \\ \left. +|\Pi_{i,\textbf{k}}|^2 \right],\label{Hg}
\end{multline}
with the canonical commutation relations for the fields in momentum space becoming,
\begin{equation}
[\phi_j(\textbf{k}),\Pi_l(\textbf{q})]=i(2\pi)^d\delta_{jl}\delta^{(d)}(\textbf{k}+\textbf{q}).
\end{equation}
The large-momentum cutoff \(\Lambda\) in Eqs.~\eqref{phik} and \eqref{Hg} is another microscopic parameter of the model. Both in thermal 
equilibrium and in the driven model \cite{Chandran2016} its specific value may affect the stability of the resulting phases of the model: further below we revisit this dependence in the case of the driven model.

\subsection{Quench protocol}

As anticipated, we study the dynamics of the system after a quench 
\cite{calabrese2007,Mitra2018}, where the initial state is  
a mixed state corresponding to the thermal equilibrium state
of the undriven model, i.e., \(r_1=0\), with a positive value \(r_0 >0\) of the detuning parameter $r$. 
This initial state is evolved under the  periodically driven model in Eq.~\eqref{Hg}. We choose the initial value \(r_0\gg r>0\) so that the initial state is deep in the paramagnetic phase with short-range spatial correlations.

Defining \(a_\textbf{k}^\dagger\) and  \(a_\textbf{k}\) as the creation and annihilation operators which diagonalize the initial undriven model
\begin{equation}
H_0=H(r_1=0)=\sum_{i=1}^{N}\int^\Lambda \frac{d^d k}{(2\pi)^d} \ {\omega_0}_k a_{i,\textbf{k}}^{\dagger} a_{i,\textbf{k}},
\end{equation}
with dispersion
\begin{equation}
{\omega_0}_k=\sqrt{r_0+k^2}\label{om0k},
\end{equation}
the initial fields obey
\begin{align}
\phi_{i,\textbf{k}}(t=0)=\frac{1}{\sqrt{2\omega_0}_k}(a_{i,\textbf{k}}+a_{i,-\textbf{k}}^{\dagger}),\\
\Pi_{i,\textbf{k}}(t=0)=-i\sqrt\frac{{\omega_0}_k}{2}(a_{i,\textbf{k}}-a_{i,-\textbf{k}}^{\dagger}).
\end{align}
As mentioned above, the initial state is the thermal equilibrium state of the pre-quench Hamiltonian \(H_0\), where the statistical average of an operator \(\hat{O}\) at temperature \(\beta^{-1}\) is defined as
\begin{equation}
\langle\hat{O}\rangle = \frac{\text{tr}\left(\hat{O}e^{-\beta H_0}\right)}{\text{tr}\left(e^{-\beta H_0}\right)}.
\end{equation}
The expectation values of the relevant operators in the above initial state are $\langle \Pi_{i,\textbf{k}}(0) \rangle = \langle \phi_{i,\textbf{k}}(0) \rangle =0$, with
\begin{align}
\langle \Pi_{i,\textbf{k}}(0) \Pi_{j,\textbf{q}}(0) \rangle &= \delta_{i,j}\delta_{\textbf{k},-\textbf{q}} \frac{\omega_{0k}}{2} \coth(\beta \omega_{0k}/2),
\label{corrPP0}\\
\langle \phi_{i,\textbf{k}}(0) \phi_{j,\textbf{q}}(0) \rangle &= \delta_{i,j}\delta_{\textbf{k},-\textbf{q}} \frac{1}{2\omega_{0k}} \coth(\beta \omega_{0k}/2),
\label{corrpp0}\\
\langle \left\{\phi_{i,\textbf{k}}(0),\Pi_{j,\textbf{q}}(0) \right\}\rangle &= 0, \label{corrpP0}
\end{align}
where we introduce the short-hand notation $\delta_{\textbf{k},-\textbf{q}} = (2\pi)^d \delta(\textbf{k}+\textbf{q})$.
In particular, we  choose \(\beta r_0 \gg 1\) in order to ensure short-range correlations in the thermal initial state.

Since both the pre-quench and post-quench Hamiltonians are symmetric in the field component \(i\),
and since we focus below on the phase without spontaneous symmetry breaking,
the initial conditions and the dynamics of all the field components are identical. 
Accordingly, in our analysis, we can conveniently omit the index of the field component.
In addition, within the Gaussian approximation,
the momentum modes evolve independently according to
\begin{equation}
\left\{
\begin{aligned}
&i\dot\phi_\textbf{k}=[\phi_\textbf{k},H]=i\Pi_{\textbf{k}},\label{Pi1} \\
&i\dot\Pi_\textbf{k}=[\Pi_\textbf{k},H]=-i(r+k^2-r_1 \cos{\omega t})\phi_{\textbf{k}}.
\end{aligned}
\right.
\end{equation}
Combining the above two equations gives
\begin{equation}
\ddot\phi_\textbf{k}=-(r+k^2-r_1 \cos{\omega t})\phi_\textbf{k},\label{phieom}
\end{equation}
the solution of which can be written in the form
\begin{equation}
\phi_\textbf{k}(t) = M_{c,k}(t) \phi_\textbf{k}(0) + M_{s,k}(t) \Pi_\textbf{k}(0), \label{Phi1}
\end{equation}
where the functions $M_{c,k}(t)$ and $M_{s,k}(t)$ obey
\begin{equation}
\frac{d^2}{dt^2}\left(
\begin{tabular}{c}
\(M_{c,k}(t)\)  \\
\(M_{s,k}(t)\)
\end{tabular}
\right) = -(r+k^2-r_1 \cos(\omega t)) \left(
\begin{tabular}{c}
\(M_{c,k}(t)\)  \\
\(M_{s,k}(t)\)
\end{tabular}
\right),\label{Mcseq}
\end{equation}
with initial conditions
\begin{equation}
\left(
\begin{tabular}{cc}
\(M_{c,k}(0)\) & \(M_{s,k}(0)\) \\
\(\dot{M}_{c,k}(0)\) & \(\dot{M}_{s,k}(0)\)
\end{tabular}
\right) = \left(
\begin{tabular}{cc}
\(1\) & \(0\) \\
\(0\) & \(1\)
\end{tabular}
\right).\label{Mcsin}
\end{equation}
Equation \eqref{Pi1} implies
\begin{align}
\Pi_k(t) = \dot{M}_{c,k}(t)\phi_k(0) + \dot{M}_{s,k}(t)\Pi_k(0),
\end{align}
and the canonical commutation relations between $\phi(t)$ and $\Pi(t)$ are obeyed because
\begin{align}
1 = M_{c,k}(t) \dot{M}_{s,k}(t) - M_{s,k}(t) \dot{M}_{c,k}(t).
\label{Mcan}
\end{align}
This can be explicitly checked by noting that the initial conditions in Eq.~\eqref{Mcsin} obey Eq.~\eqref{Mcan} at \(t=0\) and that the equations of motion \eqref{Mcseq} imply that the r.h.s.~of Eq.~\eqref{Mcan} is a constant of motion.

\subsection{Phase Diagram}

The equations of motion \eqref{Mcseq} for $M_{c,k}$ and $M_{s,k}$ are known to have the Mathieu functions \cite{Nist,McLachlin1947,Richards1983} as solutions. 
The quantum aspects of the problem only enter upon imposing the canonical commutation relations \eqref{Mcan}; before imposing them, the behavior of the classical solutions provide a first indication of the conditions under which stable solutions exist.  

\begin{figure}[ht]
\includegraphics[width = 0.5\textwidth]{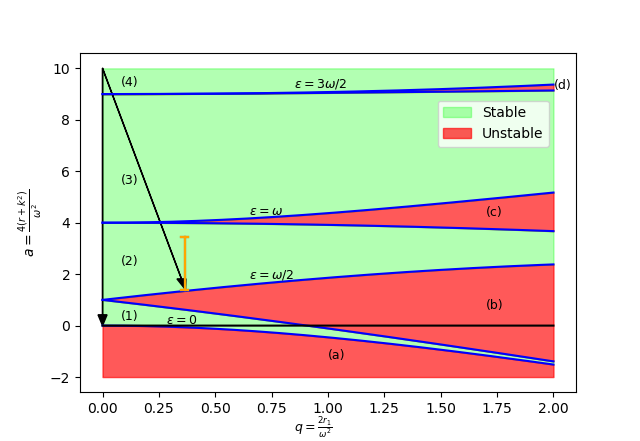}
\caption{\label{fig1}
Stability phase diagram of the Mathieu equation \eqref{Mcseq} depending on the dimensionless parameters in Eqs.~\eqref{qdef} and \eqref{eq:def-a}, which applies also 
to a fluctuation mode with wavevector $k$ of the periodically driven Gaussian model.
The arrows indicate two different kinds of quenches: The vertical one denotes a quench from an initial paramagnetic phase to the
critical point of the undriven (\(r_1=0\)) model \cite{Sotiriadis2009}. The tilted arrow, instead, denotes a quench from an initial paramagnetic phase of the undriven model, to the critical point of a FTC phase.
While there are many period-doubled FTC phases, each corresponding to an integer \(n\) such that the band-edges of the stable region are characterized by having \(\epsilon_{k=0}= (n+1/2)\omega\) (regions \((2)\) and \((4)\)),
here we study the one where the band-edge is at
half the drive frequency \(\epsilon_{k=0}=\omega/2\) (region \((2)\)).
In order for the model to have a stable solution,
it is necessary that all the fluctuation modes with $k \in [0,\Lambda]$ -- which correspond to the points belonging to a vertical segment in this phase diagram, highlighted in yellow -- are within the stable region.
}
\end{figure}

For a given mode \(k\), the ``phase diagram'' indicating the stable and unstable
regions of the parameter space is shown in Fig.~\ref{fig1}, where the horizontal axis is the dimensionless strength 
\begin{align}
q= 2r_1/\omega^2 \label{qdef}
\end{align}
of the driving field while the vertical axis corresponds to the dimensionless parameter
\begin{align}
a = 4 (r+k^2)/\omega^2,
\label{eq:def-a}
\end{align}
associated with the time-independent coefficient on the r.h.s.~of Eq.~\eqref{Mcseq}.

The red regions in Fig.~\ref{fig1} are unstable because the corresponding modes $M_{c,k}$ and $M_{s,k}$ grow exponentially in time without bound. 
Accordingly, these regions of parameters are not allowed, leading to band gaps. 
The green regions instead, correspond to stable solutions.
In Fig.~\ref{fig1} there are four stable regions labeled by (1), (2), (3), (4), and four
unstable regions $(a)$, $(b)$, $(c)$, and $(d)$.
For a choice of the driving protocol (specified by $r$, $r_1$, and $\omega$) the dynamics of the model is stable if all its fluctuation modes with $k \in [0,\Lambda]$ correspond to stable points in Fig.~\ref{fig1}. This means that, for a specified value of $q$, the vertical segment with $a \in [4r/\omega^2,4(r+\Lambda^2)/\omega^2]$ has to fall within the green region~\cite{Chandran2016}, as exemplified by the vertical yellow segment in Fig.~\ref{fig1}.

Since the system is driven periodically, the mode energies are conserved only up to integer multiples of the drive frequency and therefore they qualify as quasienergies rather than energies (see, c.f.,  Sec.~\ref{FBth}).
Let us denote by \(\epsilon_k\) the quasienergy at which the modes with a certain
$k$ oscillates.
The edges of the various bands in Fig.~\ref{fig1} are determined by the condition that $\epsilon_k = n\omega/2$, $n$ being an integer. 

One way to understand why the band edges are located at $\epsilon_k=n\omega/2$ is to consider the limit of weak driving \(q\ll 1\), 
because, then, the condition $\epsilon_k = n\omega/2$ coincides with that for the occurrence of parametric resonances in the model: integer multiples of the drive frequency become resonant with
the frequency at which the quantity in the Hamiltonian coupled to the external driving field would oscillate in the undriven model. 
In the present case, this quantity is $|\phi_{i,{\bf k}}|^2$ (see Eq.~\eqref{Hg}) and 
since the dispersion of the undriven model is $\sqrt{k^2+r}$,
the quantity coupled to the external drive oscillates, for weak drive, at the frequency $2\epsilon_k (q\to 0)=2\sqrt{r+k^2}$, yielding the resonant condition \({\text{integer}} \times \omega = 2 \sqrt{r + k^2}\). 
Accordingly, as it is clearly shown in Fig.~\ref{fig1}, the $n$-th band edge touches the vertical axis for $q\to 0$ at $a = a_n(q\to 0) = n^2$, where $a$ is defined in Eq.~\eqref{eq:def-a}. Since the most unstable mode corresponds to the spatially homogeneous one (which determines the lowermost point of the vertical segment in Fig.~\ref{fig1}), the above resonance conditions should be applied to the $k=0$ mode.

While the argument presented above was given in the limit of weak drive $q\to 0$, the fact that the band edges are pinned at $\epsilon_{k} = n\omega/2$ for generic values of $q$ follows also from noting that Eq.~\eqref{Mcseq} being a homogeneous differential equation, the slowest oscillating modes are of two kinds: those which return to themselves after a drive cycle, i.e, are periodic (even $n$) and those that flip their overall sign after a drive cycle, i.e, are anti-periodic (odd $n$).
From Floquet theory, in order to avoid over-counting the modes, the quasienergies $\epsilon_k$ for $q\neq 0$ must be restricted within the interval $\left[-\omega/2, \omega/2\right]$. Accordingly, the possible slowest oscillating modes are those at quasienergies $\epsilon_{k}=\omega/2$ and  0.

We can now distinguish two cases: when the integer $n$ leading to the resonance is even, the longest wavelength mode, i.e., that with $k=0$, oscillates at
integer multiples of the drive frequency $\omega$.  
When $n$ is odd, instead,
the longest wavelength mode oscillates at half the drive frequency, and therefore
shows period-doubling. 
Note that a periodic driving of the coefficients of
higher powers of the position or momentum operators will lead to more complex dynamics~\cite{Guo13}.

Since the \(k=0\) mode is nothing but the order-parameter of the model,
Fig.~\ref{fig1} implies that the non-trivial phase comes in two
varieties. One in which the order parameter oscillates at integer multiples of the drive frequency, including zero: this can be identified with the conventional ferromagnetic phase because the average of the order parameter over \emph{one} drive cycle is non-zero.
The other phase, instead, is characterized by the fact that the order parameter is
period doubled and it can be identified with the FTC because the average of the order parameter over \emph{two} drive cycles vanishes.

While strictly speaking, a ferromagnetic or FTC phase cannot be defined for a free system, we expect that the red unstable regions become
stable in the presence of interactions, which turn the regions marked by \((a)\) and \((c)\)  into a ferromagnet, while those marked by
\((b)\) and \((d)\) into an FTC phase.
Accordingly, the stability phase diagram in Fig.~\ref{fig1} translates into a bona fide phase diagram~\cite{Chandran2016}, with the precise microscopic values
at which the transition from the stable to the unstable regions occur in Fig.~\ref{fig1} being modified by the Hartree corrections introduced by the interactions. 
At even longer times, heating will set in, but this time can be made to approach
infinity as \(N\rightarrow \infty\).

However, even for \(N\rightarrow \infty\), there is a subtlety related to the value of the
cut-off \(\Lambda\). 
In fact,  \(\Lambda\rightarrow\infty\) in the continuum and therefore there will always be some modes in Fig.~\ref{fig1} which fall within a gap (red regions), and the solution will be unstable in the presence of the drive.
However, the gaps are rather narrow for the large values of $a$ induced by a large $\Lambda$,
as shown in Fig.~\ref{fig1}, so that the
time scales after which the FTC becomes unstable, which are related to the inverse of the gap, are also long. Accordingly, while the FTC is not expected to be completely stable for \(\Lambda \rightarrow \infty\), it is quasi-stable.

In Fig.~\ref{fig1}, from bottom to top, the ferromagnetic phases (\((a)\) and \((c)\)) and period doubled FTC phases (\((b)\) and \((d)\)) alternate with one another, with region \((a)\) being simply the driven 
version of the ferromagnetic phase of the static \(O(N)\) model.  All the other phases only arise due to a resonant drive.

It is interesting to note that, in the presence of the drive, large regions of parameter space with \(r>0\) become unstable, whereas without drive, these same regions would remain paramagnetic. A heuristic way to understand this is that as the parameter \(r_{\rm eff} = r - r_1 \cos(\omega t)\) oscillates, it can become momentarily negative, causing the development of an instability. 
A similar heuristic argument can be used in order to understand why stable (green) regions appear for \(r<0\) and sufficiently large \(r_1\).

We are interested in the properties of the critical line separating the paramagnetic phase from the FTC phase. In this paper we focus on the FTC phase corresponding to region \((b)\) in Fig.~\ref{fig1}, and in particular on the behavior of the system in the vicinity of the critical line labeled by $\epsilon = \omega/2$
between regions \((2)\) and \((b)\).
Our choice is a matter of
convenience as the same coarse-grained behavior is expected to occur at all the other critical lines separating a trivial from a FTC phase, such as the boundary marked by $\epsilon = 3\omega/2$ in Fig.~\ref{fig1}. Since quasienergies are defined modulo the drive frequency $\omega$, it is clear that both these band-edges correspond to an order-parameter that shows period doubling.
In a similar manner, we expect the coarse-grained features to be common to all the critical lines separating a paramagnet from a ferromagnet. This corresponds to lines labeled by $\epsilon= 0$ and $\epsilon=\omega$ in Fig.~\ref{fig1}.

Although Mathieu function solutions are well-known, we derive them below by using Floquet-Bloch theory, briefly recalled in Appendix~\ref{appA}. 
This is because we are interested in the vicinity of the above-mentioned critical line where standard Mathieu function solutions found in textbooks (see, e.g., Refs.~\cite{Nist,McLachlin1947,Richards1983}) are not easily generalizable. In addition, once the modes \(M_{c,k}\) and \(M_{s,k}\) in Eq.~\eqref{Phi1} are obtained, the solution of the quantum problem requires imposing the canonical commutation relations \eqref{Mcan}.

\section{Floquet-Bloch Solution} 
\label{FBth}

The dynamics of the (quantum) system is determined by the solution of Eq.~\eqref{Mcseq}, which can be cast generically in the following form:
\begin{equation}\label{mot}
\ddot{f_k} = -[r+k^2-r_1 \cos(\omega t)]f_k.
\end{equation}
The initial conditions for this equation will be specified further below in this section.
According to the Floquet-Bloch theorem summarized in Appendix~\ref{appA}, the solutions of 
Eq.~\eqref{mot} can be written as
\begin{equation}
\label{ukdef}
f_k(t)=u_k(t)\exp(i\epsilon_k t), \quad {\text{with}} \quad  u_k(t+T)=u_k(t),
\end{equation}
where \(\epsilon_k\) is the quasienergy, $T=2\pi/\omega$ the period of the drive, and \(u_k\) the quasimodes.
The periodicity in time of the quasimodes
allows their Fourier expansion, i.e.,
\begin{equation}
u_k(t)=\sum_{m=-\infty}^{\infty}c_{m}\ e^{im\omega t}.\label{ftu}
\end{equation}
The quasienergies $\{\epsilon_k\}_k$ are defined up to integer multiples of the drive frequency \(\omega\), because any shift of the quasienergy by these amounts can always be absorbed by a redefinition of \(u_k\).
Accordingly --- as it happens to the wavevectors of a wavefunction of a particle in a spatially periodic potential --- one can restrict the quasienergies \(\{\epsilon_k\}_k\) to be within a Floquet Brillouin zone (FBZ) defined by having  \(-\omega/2 < \epsilon_k \le \omega/2\).

Note that \(f_k\) and \(f_k^*\) or, equivalently, \(\text{Re}(f_k)\) and \(\text{Im}(f_k)\) are actually two independent solutions of Eq.~\eqref{mot}.
Substituting Eq.~\eqref{ftu} in Eq.~\eqref{ukdef} and then in Eq.~\eqref{mot} 
one obtains the conditions which have to be satisfied by the coefficients $\{ c_m\}_m$:
\begin{multline}\label{cms}
\left[r+k^2-(\epsilon_k+m\omega)^2\right]c_{m}-\frac{r_1}{2}[c_{m-1}+c_{m+1}]=0.
\end{multline}
In order to highlight the structure of the
infinite-dimensional space of these solutions, i.e., the so-called Sambe space \cite{Shirley65,Sambe73}, we rewrite the above equation as follows,
\begin{widetext}
\begin{equation}\label{matr}
\left(
\begin{tabular}
{cccccc} \(\ddots\)&&&&&\(\reflectbox{\(\ddots\)}\) \\ &\(r+k^2-(\epsilon_k-2\omega)^2\)&\(-r_1/2\)&0&0& \\ &\(-r_1/2\)&\(r+k^2-(\epsilon_k-\omega)^2\)&\(-r_1/2\)&0 \\&0&\(-r_1/2\)&\(r+k^2-\epsilon_k^2\)&\(-r_1/2\) \\&0&0&\(-r_1/2\)&\(r+k^2-(\epsilon_k+\omega)^2\) \\ \(\reflectbox{\(\ddots\)}\)&&&&&\(\ddots\)
\end{tabular}
\right)
\times \left(
\begin{tabular}{c}
\(\vdots\)\\\(c_{-2}\)\\\(c_{-1}\)\\\(c_{0}\)\\\(c_{1}\)\\\(\vdots\)
\end{tabular}
\right)=\left(
\begin{tabular}{c}
\(\vdots\)\\0\\0\\0\\0\\\(\vdots\)
\end{tabular}
\right).
\end{equation}
\end{widetext}
For \(f_k\) to have a non-vanishing solution, 
the determinant of the above symmetric and tridiagonal matrix has to vanish. 
This condition determines
the quasienergy \(\epsilon_k\) as a function of \(r\), \(k\), \(\omega\) and \(r_1\). A complex \(\epsilon_k\) corresponds to an unstable
solution (red regions in Fig.~\ref{fig1}), leading to forbidden states or gaps in the parameter space 
spanned by the dimensionless variables
$a$ and $q$ introduced in Eqs.~\eqref{qdef} and \eqref{eq:def-a}. 
As anticipated, we are interested in the
solution near the upper boundary of region \((b)\) in Fig.~\ref{fig1}. This boundary is also the
boundary of the FTC phase, and
is characterized by having $\epsilon_{k=0}=\omega/2$ along the curve $a = a_1(q)$ in Fig.~\ref{fig1},
which corresponds to $r = (\omega/2)^2 a_1(q)$.

In order to proceed with the analysis, we assume that the drive amplitude is small, i.e., \(q\ll 1\).
Accordingly, solving the linear system of equations \eqref{matr} perturbatively in \(q\), the zeroth order
solution with \(q \to 0\) corresponds to \(r=(\omega/2)^2\) and non-zero \(c_{0,-1}\), while the rest of the \(c_{m}\) vanish.
To find the first-order correction in \(q\), it is sufficient to truncate the matrix such that we only keep the \(2\times2\) matrix
corresponding to \(c_0\) and \(c_{-1}\).
By inspecting Eq.~\eqref{matr} it is straightforward to show that the remaining coefficients \(c_{m}\) with \( m\neq 0\), \(-1\)
are smaller than \(c_{0,-1}\) because
\begin{equation}
\begin{split}
c_{-m}&=\frac{O(q)}{O(1)}c_{-(m-1)} = O(q^{m-1})c_0, \quad {\text{for}}\quad m>1,\\
c_{m}&=\frac{O(q)}{O(1)}c_{(m-1)} = O(q^{m})c_0, \quad {\text{for}}\quad m>0.
\end{split}
\label{eq:app-orders}
\end{equation}
Accordingly, at the lowest non-trivial order, one can assume that $c_m=0$ for $m\neq 0, -1$, such that Eq.~\eqref{matr} becomes
\begin{equation}
\left(
\begin{tabular}{cc}
\(r+k^2-(\epsilon_k-\omega)^2\)&\(-r_1/2\)\\\(-r_1/2\)&\(r+k^2-\epsilon_k^2\)
\end{tabular}
\right)\times
\left(\begin{tabular}{c}
\(c_{-1}\)\\\(c_0\)
\end{tabular}\right)
=
\left(\begin{tabular}{c}
\(0\)\\\(0\)
\end{tabular}\right),
\end{equation}
and a non-trivial solution exists only if the determinant of the matrix on the l.~h.~s.~of this equation vanishes. 

There are four values of $\epsilon_k$ which satisfy this condition: 
two of them correspond to \(\epsilon_{k=0}=-\omega/2\) and  \(3\omega/2\) for \(q \to 0\)
and therefore they are relevant only when one enlarges the
matrix in Sambe space in order to account also for these resonances.
The two remaining solutions are related to each other by the simultaneous exchange
\(\epsilon_k \rightarrow -\epsilon_k +\omega\) and \(c_{0}\leftrightarrow c_{-1}\), and hence
they actually represent the same state. 
Accordingly, of the four solutions, only one is physical, and it is given by
\begin{equation}\label{dism}
\epsilon_k=\frac{\omega}{2}+\sqrt{\left(\frac{\omega}{2}\right)^2+r+k^2-\omega\sqrt{r+k^2+\left(\frac{r_1}{2\omega}\right)^2}}.
\end{equation}
Requiring
\(\epsilon_{k=0}=\omega/2\) in order to determine the critical line separating region \((2)\) from region \((b)\)
in Fig.~\ref{fig1}, one finds that such a line corresponds to
\begin{equation}
r =r_c=(\omega/2)^2 a_1(q) \quad\mbox{with}\quad
a_1(q)=1+q+O\left(q^2\right).
\label{critp}
\end{equation}
In fact, one can easily verify that $\epsilon_k$ in Eq.~\eqref{dism} acquires an imaginary part when the parameter $a$ in Fig.~\ref{fig1} is within the interval
$1-q+O\left(q^2\right) < a < 1+q+O\left(q^2\right)$, indicating that region $(b)$ opens up symmetrically and linearly around $a=1$.
This procedure can be systematically generalized to higher-orders of the expansion in $q$ by searching for  solutions of Eq.~\eqref{matr} in terms of an increasing number of non-vanishing coefficients (i.e., of increasingly larger matrices), expected to be of increasing order in $q$ according to Eq.~\eqref{eq:app-orders}. In doing so, for example, one systematically recovers the well-know results (see, e.g., \S2.151 of Ref.~\cite{McLachlin1947}) that the boundaries of region $(b)$ are approximated by $1  - q  - q^2/8 + q^3/64 + O\left(q^4\right) < a < 1  + q  - q^2/8 - q^3/64 + O\left(q^4\right) = a_1(q)$, those of region  
$(c)$ by $4 - q^2/12 + O\left(q^4\right)< a < 4 + 5 q^2/12 + O\left(q^4\right) = a_2(q)$, while those of region $(d)$ by $9 + q^2/16 - q^3/64 + O\left(q^4\right)  < a < 9 + q^2/16  + q^3/64 + O\left(q^4\right) = a_3(q)$, in qualitative agreement with Fig.~\ref{fig1}.

Note that, in the vicinity of the band-edge with $r\simeq r_c$ and for a weak drive \(q\ll 1\),
we can identify several energy scales. These are naturally determined by $k$, $\omega$, and
$\sqrt{r_1}$ where the latter, in terms of the dimensionless variables, can also be expressed as
$\sqrt{2 r_1} = \sqrt{q}\omega$.
Substituting \(r=r_c\) in Eq.~\eqref{dism} and expanding for small momenta \(k\ll\omega\), two natural regimes of values of $k$
emerge. One for \( \sqrt{q} \omega \ll k\ll \omega\), and the other for \(k\ll \sqrt{q} \omega \ll \omega\).
In these two cases, in terms of
the renormalized momentum
\begin{align}
\bar{k} = \sqrt{\frac{q}{2}}k, \label{kbar}
\end{align}
the following dispersion emerges (see Appendix \ref{disD} for details), 
\begin{subequations}\label{dis1}
\begin{align}
\epsilon_k 
&\simeq
\frac{\omega}{2} + \bar{k} &\mbox{for} \quad k\ll \sqrt{q} \omega \ll \omega,\label{dis1a}\\
\epsilon_k &\simeq \frac{\omega}{2} +\frac{k^2}{\omega} &\mbox{for} \quad \sqrt{q}\omega \ll k\ll \omega,
\label{dis1b}\\[2mm]
 \epsilon_k &\simeq k &\mbox{for} \quad \sqrt{q}\omega \ll \omega \ll k. \label{dis1c}
\end{align}
\end{subequations}
Further below, in Sec.~\ref{corr}, we will use these expressions in order to determine
the correlation function in the long-wavelength limit \(k\ll \sqrt{q} \omega \ll \omega\).
We therefore substitute the value of \(\epsilon_k\)
in Eq.~\eqref{dis1} into Eq.~\eqref{matr}, and solve for \(c_{m}\), obtaining,
\begin{subequations}
\begin{align}
c_{-1}&= \frac{r_1/2}{(\omega/2)^2+r_1/2+k^2-(\epsilon_k-\omega)^2}c_0 \label{c10gen}\\
&\simeq  \left[1-\frac{4\bar{k}}{q\omega}+\frac{4(2+q)\bar{k}^2}{q^2\omega^2}+O\left((\bar{k}/q\omega)^3\right)\right]c_0,\label{ksm}\\
c_{1}&= \frac{r_1/2}{(\omega/2)^2+r_1/2+k^2-(\epsilon_k+\omega)^2}c_0\nonumber \\ &=\left[-\frac{q}{8}+O\left(q^2\right)\right]c_0,\label{cr1}\\
c_{-2}&= \frac{r_1/2}{(\omega/2)^2+r_1/2+k^2-(\epsilon_k-2\omega)^2}c_{-1}\nonumber \\
&= \left[-\frac{q}{8}+O\left(q^2\right)\right]c_0. \label{cr2}
\end{align}
\end{subequations}
While Eq.~\eqref{c10gen} holds for generic momenta, Eq.~\eqref{ksm} assumes long wavelengths, i.e.,  \(k\ll \sqrt{q}\omega\ll \omega\), 
corresponding to $\bar k/(q\omega) \ll 1$.

The equation of motion \eqref{mot} is real and therefore, up to a multiplicative factor, we can choose the real and imaginary parts of \(f_k\) as its two independent solutions. Accordingly, $M_{c,k}(t)$ and  $M_{s,k}(t)$ in  Eq.~\eqref{Phi1} can be taken proportional to these functions, i.e.,
\begin{subequations}\label{Mcsdef}
\begin{align}
M_{c,k}(t)=2\alpha_k \textrm{Re}\left[f_k(t)\right],\\
M_{s,k}(t)=2\beta_k \textrm{Im}\left[f_k(t)\right],
\end{align}
\end{subequations}
with the initial condition given in Eq.~\eqref{Mcsin}. The real coefficients $\alpha_k$ and $\beta_k$ 
in this expression are going to be determined explicitly in Appendix~\ref{microAp}. For the discussion below their actual expressions are not needed. 

From Eqs.~\eqref{Mcsdef}, \eqref{ukdef}, and \eqref{ftu} we can write
\begin{subequations}\label{McMs}
\begin{align}
M_{c,k}(t)&=\sum_{m=-\infty}^{\infty} b_{m} \cos{((\epsilon_k+m\omega)t}), \label{Mc}\\
M_{s,k}(t)&=\sum_{m=-\infty}^{\infty} d_{m} \sin{((\epsilon_k+m\omega)t}),
\label{Ms}
\end{align}
\end{subequations}
where
\begin{equation}
b_{m}=2\alpha_k c_{m} \quad \textrm{and} \quad d_{m}=2\beta_k c_{m}.
\end{equation}

Ignoring terms with \(m\neq 0,-1\), as implied by Eqs.~\eqref{cr1} and \eqref{cr2}, we obtain
\begin{align}
M_{c,k}(0)&=b_0+b_{-1}\approx\left[ 1+\left(1-\frac{4\bar{k}}{q\omega}\right) \right]b_0,
\end{align}
and thus the initial condition $M_{c,k}(0)=1$ can be used to determine $b_0$ and, via 
Eq.~\eqref{ksm}, \(b_{-1}\) as
\begin{align}
& b_0 \approx\frac{1}{2}+\frac{\bar{k}}{q\omega},\label{b0e}\\
\mbox{and}\quad & b_{-1} \approx\frac{1}{2}-\frac{\bar{k}}{q\omega}.
\label{b1e}
\end{align}
Similarly, from Eq.~\eqref{Ms} one has,
\begin{align}
\dot{M}_{s,k}(0)&= \epsilon_k d_0+(\epsilon_k-\omega)d_{-1}\nonumber \\
&\approx\left[\left(\frac{\omega}{2}+\bar{k}\right)+\left(-\frac{\omega}{2}+\bar{k}\right)\right. \nonumber \\
&\ \ \ \ \ \ \ \times \left.\left(1-\frac{4\bar{k}}{q\omega}+\frac{4(2+q)\bar{k}^2}{q^2\omega^2}\right) \right]d_0,
\end{align}
and the initial condition $\dot M_{s,k}(0)=1$ (see Eq.~\eqref{Mcsin}) together with 
Eq.~\eqref{ksm} can be used in order to determine
\begin{align}
& d_0 \approx \left(\frac{1}{1+q}\right)\frac{1}{\omega}\biggl[\frac{q\omega}{ 2\bar{k}} +1 \biggr],\label{d0e}\\
\text{and}\quad  & d_{-1} \approx \left(\frac{1}{1+q}\right)
\frac{1}{\omega}\biggl[\frac{q\omega}{2\bar{k}}-1\biggr].\label{d1e}
\end{align}
In the expressions above for $d_{0,-1}$ we kept the first two terms in the expansion in $\bar{k}/(q\omega)$, while in the overall multiplicative factor we kept the complete dependence on $q$ in order to satisfy Eq.~\eqref{Mcan} at $t=0$.
Here we note that if we could solve the Floquet problem exactly, then the canonical commutation
relation for the fields, and in particular Eq.~\eqref{Mcan}, would be obeyed exactly at all times. 
Since we have solved the problem 
perturbatively in \(q\), the canonical commutation relation, which 
we imposed
exactly at \(t=0\),
is violated at longer times: for example, as shown in Appendix \ref{commr}, this violation at long wavelengths is given by \(\left[\phi(t),\Pi(t)\right] = 1 + O(q) \times \sin^2(\omega t/2)\). This violation can be reduced to higher powers of $q$ by keeping higher-order terms in Sambe space, i.e., by approximating the solution with larger matrices. 

Before continuing, let us briefly discuss the solution of Eq.~\eqref{mot} for \(k=0\)  and its 
connection with the Schr\"{o}dinger cat states.
In the absence of the drive, i.e., with \(r_1=0\), the modes at \(k=0\) are
\(\phi_1= e^{i\sqrt{r} t} = e^{i\omega t/2}\) and \(\phi_2= e^{-i\sqrt{r}t}= e^{-i\omega t/2}\).
In the presence of a weak drive \(q\ll 1\), from Eqs.~\eqref{ftu}, \eqref{cms}, and \eqref{c10gen}
we find at order $q^0$ that
\begin{align}
f_{k=0}=c_0 e^{i\omega t/2} \biggl[1+e^{-i\omega t}\biggr].
\end{align}
The two independent solutions of the equation are provided by the real and imaginary parts of \(f_k\), which, in the limit of weak drive,
are simply given by the symmetric and anti-symmetric combinations of \(\phi_{1}\) and \(\phi_2\), i.e.,
\begin{align}
  M_{c,s,k=0} = c_0
\bigl(\phi_1 \pm \phi_2\bigr).\label{fq0}
\end{align}
Now consider the many-particle problem with $N$ bosons, 
which would involve macroscopically occupying 
the two modes $\phi_{1,2}$. 
Denoting by $|N_1,N_2\rangle$ 
the 
Fock state in which $N_{1,2}$ bosons occupy the $\phi_{1,2}$ orbitals, the many-particle eigenstates in Fock space,
denoted as $|M\rangle_{c,s}$ become \cite{Sacha18}, 
\begin{align}
|M\rangle_{c,s} = \frac{|N,0\rangle \pm |0,N\rangle}{\sqrt{2}}.
\end{align}
Thus the many-particle eigenstates are Schr\"{o}dinger cat states of the unperturbed orbitals 
corresponding to symmetric and anti-symmetric combinations of $|N,0\rangle$ with $|0,N\rangle$. The broken symmetry state
therefore occurs when the many-particle system spontaneously, or as a result of a measurement, chooses to be in $|N,0\rangle$ or $|0,N\rangle$, with the limit
$N\rightarrow \infty$ stabilizing the broken symmetry state by suppressing tunneling from $|N,0\rangle$ to $|0,N\rangle$,
see Ref.~\onlinecite{Sacha18} for further details.

\section{Correlation Functions} \label{corr}

We will now present the predictions for the time-dependent correlation functions of the position and momentum fields. 
Let us briefly discuss what to expect.
While in thermal equilibrium all correlation functions are time translationally invariant (TTI), we do not expect this to be the case in the presence of the driving, because these functions will show
period-doubling in the FTC, and period synchronization in the trivial phase.
Secondly, just as in thermal equilibrium a trivial phase is characterized by the absence of long-range order and by correlations that extend across short distances in space, we expect a similar behavior here for the trivial phase. 
Thirdly, in thermal equilibrium, a broken-symmetry phase generically features 
long-range order and correlations which become long-ranged in space upon approaching the critical line separating it from the trivial phase.
Accordingly, one expects the FTC to also show long-range 
order \cite{Else2019,Khemani2019} and critical correlations.

The unexplored issue we would like to address here concerns how the transition from
the non-trivial to the FTC phase actually occurs. If this transition is continuous,
then we expect the correlations at the critical point to decay algebraically in space,
leading to scaling and universality. 
We also expect that detuning the system slightly away from the critical point and towards the trivial phase will introduce another length scale into the system which will cut off the critical power-law spatial decays.
We explore this physics below in the vicinity of the transition between the FTC and trivial phase, within the Gaussian approximation.

To this end, in this section we shall first derive the expressions of the correlations at the critical line, i.e., along the line
\(r=r_c\) which bounds the edge of band \((2)\) in Fig.~\ref{fig1}.
Following this, we shall study how the correlations decay at large distances for a non-zero detuning away from the critical line, within the trivial phase (green region (2) in Fig.~\ref{fig1}).

\subsection{Correlation functions along the critical line}
\label{subs:critical-corr-fun}

Using Eqs.~\eqref{Pi1}, \eqref{Phi1}, and the solution for \(M_{c,k}\) and \(M_{s,k}\) obtained in the previous section, we find that at the critical line, for small drive amplitude \(q\ll 1\), the longest wavelength modes (\(k\ll \sqrt{q}\omega\ll \omega \))
of the fields $\phi_\textbf{k}$ and $\Pi_\textbf{k}$ 
evolve as follows:
\begin{align}
  \phi_\textbf{k}(t)=& \biggl[ \cos\left(\frac{\omega}{2}t\right)\cos\left(\bar{k} t\right)\biggr] \phi_\textbf{k}(0)\nonumber\\ &
  +\biggl[\frac{q}{\bar{k}}\cos\left(\frac{\omega}{2}t\right)\sin\left(\bar{k} t\right)\biggr]\Pi_\textbf{k}(0),\label{phi_s}\\
\Pi_\textbf{k}(t)=& \biggl[ -\frac{\omega}{2}\sin\left(\frac{\omega}{2}t\right)\cos\left(\bar{k} t\right)\biggr]\phi_\textbf{k}(0)\nonumber\\
&+\biggl[ -\frac{\omega}{2}\frac{q}{\bar{k}}\sin\left(\frac{\omega}{2}t\right)\sin\left(\bar{k} t\right)\biggr]\Pi_\textbf{k}(0).
\label{pi_s}
\end{align}
For a deep quench with \(\beta \sqrt{r_0} \gg 1\) and for long wavelengths, i.e.,
\(k\ll \sqrt{r_0}\), the initial correlations are, from Eqs.~\eqref{corrPP0}, \eqref{corrpp0}, and \eqref{corrpP0},
\begin{align}\label{ini}
\langle \Pi_{i,\textbf{k}}(0) \Pi_{j,\textbf{q}}(0) \rangle &= \delta_{i,j}\delta_{\textbf{k},-\textbf{q}} \frac{\omega_{0k}}{2} \approx \delta_{i,j}\delta_{\textbf{k},-\textbf{q}} \frac{\sqrt{r_0}}{2},\nonumber\\
\langle \phi_{i,\textbf{k}}(0) \phi_{j,\textbf{q}}(0) \rangle &= \delta_{i,j}\delta_{\textbf{k},-\textbf{q}} \frac{1}{2\omega_{0k}} \approx \delta_{i,j}\delta_{\textbf{k},-\textbf{q}} \frac{1}{2\sqrt{r_0}},\nonumber\\
\langle \left\{\phi_{i,\textbf{k}}(0),\Pi_{j,\textbf{q}}(0) \right\}\rangle &= 0.
\end{align}
The lack of momentum dependence in the correlations reported above implies that they are very short-ranged in position space, essentially \(\delta\)-functions.

The dynamics of the model is fully characterized in terms of the following 
Keldysh and retarded Green's functions \cite{Kamenevbook}:
\begin{align}
\delta_{lj}\delta_{\textbf{k},-\textbf{q}}iG_{K}^{\phi \phi}(k,t,t') &= \langle \left\{\phi_{l,\textbf{k}}(t),\phi_{j,\textbf{q}}(t') \right\}\rangle,\\
\delta_{lj}\delta_{\textbf{k},-\textbf{q}}iG_{K}^{\Pi \Pi}(k,t,t') &= \langle \left\{\Pi_{l,\textbf{k}}(t),\Pi_{j,\textbf{q}}(t') \right\}\rangle,\\
\delta_{lj}\delta_{\textbf{k},-\textbf{q}}iG_{K}^{\phi \Pi}(k,t,t') &= \langle \left\{\phi_{l,\textbf{k}}(t),\Pi_{j,\textbf{q}}(t') \right\}\rangle,\\
\delta_{lj}\delta_{\textbf{k},-\textbf{q}}iG_{R}^{\phi \phi}(k,t,t') &= \theta(t-t') \langle \left[\phi_{l,\textbf{k}}(t),\phi_{j,\textbf{q}}(t') \right]\rangle,\\
\delta_{lj}\delta_{\textbf{k},-\textbf{q}}iG_{R}^{\Pi \Pi}(k,t,t') &= \theta(t-t') \langle \left[\Pi_{l,\textbf{k}}(t),\Pi_{j,\textbf{q}}(t') \right]\rangle,\\
\delta_{lj}\delta_{\textbf{k},-\textbf{q}}iG_{R}^{\phi \Pi}(k,t,t') &= \theta(t-t') \langle \left[\phi_{l,\textbf{k}}(t),\Pi_{j,\textbf{q}}(t') \right]\rangle,
\end{align}
which can be easily determined by substituting Eqs.~\eqref{phi_s} and \eqref{pi_s} in the expressions above and by using the explicit expressions for the correlation functions in the initial state. In particular, for the initial conditions in Eq.~\eqref{ini} and for the longest wavelength modes with \(k\ll \sqrt{q}\omega\ll \omega \), one finds the following Keldysh Green's functions:
\begin{align}
iG_{K}^{\phi \phi}(k,t,t') =& q^2\frac{\sqrt{r_0}}{2\bar{k}^2} \cos\left(\frac{\omega}{2}t\right) \cos\left(\frac{\omega}{2}t'\right)  \nonumber\\ &\times [\cos(\bar{k}(t-t'))-\cos(\bar{k}(t+t'))],\label{gkpp}\\
iG_{K}^{\Pi \Pi}(k,t,t') =& q^2\left(\frac{\omega}{2}\right)^2\frac{\sqrt{r_0}}{2\bar{k}^2} \sin\left(\frac{\omega}{2}t\right) \sin\left(\frac{\omega}{2}t'\right)  \nonumber\\
&\times [\cos(\bar{k}(t-t'))-\cos(\bar{k}(t+t'))],\\
iG_{K}^{\phi \Pi}(k,t,t') =&- q^2\frac{\omega}{2}\frac{\sqrt{r_0}}{2\bar{k}^2} \cos\left(\frac{\omega}{2}t\right) \sin\left(\frac{\omega}{2}t'\right)  \nonumber\\
&\times [\cos(\bar{k}(t-t'))-\cos(\bar{k}(t+t'))].
\end{align}
Note that for equal times \(t=t'\), the Keldysh Green's functions $G_K$'s become synchronized with the drive frequency. In order to observe period-doubling, these 
functions have to be evaluated at unequal times \(t\neq t'\).

Similarly, the retarded Green's functions turn out to be:
\begin{align}
G_{R}^{\phi \phi}(k,t,t') =&-\theta(t-t')q \cos\left(\frac{\omega}{2}t\right) \cos\left(\frac{\omega}{2}t'\right) \nonumber\\
&\times \frac{\sin(\bar{k} (t-t'))}{\bar{k}}, \label{grpp}\\
G_{R}^{\Pi \Pi}(k,t,t') =&-\theta(t-t')q\left(\frac{\omega}{2}\right)^2  \sin\left(\frac{\omega}{2}t\right) \sin\left(\frac{\omega}{2}t'\right)
\nonumber\\
& \times \frac{\sin(\bar{k} (t-t'))}{\bar{k}},\\
G_{R}^{\phi \Pi}(k,t,t') =&\theta(t-t')q\frac{\omega}{2} \cos\left(\frac{\omega}{2}t\right)  \sin\left(\frac{\omega}{2}t'\right)
\nonumber\\
&\times \frac{\sin(\bar{k} (t-t'))}{\bar{k}}.
\end{align}
These quantities also show period-doubling at unequal times \(t> t'\) while, due to causality, they vanish at $t\leq t'$.
Appendix \ref{ON} provides the corresponding expressions for a critical quench of the undriven \(O(N)\) model. For convenience, we report here only the correlators of the \(\phi\) fields:
\begin{align}
iG_{K,u}^{\phi \phi}(k,t,t') &= \frac{\sqrt{r_0}}{2k^2} [\cos(k(t-t'))-\cos(k(t+t'))],\label{GKst}\\
G_{R,u}^{\phi \phi}(k,t,t') &=-\theta(t-t')\ \frac{\sin(k (t-t'))}{k},\label{GRst}
\end{align}
where the subscript \(u\) here and below denotes that the quantity has been calculated for the undriven model.

Comparing the driven with the undriven case, one finds that they are related via 
\begin{align}
  G_{K}^{\phi \phi}(k,t,t') = q^2
  \cos\left(\frac{\omega}{2}t\right) \cos\left(\frac{\omega}{2}t'\right)G_{K,u}^{\phi \phi}(\bar{k},t,t'),\label{gkcom}\\
  G_R^{\phi \phi}(k,t,t') = q\cos\left(\frac{\omega}{2}t\right) \cos\left(\frac{\omega}{2}t'\right)
  G_{R,u}^{\phi \phi}(\bar{k},t,t').\label{grcom}
\end{align}
Note that the driven correlators cannot be obtained from the undriven ones by simply setting the drive
amplitude \(q\) to zero. This signals that the parametric resonance generated by the drive is in fact a non-analytic effect in the drive
amplitude as setting the drive amplitude to zero, does not render the correlators of the undriven problem. Moreover, the temporal behavior in the presence of the drive is more complicated than in its absence, due to the appearance of the energy scale \(\omega/2\), as seen explicitly in Eqs.~\eqref{gkcom} and \eqref{grcom}.
However, both the driven and undriven \(G^{\phi \phi}_{K,R}\) correlators feature an algebraic prefactor of the form
\(1/k^{2},1/k\) in momentum space, with the difference being that the momenta for the driven case become renormalized according to
\(k \rightarrow \bar{k}\), see Eq.~\eqref{kbar}. 
As we discuss in detail further below in connection with the emergence of a light-cone in the 
dynamics, this algebraic dependence in momentum space results in a power-law decay of spatial correlations at large distances.

Despite these similarities in the spatial behavior of \(G^{\phi\phi}\), those of
\(G^{\Pi \Pi}\) and \(G^{\phi \Pi}\) are markedly different for the driven and undriven cases.
In particular, the drive makes \(G^{\Pi \Pi}\) and \(G^{\phi \Pi}\) more singular as \(k \rightarrow 0\), i.e., they 
 diverge more rapidly upon decreasing 
 $k$ towards zero, 
 implying that their behavior is more long-ranged in space compared to the undriven case.
The reason for this difference is that the \(\langle \phi \Pi\rangle, \langle \Pi \Pi\rangle\) correlators are obtained from the \(\langle \phi \phi\rangle\) correlator by taking time derivatives since \(\Pi=\dot{\phi}\).
Because the driven case has time-dependent oscillations at the momentum-independent scale \(\omega/2\), this leads to \(\langle \phi \Pi\rangle, \langle \Pi \Pi\rangle\) correlators which
are as singular as the \(\langle \phi \phi\rangle\) correlator. 
The physical reason of this longer-range order in the presence of the drive in comparison to the undriven case is the non-trivial spatio-temporal order of the FTC. The latter has
long-range order in space which is accompanied by precise period-doubled dynamics.

In the absence of driving, the correlation functions after a quench onto a critical point are know to feature a universal temporal behavior \cite{Tavora2015,Maraga2015,Chiocchetta2016}. In order to explore the possible similarities with that case, let us consider here the limits of short and long times, focusing on the \(\langle \phi \phi\rangle\) correlators.
In particular, let us first consider short times
\(t,t'\ll k^{-1}\), at which Eqs.~\eqref{gkpp} and \eqref{grpp} give
\begin{align}
iG_{K}^{\phi\phi}(k,t,t')&=\cos\left(\frac{\omega}{2}t\right) \cos\left(\frac{\omega}{2}t'\right)q^2\sqrt{r_0} tt',\\
G_{R}^{\phi \phi}(k,t,t')&=-\theta(t-t')\cos\left(\frac{\omega}{2}t\right) \cos\left(\frac{\omega}{2}t'\right) q (t-t').
\end{align}
When compared with the results for the undriven case, the difference is the appearance of the prefactors as summarized in Eqs.~\eqref{gkcom} and \eqref{grcom}.

At this point we can speculate on the effects of accounting for interactions, based on
our knowledge of how they affect the short-time behavior in the undriven case~\cite{Tavora2015,Maraga2015,Chiocchetta2016}.
We expect that for \(t/t'\gg 1\) algebraic behaviors \(G_K \propto (tt')^{1-\theta}\) and \(G_R\propto t(t'/t)^{\theta}\) will emerge in these two quantities, where \(\theta\) is a universal initial-slip exponent, which vanishes in the absence of interactions. 
The Gaussian results presented above are consistent with these limiting forms of the Green's functions. Accordingly, as long as the correlators for the driven and undriven cases are related as in 
Eqs.~\eqref{gkcom} and \eqref{grcom}, we speculate that interactions will anyhow lead to the appearance of an initial-slip exponent. 

Drawing further analogies between the driven and undriven problem, this initial-slip exponent $\theta$ is also expected to modify the steady-state behavior of \(G_K^{\phi \phi}\) in Eq.~\eqref{gkpp} by changing the algebraic prefactor \(\bar{k}^{-2}\) into \(\bar{k}^{-2+2\theta}\) in the presence of interactions.
The analysis of the effects of interactions will be reported elsewhere \cite{Natsheh2020b}.

\subsection{Average dynamics along the critical line}

To further emphasize the difference between the undriven and driven critical points, 
we now discuss the long-time limit of the dynamics, focusing on $G_{K,R}^{\phi \phi}$. 
In order to access this limit, let us define the time difference \(\tau = t-t'\) and the mean time \(T_m=\frac{1}{2}(t+t')\). Then, from Eq.~\eqref{gkpp}, we find
\begin{align}
iG_{K}^{\phi \phi}(k,\tau,T_m) &= \biggl[\cos\left(\frac{\omega}{2} \tau\right)+\cos(\omega T_m)\biggr]\nonumber\\
&\times q^2\frac{\sqrt{r_0}}{4\bar{k}^2}\biggl[\cos(\bar{k}\tau)+\cos(2\bar{k} T_m)\biggr].
\end{align}
Similarly, from Eq.~\eqref{grpp} for the retarded Green's function we obtain,
\begin{align}
G_{R}^{\phi \phi}(k,\tau,T_m) &= -\theta(\tau)\left[\cos\left(\frac{\omega}{2} \tau\right)+\cos(\omega T_m)\right]\nonumber\\
&\times \frac{q}{2}\frac{\sin(\bar{k}\tau)}{\bar{k}}.
\end{align}
Due to the presence of the drive, these expressions are generically not TTI, indicating that the long-time limit of the dynamics is necessarily non-stationary. However, if one is interested in the behavior of the system at time scales much longer than the period of the drive a sort of average behavior can be identified by time-averaging $G_{K}^{\phi \phi}(k,\tau,T_m)$ and $G_{R}^{\phi \phi}(k,\tau,T_m) $ over the mean time $T_m$. The respective averages \(\bar{G}^{\phi \phi}_{K}(k,\tau)\) and \(\bar{G}^{\phi \phi}_{R}(k,\tau)\) turn out to be
\begin{align}
  i\bar{G}_{K}^{\phi \phi}(k,\tau)
  &
  = q^2 \frac{\sqrt{r_0}}{8\bar{k}^2} \biggl[\cos\left(\left(\bar{k}+\frac{\omega}{2}\right)\tau\right)
  \nonumber\\ &    
+\cos\left(\left(\bar{k}-\frac{\omega}{2}\right)\tau\right)\biggr],
\end{align}
and
\begin{align}
  \bar{G}_{R}^{\phi \phi}(k,\tau)
  &=-\theta(\tau)\frac{q}{4} \frac{1}{\bar{k}}\biggl[\sin\left(\left(\bar{k}+\frac{\omega}{2}\right)\tau\right)\nonumber\\
&    +\sin\left(\left(\bar{k}-\frac{\omega}{2}\right)\tau\right)
\biggr].
\end{align}
By Fourier transforming $\bar{G}_{K}^{\phi \phi}(k,\tau)$ in the time difference $\tau$ we obtain,
\begin{align}
\bar{G}_{K}^{\phi \phi}(k,\nu)&=\int d\tau e^{i\nu \tau} \bar{G}_{0K}(k,\tau)\nonumber\\
&=-iq^2\frac{2\pi\sqrt{r_0}}{16\bar{k}^2}\biggl[\delta\left(\bar{k}-\frac{\omega}{2}-\nu\right)+\delta\left(\bar{k}-\frac{\omega}{2}+\nu\right)
\nonumber\\
&+\delta\left(\bar{k}+\frac{\omega}{2}-\nu\right)+\delta\left(\bar{k}+\frac{\omega}{2}+\nu\right)\biggr].\label{GKw}
\end{align}
Similarly, by taking the Fourier transform of $\bar{G}_{R}^{\phi \phi}(k,\tau)$, one can calculate   its imaginary part as
\begin{align}
\bar{G}_{R}^{\phi \phi}(k,\nu)-\bar{G}_{R}^{\phi \phi}(k,-\nu)&=-iq \frac{2\pi}{8\bar{k}}
\biggl[\delta\left(\bar{k}-\frac{\omega}{2}-\nu\right)
\nonumber\\
&-\delta\left(\bar{k}-\frac{\omega}{2}+\nu\right)\nonumber\\
&+\delta\left(\bar{k}+\frac{\omega}{2}-\nu\right)\nonumber\\
&-\delta\left(\bar{k}+\frac{\omega}{2}+\nu\right)\biggr].\label{GRw}
\end{align}
The \(\delta\) functions in the previous expression show that while for the undriven case, dissipation occurs when the external frequency \(\nu\)
is resonant with the single-particle
excitation energy \(\epsilon_k \simeq k\), for the driven problem this condition is shifted by \(\pm \omega/2\), as expected.

The fluctuation-dissipation theorem states that in thermal equilibrium at temperature $\beta^{-1}$, the Keldysh Green's function $G_K$ (quantifying fluctuations)
and the imaginary part of the
retarded Green's function $G_R$ (quantifying dissipation),
are related to the temperature \(\beta^{-1}\) as
\begin{align}
G_{K}(k,\nu) &= \coth\left(\frac{\beta\nu}{2}\right)\left[G_{R}(k,\nu)-G_{R}(k,-\nu)\right]\nonumber\\
&\approx \frac{2}{\nu \beta}\left[G_{R}(k,\nu)-G_{R}(k,-\nu)\right].\label{FDT}
\end{align}
On the second line, we assumed the frequency $\nu$ to be small compared with the temperature $\beta^{-1}$, i.e., \(\beta \nu \ll 1\). 

Since our system is inherently out of equilibrium and has no actual stationary state, there is no well-defined temperature in the problem.
However, as it happens in a number of classical and quantum statistical systems out of equilibrium \cite{Cugliandolo2011,Foini2011b,Foini2012}, 
effective temperatures may emerge under certain limits. For example
in the undriven problem \cite{Tavora2015,Chiocchetta2016}, an effective temperature which equals the energy injected during the initial quench, indeed emerges when the system is probed at low frequencies and long wavelengths. 
However, in the driven problem, no effective temperature clearly emerges in the long-wavelength limit (although an effective temperature may emerge at shorter wavelengths).
Studies of driven systems often show a behavior in which the nonequilibrium steady-state is  characterized better as a state with net entropy production \cite{Dehghani16} than in terms of an effective temperature.

\subsection{Magnetization dynamics along the critical line}
\label{magdyn}

In the previous sections, we studied the quench dynamics when the system is initially prepared in the thermal state of a
Hamiltonian which is symmetric in the field components. As a consequence, 
the one-point correlation function of the order parameter, i.e., the magnetization, vanishes initially and therefore it does so also at subsequent times during the time-evolution.

In this section, we will study the dynamics of the magnetization when we explicitly break the \(O(N)\) symmetry by applying an initial external field
\(h_0\) in the direction of a field component, e.g., the one corresponding to \(i=1\). 
Accordingly, the pre-quench Hamiltonian is the static \(O(N)\) model with a large detuning $r_0$ as before (i.e., a short correlation length) and, in addition, also a non-zero magnetic field:
\begin{multline}
H_0=\sum_{i=1}^{N} \int d^d x \frac{1}{2}\Big[r_0\phi_i^2(\textbf{x}) +(\vec{\nabla}\phi_i)^2 +\Pi^2_i(\textbf{x})  \\
- 2h_0 \delta_{1i} \phi_i(\textbf{x}) \Big].
\end{multline}
Defining the magnetization as
\begin{align}
M(t) = \langle\phi_{i=1}(\textbf{x},t)\rangle=\frac{1}{V}\langle\phi_{i=1,\textbf{k}=0}(t)\rangle
\end{align}
where \(V\) is the volume, its initial value is therefore given by
\begin{equation}
M(0) = m_0 = h_0/r_0.
\end{equation}
The time-evolution of all the \(N\) field components obey Eq.~\eqref{Phi1} and
here we focus on the case in which \(H\) is tuned near the critical line.
Using the \(k\to 0\)
limits of the expressions in Eqs.~\eqref{dis1}, \eqref{b0e}, \eqref{b1e}, \eqref{d0e}, and \eqref{d1e} we obtain
\begin{align}
M_{c,k=0}= \cos(\omega t/2),
\end{align}
while
\begin{align}
M_{s,k=0} =
\frac{2}{\omega\left(1+q\right)} \left[\sin\left(\frac{\omega t}{2}\right) 
+q\frac{\omega t}{2} \cos\left(\frac{\omega t}{2}\right)\right].
\end{align}
These expressions can be used to derive the time evolution of the magnetization:
\begin{align}
M(t) = m_0 M_{c,k=0}(t) \approx m_0 \cos(\omega t/2).
\end{align}
Accordingly, we find that the initial non-zero magnetization \(m_0\) evolves in time and features period-doubling at the critical line. 
The corresponding dynamics near the
critical point of the undriven model is easily deduced from, c.f.,  
Eq.~\eqref{Phi1s}, finding that \(M(t) = m_0\), i.e., the order parameter of the Gaussian model in the absence of drive does not evolve after a quench to the critical point.

\subsection{Correlation functions close to the critical line}

In the previous sections we studied the quench dynamics where the parameters of the post-quench Hamiltonian were tuned to be exactly on the critical line $r=r_c$
parameterized by Eq.~\eqref{critp}.
In this section we consider the case of a slight detuning, i.e., \(r = r_c + \Delta_r\), where \(0<\Delta_r \ll r_1\) so that the system is anyhow in the stable phase corresponding to region (2) of Fig.~\ref{fig1}. We also assume that the whole set of fluctuation modes from $k=0$ to $k=\Lambda$ are within the same stable region.
The dispersion relation corresponding to this slight detuning can be determined as explained in Sec.~\ref{FBth} for the case $\Delta_r=0$, finding
\begin{equation}
\epsilon_k = \frac{\omega}{2} + \overline{\omega_k}, 
\end{equation}
where
\begin{equation}
\overline{\omega_k} = \left(\frac{q}{2}\right)^{1/2}\omega_k\quad \mbox{with}\quad \omega_k= \sqrt{\Delta_r +k^2}.
\label{eq:def-w-k-det}
\end{equation}
Above we have also assumed $k \ll \sqrt{q}\omega \ll \omega$.
Note that \(\overline{\omega_k} \rightarrow \bar{k}\) for \(\Delta_r=0\), as expected. 

Similarly, the coefficients entering Eqs.~\eqref{Mc} and \eqref{Ms} for $m=0$ and $-1$ are found to be
\begin{align}
b_0 \approx\frac{1}{2}+\frac{\overline{\omega_k}}{q\omega};\quad\quad
b_{-1} \approx\frac{1}{2}-\frac{\overline{\omega_k}}{q\omega},\\
d_0 \approx \frac{q}{2\overline{\omega_k}};\quad\quad
d_{-1}\approx \frac{q}{2\overline{\omega_k}}-\frac{2}{\omega}.
\end{align}
Using these expressions and by repeating the analysis outlined in Sec.~\ref{subs:critical-corr-fun}, the Keldysh Green's function $G_{K}^{\phi \phi}$ turns out to be
\begin{multline}
iG_{K}^{\phi \phi}(k,t,t') = \cos\left(\frac{\omega t}{2}\right)\cos\left(\frac{\omega t'}{2}\right)\\ \times 
\biggl[\frac{\overline{K_{+}} \cos(\overline{\omega_k}(t-t'))+\overline{K_{-}} \cos(\overline{\omega_k}(t+t'))}{\overline{\omega_k}}
\biggr],
\end{multline}
while the retarded Green's function $G_{R}^{\phi \phi}$ is
\begin{multline}
G_{R}^{\phi \phi}(k,t,t') = -\theta(t-t') q\cos\left(\frac{\omega t}{2}\right)\cos\left(\frac{\omega t'}{2}\right) \\ \times
\frac{\sin(\overline{\omega_k}(t-t'))}{\overline{\omega_k}},
\end{multline}
with
\begin{equation}
\overline{K_{\pm}} = \frac{1}{2}\left(\frac{\overline{\omega_k}}{\omega_{0k}} \pm q^2 \frac{\omega_{0k}}{\overline{\omega_k}}\right),
\end{equation}
where \(\omega_{0k}\) is the pre-quench dispersion defined in Eq.~\eqref{om0k}. 
We emphasize that the expressions above are obtained for long wavelengths \(k \ll \sqrt{r_1}  \ll \omega\).

For comparison, consider again the undriven case for which the corresponding correlators are \cite{Tavora2015,Chiocchetta2016}
\begin{align}
iG_{K,u}^{\phi \phi}(k,t,t') &=  \frac{K_{+} \cos(\omega_{k}(t-t'))+K_{-} \cos(\omega_{k}(t+t'))}{\omega_{k}},\\
G_{R,u}^{\phi \phi}(k,t,t') &= -\theta(t-t') \frac{\sin(\omega_{k}(t-t'))}{\omega_{k}},\\
K_{\pm} &= \frac{1}{2}\left(\frac{\omega_{k}}{\omega_{0k}} \pm \frac{\omega_{0k}}{\omega_{k}}\right),
\end{align}
where $\omega_k$ is defined in Eq.~\eqref{eq:def-w-k-det} and $\omega_{0k}$ in Eq.~\eqref{om0k}. 
Comparing the driven with the undriven correlators, we see that the main difference between them is the period-doubled behavior in the unequal time
correlators of the former. 
However, both of them show the emergence of a length scale corresponding to the
inverse detuning, i.e., to \(\Delta_r^{-1}\), which is responsible for
cutting off the algebraic decay at large distances.

\subsection{Light-cone dynamics along the critical line}

In this section we discuss the real-space and real-time behavior of the critical correlation functions.
Performing a Fourier transform of their expression $G_{R,K}(k,t,t')$ in momentum space, the correlators in  real space with \(d\) dimensions are given by~\cite{Chiocchetta2016}
\begin{align}
&G_{R,K}(x,t,t') =
  \frac{1}{(2\pi)^{d/2} x^{d/2-1}}\nonumber\\
&\quad\quad \times \int_0^{\Lambda} \!\!dk\, k^{d/2}  J_{d/2-1}(kx)\ G_{R,K}(k,t,t'),
\label{eq:k-to-x}
\end{align}
where \(J_{\alpha}\) is the Bessel function of the first kind arising from the angular integration.

We focus below on the \(\langle \phi \phi \rangle\) correlators $G_{K,R}^{\phi\phi}(x,t,t')$ as the other relevant correlators $G_{K,R}^{\Pi\phi}$ and $G_{K,R}^{\Pi\Pi}$ involving the field \(\Pi\) can be obtained by taking suitable time derivatives. 
As discussed above, the undriven and the driven correlators in momentum space, at the critical line and for long wavelengths \(k \ll \sqrt{q} \omega\), are related by Eqs.~\eqref{gkcom} and \eqref{grcom}.
In turn, after defining
\begin{align}
  \bar{x} = \sqrt{\frac{2}{q}}x,
\end{align}
they imply the following relationship between the real-space correlators of the driven and undriven model:
\begin{align}
 G_{K}^{\phi \phi}(\sqrt{q} \omega x\gg 1,t,t') =& 4 \left(\frac{q}{2}\right)^{2-d/2}  \cos\left(\frac{\omega}{2}t\right) \cos\left(\frac{\omega}{2}t'\right)\nonumber\\
&\quad\quad\times G_{K,u}^{\phi \phi}(\bar{x},t,t'), \label{gkcomr2}\\
  G_R^{\phi \phi}(\sqrt{q} \omega x\gg 1,t,t') =& 2 \left(\frac{q}{2}\right)^{1-d/2} \cos\left(\frac{\omega}{2}t\right) \cos\left(\frac{\omega}{2}t'\right)\nonumber\\
&\quad\quad\times   G_{R,u}^{\phi \phi}(\bar{x},t,t'). \label{grcomr2}
\end{align}
The reason for the condition \(\sqrt{q} \omega x \gg 1\) is that the relations \eqref{gkcom} and \eqref{grcom} between the driven and undriven correlators hold for the longest wavelength modes for which the dispersion is given by Eq.~\eqref{dis1}. This places constraints on the spatial distances at which driven and undriven correlators will show the same algebraic decays at large distances,  as summarized in Eqs.~\eqref{gkcomr2} and \eqref{grcomr2}.

The behavior of the undriven correlators \(G_{K,u}^{\phi \phi}\) and \(G_{R,u}^{\phi \phi}\) were discussed in Ref.~\cite{Chiocchetta2016}, where ballistically propagating quasiparticles with a certain velocity \(v\) were shown to give rise, as expected~\cite{Calabrese2005,Foini2012,Marcuzzi2014}, to a light- cone. 
For \(G_{R,u}^{\phi \phi}(x,t,t')\), the light-cone occurs at \(x\approx v|t-t'|\), while for
\(G_{K,u}^{\phi \phi}(x,t,t')\), the light cone occurs at \(x \approx v(t+t')\) and \(x \approx v|t-t'|\), where $v=1$ within the present model. 
In particular, \(G_{K,u}^{\phi \phi}(x,t,t)\) shows a single light-cone at $x\approx 2 v t$.
In addition, the correlators for large distances $x$ show qualitatively different
power-law decays outside, on, and inside the light-cone.

From Eqs.~\eqref{gkcomr2} and \eqref{grcomr2}, the driven problem also shows a similar light-cone
behavior with the difference that the velocity at which the light-cone occurs 
is significantly reduced from \(v\) to \(\sqrt{q/2} v \ll v\).
In particular, \(G_K\) at equal times behaves as follows
\begin{align}
  iG_K^{\phi \phi}(\bar{x} \gg 2t) &\approx 0,\nonumber\\
  iG_K^{\phi \phi}(\bar{x}=2t) &\propto q^{2-d/2}
\cos^2\left(\frac{\omega}{2}t\right)\frac{1}{\left(\bar{\Lambda} \bar{x}\right)^{(d-1)/2}} ,\nonumber\\
  iG_K^{\phi\phi}(\bar{x}\ll 2t) & \propto  q^{2-d/2}
\cos^2\left(\frac{\omega}{2}t\right)\frac{1}{\left(\bar{\Lambda}\bar{x}\right)^{d-2}}.\label{gkx}
\end{align}
Note that \(G_K\) at equal times does not show period doubling, but it is synchronized with the drive. 
In the previous equation we introduced
\(\bar{\Lambda} = \sqrt{q/2}\Lambda\) such that \(\bar{\Lambda}\bar{x}=\Lambda x\). 
Thus, while the algebraic decays at large distances are the same for the
undriven and the driven \(\langle \phi \phi \rangle\) correlators, 
the transition between the various regions of the light-cone is characterized by the renormalized velocity \(\sqrt{q/2}\).

The light-cone behavior for \(G_R^{\phi \phi}\) is, instead, 
\begin{align}
iG_R^{\phi \phi}(\bar{x} \gg |t-t'|) &\approx 0,\nonumber\\
iG_R^{\phi \phi}(\bar{x}=|t-t'|) & \propto
q^{1-d/2}\cos\left(\frac{\omega}{2}t\right) \cos\left(\frac{\omega}{2}t'\right)\nonumber\\
&\quad \quad\times \frac{1}{\left(\bar{\Lambda}\bar{x}\right)^{(d-1)/2}},\nonumber\\
  iG_R^{\phi\phi}(\bar{x}\ll |t-t'|) & \approx 0.\label{grx}
\end{align}
In analogy with the undriven problem \cite{Chiocchetta2016}, we expect that the presence of interactions will modify the exponents of the various algebraic decays. For example, we expect that \(G_K(x,t,t)\) will decay inside the light-cone \(\bar{x}\ll 2t\) with an exponent which involves the initial-slip exponent \(\theta\).

While the above analytical expressions assumed the dispersion relation in Eq.~\eqref{dis1a}, we now discuss the effects of having the actual 
dispersion in Eq.~\eqref{dism} (in the limit \(q\ll 1\)) of which Eq.~\eqref{dis1a} is a special case. In fact, Eq.~\eqref{dism} implies that quasiparticles propagate with various velocities, which span a certain range of values. 
In particular, note that for \(k, \omega \gg \sqrt{r_1}\), \(\epsilon_k = \sqrt{(\omega/2)^2 + k^2}\) and therefore \(\epsilon_k \approx k \) at large momenta  \(k\gg \omega\), as summarized in  Eq.~\eqref{dis1c}. This implies that  the fastest velocity is in fact \(v=1\) when the entire range of momenta \(k \in  \left[0,\Lambda\right]\) is considered.

\begin{figure}[ht]
\includegraphics[width = 0.5\textwidth]{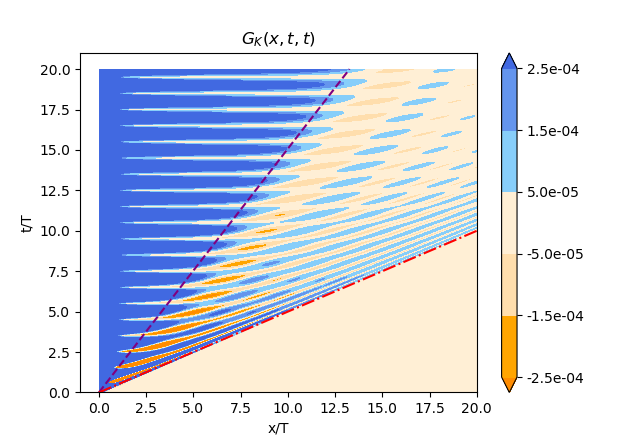}
\caption{\label{figGKeq}
Equal-time correlation function \(G_K^{\phi \phi}(x,t,t)\) as a function of $x$ and $t$ (in units of the period $T$ of the drive), 
in three spatial dimensions \(d=3\), with a dimensionless drive amplitude \(q=0.22\), and along the critical line.
Space and time are measured here in units of the period $T=2\pi/\omega$ of the drive. 
The other parameters are \(\omega= 2\), \(r_1 =0.44\), and the cut-off \(\Lambda = 2\pi\), while
\(r\) is chosen according to Eq.~\eqref{critp}.
The dash-dotted line corresponds to the light-cone of the fastest 
quasiparticles moving with velocity \(v=1\), while the dashed line corresponds to the light-cone of the quasiparticles with the slower velocity \(\sqrt{q/2}\simeq 0.33\).
The equal time correlator is synchronized with the drive, as it is clearly shown as a function of $t$ for a fixed value of $x$.
A different choice of the various parameters does not affect the qualitative features observed here. 
}
\end{figure}
%

%
%
%
%
\begin{figure}[ht]
\includegraphics[width = 0.5\textwidth]{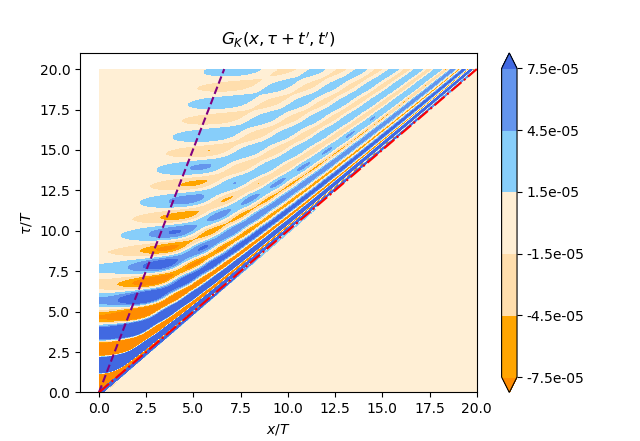}
\caption{\label{figGKneq}
Unequal-time correlator \(G_K^{\phi \phi}(x,\tau+t',t')\) with \(t'=0.1\) as a function $x$ and $\tau$ (in units of the period $T$ of the drive) in three spatial dimensions \(d=3\), with a dimensionless drive amplitude \(q=0.22\) and along the critical line. The remaining microscopic parameters are the same as in Fig.~\ref{figGKeq}. 
The dash-dotted line corresponds to the light-cone of the fastest 
quasiparticles which move with velocity \(v=1\), while the dashed line indicate the light-cone of
the slower quasiparticles with velocity \(\sqrt{q/2}\simeq 0.33\). 
The correlator shows period doubling for a fixed value of $x$ as a function of $\tau$.
A different choice of the various parameters and of $t'$ does not affect the qualitative features observed here.}
\end{figure}
%
%
%
%

\begin{figure}[ht]
\includegraphics[width = 0.5\textwidth]{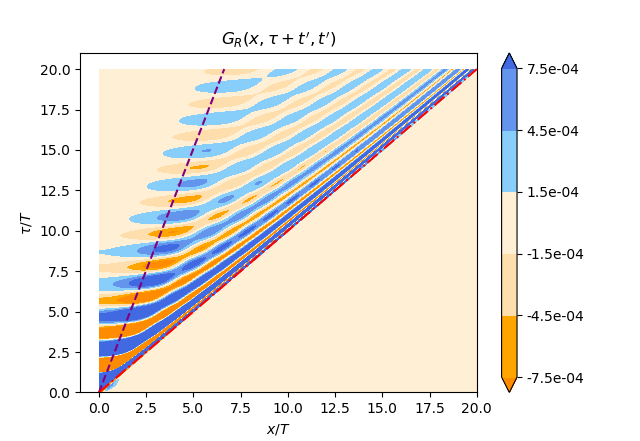}
\caption{\label{figGR}
Retarded correlator \(G_R^{\phi \phi}(x,\tau+t',t')\)  with \(t'=0.1\) as a function $x$ and $\tau$ (in units of the period $T$ of the drive) in three spatial dimensions \(d=3\), with a dimensionless drive amplitude \(q=0.22\) and along the critical line. The remaining microscopic parameters are the same as in Figs.~\ref{figGKeq} and \ref{figGKneq}. 
The dash-dotted line corresponds to the light-cone of the fastest quasiparticles which move with velocity \(v=1\), while the dashed line indicate the light-cone of
the slower quasiparticles with velocity \(\sqrt{q/2}\simeq 0.33\). 
The correlator shows period doubling for a fixed value of $x$ as a function of $\tau$.
A different choice of the various parameters and of $t'$ does not affect the qualitative features observed here. %
}
\end{figure}

Figures \ref{figGKeq}, \ref{figGKneq}, and \ref{figGR} show the contour plots of $G_{K,R}$ in spatial dimension \(d=3\), calculated by taking into account the full dispersion relation in Eq.~\eqref{dism} and by determining the corresponding coefficients \(M_{c,k}\) and \(M_{s,k}\) according to  Eq.~\eqref{c10gen}. 
The momentum integral in Eq.~\eqref{eq:k-to-x} is performed by assuming a Gaussian cutoff function defined by
\begin{align}
\int_0^\Lambda\!\! dk\ \ldots \longrightarrow  \int_0^{+\infty}\!\!dk\, e^{-k^2/(2\Lambda^2)}\ \ldots.
\end{align}
Note that the solution of the dynamics obtained by truncating the Sambe space in the vicinity of a certain critical line (in the present case, the one corresponding to $\epsilon_{k=0} = \omega/2$) is actually stable for all possible real values of $k$. Accordingly, the extension of the integral to values of $k$ beyond the original cutoff $\Lambda$ (see the discussion in the paragraph after Eq.~\eqref{eq:def-a}) implied by the Gaussian cutoff above is legitimate.
In the figures mentioned above, for concreteness, we choose the following values of the various parameters: pre-quench detuning \(r_0=1\), drive frequency \(\omega = 2\), drive amplitude \(r_1 =0.44\), dimensionless drive amplitude \(q = 0.22\), and the cut-off \(\Lambda = 2\pi\). 
In addition, the detuning parameter \(r\) is chosen to be on the critical line, i.e., according to Eq.~\eqref{critp} which ensures \(\epsilon_{k=0}=\omega/2\).\
Note that the Keldysh correlations also assume a deep quench which corresponds to accounting for only the momentum-momentum average of the initial state in Eq.~\eqref{ini}.  

In particular, \(G_K(x,t,t')\) is shown in Fig.~\ref{figGKeq} for $t'=t$ as a function of $x$ and $t$
while in Fig.~\ref{figGKneq} for $t = \tau + t'$ as a function of $x$ and $\tau$ with fixed $t'$. The retarded function,  \(G_R(x,t,t')\), instead, is shown in Fig.~\ref{figGR} for $t = \tau + t'$ with $\tau>0$ (as it vanishes for $\tau\le 0$) as a function of $x$ and $\tau$ with fixed $t'$.
All the three plots clearly feature the emergence of two light-cones.
One of them is indicated by the dot-dashed line and corresponds to quasiparticles moving at the fastest speed \(v=1\). The second light-cone is indicated by the dashed line and corresponds to 
quasiparticles moving at the renormalized velocity \(\sqrt{q/2} v
= \sqrt{q/2}\), which corresponds to $\simeq 0.33$ with the parameters of the plot. 
One also sees a clear period doubling in the unequal-time correlators in Figs.~\ref{figGKneq} and \ref{figGR}.
The equal-time correlator in Fig.~\ref{figGKeq} is, instead, synchronized with the drive. The analytic expressions for the power-law decays given in Eqs.~\eqref{gkx} and \eqref{grx} assume the simpler dispersion and therefore does not capture the more complex behavior observed between the two light-cones.
%
%
%
%
%

\section{Floquet Unitary} \label{microm}

In this section we reconsider
the dynamics of the driven model by constructing the time-evolution operator
in the vicinity of the critical line for generic times, including the stroboscopic ones. 
Floquet unitaries are usually studied numerically but the present case of the Gaussian model allows us to construct this operator analytically and therefore we are in the position to explore how its structure depends on the resonant nature of the drive. 
The expectation is that when the drive is effectively off-resonant, the Floquet unitary is essentially the unitary time evolution controlled by the undriven model with parameters which are renormalized by the drive. 
When the drive is resonant, instead, the Floquet unitary is expected to be qualitatively different from the time-evolution operator of the undriven case.

According to Floquet theory, briefly reviewed in Appendix \ref{appA}, the time-evolution operator $U$ for a periodic Hamiltonian $H(t)$ with period $T$ can be written as~\cite{Eckardt2015}
\begin{equation}
\begin{split}
U(t_2,t_1) &=  {\mathbb T}\exp\left(-i\int_{t_1}^{t_2}d\tau H(\tau)\right)\\
&= U_F(t_2) e^{-i(t_2-t_1) H_F} U_F^{\dagger}(t_1),
\label{eq:FU-gen}
\end{split}
\end{equation}
where \(H_F\) is the time-independent Floquet Hamiltonian. The operator \(U_F(t)\) is the so-called micro-motion operator (also sometimes referred to as the kick-operator), at time \(t\) and has the property of being time periodic, i.e., \(U_F(t+T) = U_F(t)\). 
The Floquet unitary is defined as the time-evolution operator over one period, i.e., \(U(t+T,t)\), and determines the stroboscopic time evolution.
In particular, it can be written in the form
\begin{equation}
U(t+T,t) = e^{-i T \widetilde{H}_F(t)}, 
\end{equation}
where, from Eq.~\eqref{eq:FU-gen},  
\begin{equation}
\widetilde{H}_F(t)= U_F(t) H_F U_F^{\dagger}(t).\label{Hft}
\end{equation}
This relationship shows that the combined effect of \(U_F\) and \(H_F\) --- which we construct explicitly below --- actually corresponds to an effective rotation of \(H_F\) by \(U_F\).

The Floquet Hamiltonian \(H_F\) for the Gaussian model we are interested in can be constructed straightforwardly because its eigenvalues are the quasienergies \(\epsilon_k\) determined in Eq.~\eqref{dis1} of Sec.~\ref{FBth}; accordingly, 
\begin{equation}
H_F = \frac{1}{2} \sum_k \left( |\Pi_\textbf{k}|^2 + \epsilon_k^2 |\phi_\textbf{k}|^2\right). \label{Hf}
\end{equation}
In what follows we explore the structure of \(U_F\). Due to Eq.~\eqref{Hft}, a non-trivial
\(U_F\), e.g., one with a singular structure in momentum space,
will generate a non-trivial \(\widetilde{H}_F\) and hence a non-trivial Floquet unitary.

We define the matrix \(\textbf{F}_k(t)\) which captures the effect on the fields of the time evolution with the Floquet Hamiltonian $H_F$ as
\begin{equation}
\begin{split}
&e^{i t H_F} \left(
\begin{matrix}
\phi_\textbf{k}(t_1)  \\
\Pi_\textbf{{k}}(t_1)
\end{matrix}
\right) e^{-i t H_F} \\
&\quad\quad= \textbf{F}_k(t)
 \left(
\begin{matrix}
\phi_\textbf{k}(t_1)  \\[2mm]
\Pi_\textbf{{k}}(t_1)
\end{matrix}
\right).
\end{split}
\label{Fdef1}
\end{equation}
Since \(H_F\) in Eq.~\eqref{Hf} represents simple harmonic oscillators with dispersion \(\epsilon_k\), it is straightforward to see that
\begin{equation}
\textbf{F}_k(t)=\left(
\begin{tabular}{cc}
\(\cos(\epsilon_k t)\) & \(\epsilon_k^{-1}\sin(\epsilon_k t)\) \\
\(-\epsilon_k \sin(\epsilon_k t)\) & \(\cos(\epsilon_k t)\)
\end{tabular}
\right).\label{Fdef}\\[2mm]
\end{equation}
Similarly, let us define \(\textbf{V}_k(t)\) as the matrix which captures the effect of the evolution induced by the micro-motion operator, i.e., 
\begin{equation}\label{Vdef}
U_F^\dagger(t) \left(
\begin{tabular}{c}
\(\phi_\textbf{k}(t_1)\)  \\
\(\Pi_\textbf{{k}}(t_1)\)
\end{tabular}
\right) U_F(t) = \textbf{V}_k(t)
 \left(
\begin{tabular}{c}
\(\phi_\textbf{k}(t_1)\)  \\
\(\Pi_\textbf{{k}}(t_1)\)
\end{tabular}
\right),
\end{equation}
and its inverse
\begin{equation}\label{Vdefi}
U_F(t) \left(
\begin{tabular}{c}
\(\phi_\textbf{k}(t_1)\)  \\
\(\Pi_\textbf{{k}}(t_1)\)
\end{tabular}
\right) U_F^\dagger(t) = \textbf{V}^{-1}_k(t)
 \left(
\begin{tabular}{c}
\(\phi_\textbf{k}(t_1)\)  \\
\(\Pi_\textbf{{k}}(t_1)\)
\end{tabular}
\right).
\end{equation}
For the exact solution of the dynamics which does not involve the truncation of the full Sambe space discussed in Sec.~\ref{FBth}, \(\textbf{V}_k(t)\) is a matrix with unit determinant, i.e.,  \(\text{det}\, \left[\textbf{V}_k(t)\right] =1\). 
However, since we have determined the solution of the dynamical equation by truncating the Sambe space, this condition is no longer fulfilled, as discussed in Appendix~\ref{commr} and the error in the determinant turns out to be given by Eq.~\eqref{Verr1} at intermediate momenta and by Eq.~\eqref{Verr2} at small momenta $k$. 

In order to capture the effect of the complete evolution operator $U$ in Eq.~\eqref{eq:FU-gen} we introduce the matrix \(\textbf{M}_k(t_2,t_1)\) as
\begin{align}
\left(
\begin{matrix}
\phi_\textbf{k}(t_2)  \\
\Pi_\textbf{{k}}(t_2)
\end{matrix}
\right) &=U^\dagger (t_2,t_1) \left(
\begin{matrix}
\phi_k(t_1)  \\
\Pi_{k}(t_1)
\end{matrix}
\right) U(t_2,t_1) \nonumber \\
&=\textbf{M}_k(t_2,t_1)
\left(
\begin{matrix}
\phi_\textbf{k}(t_1)  \\
\Pi_\textbf{{k}}(t_1)
\end{matrix}
\right).\label{timeev}
\end{align}
By using Eqs.~\eqref{Fdef1}, \eqref{Vdef}, and \eqref{Vdefi}, it is straightforward to see that this matrix can be expressed in terms of the matrices $\mathbf{F}_k$ and $\mathbf{V}_k$ introduced above as
\begin{align}
\textbf{M}_k(t_2,t_1)
= \mathbf{V}_k(t_2) \mathbf{F}_k(t_2-t_1) \mathbf{V}^{-1}_k(t_1).
\label{eq:M-VFV}
\end{align}
In Appendix~\ref{appA} we show that the matrix \(\textbf{M}_k(t_2,t_1)\) can be written as \(\textbf{M}_k(t_2,t_1)=\textbf{M}_k(t_2,0)\textbf{M}_k(0,t_1)\) where
\begin{equation}\label{M2to1}
\textbf{M}_k(t,0)=\left(
\begin{tabular}{cc}
\(M_{c,k}(t)\) & \(M_{s,k}(t)\)  \\
\(\dot{M}_{c,k}(t)\) & \(\dot{M}_{s,k}(t)\)
\end{tabular}
\right),
\end{equation}
while \(M_{c,k}\) and \(M_{s,k}\) are the mode functions derived in Sec.~\ref{FBth}.
With this background, we are in the position to determine the matrix \(\mathbf{V}_k(t)\),
and the corresponding micro-motion operator \(U_F(t)\).

It is instructive to construct \(\mathbf{V}_k(t)\) and \( U_F(t)\) in the two limiting cases 
discussed in Sec.~\ref{FBth}, corresponding to intermediate momenta \(\sqrt{q} \omega \ll k \ll \omega\), and to small momenta \(k\ll \sqrt{q} \omega \ll \omega\), with the corresponding quasienergies reported in Eq.~\eqref{dis1}.
We show below that there is a qualitative change in the structure of the Floquet unitary when $k$ decreases from intermediate to small values because the drive goes from being effectively off-resonant in the former regime to becoming resonant in the latter.

At intermediate momenta \(\sqrt{q} \omega \ll k \ll \omega\) we find in Appendix~\ref{microAp} that, up to
\(O(q^2\omega^2/\bar{k}^2)\),
\begin{widetext}
\begin{equation}
\mathbf{V}_k(t) \approx \left(
\begin{matrix}
1 & 0 \\[2mm]
0 & 1
\end{matrix}
\right)+
\frac{1}{8} \frac{q^2\omega^2}{\overline{k}^2}
\left(
\begin{matrix}
\frac{1}{2}\cos(\omega t) &  -\frac{1}{\omega}\sin(\omega t) \\[2mm]
-\frac{\omega}{4}\sin(\omega t) & -\frac{1}{2}\cos(\omega t)
\end{matrix}
\right).\label{V1}
\end{equation}
\end{widetext}
Since the regime of intermediate momenta corresponds to having \(q^2\omega^2\ll \bar{k}^2\),
the corresponding \(U_F\) is well-approximated by the identity matrix.
Accordingly, the micromotion operator can be neglected at high (non-resonant) drive frequencies, 
and the Floquet unitary $U(t+T,t)$ is then accurately described by the sole Floquet Hamiltonian $H_F$, with \(U(t+T,t)\approx e^{-i H_F T}\) where \(H_F\) is the Gaussian Hamiltonian in Eq.~\eqref{Hf}, which is spatially short-ranged.

Next we show that the high-frequency expansion breaks down in the opposite limit of \(k \ll \sqrt{q} \omega \ll \omega\) as expected due to the resonant character of the drive at this scale.
In fact, at the leading order in \(O(\bar{k}/q\omega, q)\) we find in Appendix~\ref{microAp} that the leading term is
\begin{align}\label{V2}
\mathbf{V}_k(t) \approx \frac{1}{2} \sqrt{\frac{q\omega}{2\overline{k}}}\left(
\begin{matrix}
1+\cos(\omega t) & -\frac{2}{\omega} \sin(\omega t)\\[2mm]
-\frac{\omega}{2} \sin(\omega t) &  1-\cos(\omega t)
\end{matrix}
\right).
\end{align}
Note that this expression 
has zero determinant because the two eigenvalues have different orders of magnitude in the small momentum limit. Keeping only the leading term results in a singular matrix as it only captures one eigenvalue while effectively setting the other to zero. The next leading term in \( \textbf{V}_k \) is accounted for in Eq.~\eqref{Vkf}.

In Appendix~\ref{UFder} we show that the transformation \( \textbf{V}_k \) is generated by the micromotion operator 
\begin{equation}\label{UF1}
  U_F(t)\approx \exp\biggl[-\sum_\mathbf{k}\frac{1}{4} \ln\left(\frac{2\bar{k}}{q\omega}\right)\biggl( e^{i\omega t}
      a_\mathbf{k}^\dagger a_\mathbf{-k}^\dagger -h.c.\biggr)\biggl],
\end{equation}
according to Eq.~\eqref{Vdef}, where the operators \(a_k^{\dagger}\) and \(a_k\) are the ones which diagonalize \(H_F\).
The logarithmic dependence on the small momentum $k$ of the coefficient of the bilinear in the exponential of $U_F$ indicates that the time-evolution operator is effectively long-ranged in space and therefore it is qualitatively different from that one of the undriven model, which looks similar to Eq.~\eqref{Hf}. 

Note that if the eigenvalues of
\(U_F(t)\) were on a unit circle, then \(U_F(t)\) would simply rotate the modes. In contrast, the two eigenvalues of \(U_F(t)\) at long wavelengths (c.f., Appendix~\ref{UFder}) are actually \((q\omega/2\bar{k})^{1/2}\) and \((q\omega/2\bar{k})^{-1/2}\) (see Eq.~\eqref{eq:UF-eigenv}),
which do not lie on a unit circle, with one of the two being much larger than the other. This structure of \(U_F(t)\) where one mode is strongly amplified relative to the other in an example of mode squeezing. 
We note that, in general, the eigenvalues of \(U_F(t)\) are time-dependent, but for this example, in the limit of long wavelength and small drive, the time-dependence of
the eigenvalues turn out to be sub-leading.

In order to highlight the squeezing induced by $U_F(t)$, we evaluate the
uncertainty in the position and momentum operators $\phi_\mathbf{k}$ and $\Pi_\mathbf{k}$, respectively, in the state $|\Psi\rangle =U_F(t=0)|0\rangle$ obtained by applying $U_F$ 
to a state with no squeezing which, for convenience, we assume to be the ground-state $|0\rangle$ of the pre-quench Hamiltonian \(H_0\). 

\begin{figure}[ht]
\includegraphics[width = 0.45\textwidth]{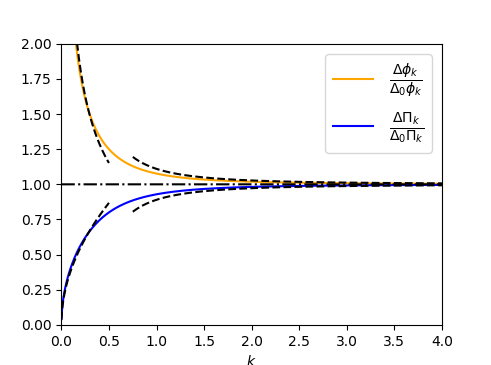}
\caption{\label{figsq}
Dependence of the normalized position and momentum uncertainties $\Delta \phi_k$ and $\Delta \Pi_k$, respectively, in Eqs.~\eqref{sq2a} and \eqref{sq2b} on the momentum $k$, which highlights the emergence of squeezing at small momenta, signalled by the deviation of these quantities from 1 (dash-dotted horizontal line). 
The curves refer to a dimensionless drive amplitude \(q=0.22\), drive frequency \(\omega =2\) on the critical line. A different choice of the parameters does not affect the qualitative features of the curves. The dashed lines at small and large values of $k$ indicate the corresponding approximations reported in Eqs.~\eqref{sq-smallk} and \eqref{sq-largek}, respectively. }
\end{figure}

In particular, we quantify the  
uncertainty on the position $\phi_\mathbf{k}$   
in the above state as
\begin{align}\label{sq1}
 \Delta \phi_k = \sqrt{\langle \Psi| 
 \phi_\mathbf{k} \phi_{-\mathbf{k}}|\Psi\rangle},   
\end{align}
with an analogous definition for the uncertainty $\Delta \Pi_k$ on the momentum $\Pi_k$. Moreover, we denote by $\Delta_0 \phi_{k} = 1/\sqrt{2\omega_{0k}}$ and $\Delta_0 \Pi_{k} = \sqrt{\omega_{0k}/2}$ the corresponding quantities in the initial state $|0\rangle$, given by the first equalities in Eq.~\eqref{ini}, where \(\omega_{0k}\) is the dispersion of the pre-quench Hamiltonian given in Eq.~\eqref{om0k}.
By using the results derived in Appendix \ref{microAp}, we find that the corresponding squeezing are given by 
\begin{align}
\frac{\Delta \phi_k}{\Delta_0 \phi_k} &= \left(1-\frac{\omega}{\epsilon_k}\frac{c_{-1}}{c_0+c_{-1}}\right)^{-1/2}, 
\label{sq2a}\\
\frac{\Delta \Pi_k}{\Delta_0 \Pi_k}
&= \left(1-\frac{\omega}{\epsilon_k}\frac{c_{-1}}{c_0+c_{-1}}\right)^{1/2},\label{sq2b}
\end{align}
where  \(\epsilon_k\) is the quasienergy given in Eq.~\eqref{dism}, while the coefficients \(c_{-1,0}\) are given in Eq.~\eqref{c10gen}.
Note that, as expected, $\Delta \phi_k \Delta \Pi_k  = \Delta_0 \phi_k \Delta_0 \Pi_k$.
These normalized uncertainties
are plotted 
in Fig.~\ref{figsq} as a function of $k$, for a given choice of the parameter $q$ of the drive, along the critical line.
The plot shows how the squeezing varies as a function of the momentum $k$, by eventually vanishing at large momenta. 
The occurrence of dynamical squeezing is signalled by the fact that the quantities reported in Fig.~\ref{figsq} deviate from the unit reference value. 
In particular, the behaviour at small momenta \(k\ll \sqrt{q}\omega \ll \omega \), which results in  the largest squeezing, is given by
\begin{subequations}\label{sq-smallk}
\begin{align}
\frac{\Delta \phi_k}{\Delta_0 \phi_k} &\approx \left(\frac{q\omega}{2\bar{k}}\right)^{1/2} ,\\
\frac{\Delta \Pi_k}{\Delta_0 \Pi_k} &\approx \left(\frac{q\omega}{2\bar{k}}\right)^{-1/2},
\end{align}
\end{subequations}
while at intermediate momenta 
\( \sqrt{q}\omega \ll k \ll \omega \) one finds
\begin{subequations}\label{sq-largek}
\begin{align}
\frac{\Delta \phi_k}{\Delta_0 \phi_k} &\approx 1+\frac{1}{16}\frac{q^2\omega^2}{\bar{k}^2},\\
\frac{\Delta \Pi_k}{\Delta_0 \Pi_k} &\approx  1-\frac{1}{16}\frac{q^2\omega^2}{\bar{k}^2}.
\end{align}
\end{subequations}
These expressions for small and large momenta $k$ are indicated in Fig.~\ref{figsq} as dashed lines and they turn out to capture rather accurately the actual behavior of these quantities. 
%
%
%
%
%
%

\section{Conclusions} \label{conclu}

The Floquet time crystal (FTC) is a non-equilibrium phase of matter which by now has been realized in numerous
theoretical models and experimental systems. 
Thus the time is ripe to understand if model-independent features of these phenomena emerge, 
possibly establishing a notion of universality in these systems.
As a first attempt in this direction, we studied in detail the dynamical and structural properties of the periodically driven \(O(N)\)
model along the critical line
separating a trivial phase from the FTC phase, within the Gaussian approximation. 
In particular, we showed the emergence of scale-invariant behaviors, within the Gaussian approximation, and highlighted that certain correlators are more long-ranged in the driven problem than in the absence of drive. 
For the latter, our point of comparison was the paramagnetic-ferromagnetic critical point of the undriven \(O(N)\) model.
Appearance of scaling in the exactly solvable Gaussian limit is the first step towards rigorously establishing universality in the presence of 
interactions, and our work paves the way for such a treatment. 

We also showed that relevant correlation functions of the model display various light-cones near the FTC critical line. %
The quasienergy dispersion relation of the problem was found to be a rather complicated function of the momentum $k$, so that no single quasiparticle velocity is associated with it. Nonetheless, the 
light-cone dynamics turns out to be dominated by a slow and a fast velocity, the ratio of which 
was found to be \(\sqrt{q/2}\), \(q\) being the dimensionless drive amplitude (see Eq.~\eqref{qdef}), assumed to be small in our analysis.

The Floquet unitary which describes the stroboscopic evolution was found to be qualitatively different at short and long wavelengths. At long wavelengths, i.e., close to the resonance condition, the Floquet unitary turns out to squeeze the modes, as in a parametrically driven oscillator. On the other hand, at shorter wavelengths, the Floquet unitary effectively rotates the modes, as in a simple harmonic oscillator.

Future work will study the effect of interactions. We expect that the power-laws which characterize the scale-invariant behaviors found here will be modified and the results of this investigation will be reported elsewhere \cite{Natsheh2020b}. Exploring the question of universality along the critical line of a FTC coupled to a bath is also an interesting open question.

\smallskip


{\sl Acknowledgements.} 
This work was supported by the US National Science Foundation Grant NSF-DMR 1607059 and partially
by the MRSEC Program of the National Science Foundation under Award Number DMR-1420073.


\appendix

\section{The Floquet-Bloch theorem and its application to the Mathieu equation} \label{appA}

In Subsec.~\ref{appA-FBT} of this Appendix we briefly review the Floquet-Bloch theorem while in Subsec.~\ref{meapp} we apply it to the Mathieu equation and also highlight some subtleties related to our model.

\subsection{The Floquet-Bloch theorem} \label{appA-FBT}
The Floquet-Bloch theorem states that a \(n\times n\) matrix \(\mathbf{\Phi}(t)\) which obeys the equation of motion
\begin{equation}\label{sys}
\frac{d\mathbf{\Phi}(t)}{dt} = \textbf{A}(t)\mathbf{\Phi}(t),
\end{equation}
where \(\textbf{A}\) is a \( n\times n\) periodic matrix with period $T$, i.e., \(\textbf{A}(t+T)=\textbf{A}(t)\), can be written as
\begin{equation}
\mathbf{\Phi}(t) = \mathbf{P}(t) e^{\mathbf{B}t}, 
\label{app-FBT-2}
\end{equation}
where \(\mathbf{P}(t)\) is an \( n\times n \)
periodic matrix with period $T$
and \(\mathbf{B}\) is a \( n\times n\) non-singular and therefore invertible matrix.
We outline here the proof of the theorem. 
Since both \(\mathbf{\Phi}(t+T)\) and \(\mathbf{\Phi}(t)\) obey Eq.~\eqref{sys}, one
can be written as a linear combination of the other. Thus one may
define \(\mathbf{C}\), a non-singular \(n\times n\) matrix, such that
\begin{equation}
\mathbf{\Phi}(t+T) = \mathbf{\Phi}(t)\mathbf{C}.\label{def-C}
\end{equation}
Now we use the fact that the matrix logarithm of a non-singular matrix exists in order to introduce the matrix \(\mathbf{B}\) such that
\begin{equation}
\mathbf{C} = e^{\mathbf{B}T}.
\label{CB-rel}
\end{equation}
Introducing \(\mathbf{P}(t)=\mathbf{\Phi}(t)e^{-\mathbf{B}t}\) it is then easy to show, using Eq.~\eqref{def-C}, that \(\mathbf{P}(t+T)=\mathbf{P}(t)\), which proves the theorem.

Below we discuss two special cases, both of which emerge in the periodically driven \(O(N)\) model Eq.~\eqref{Hg}, and which depend on the diagonalizability of the matrices introduced above.

\subsubsection{Special Cases} 
\label{appA-sc}

We begin by recalling that a diagonalizable matrix is characterized by having a linearly independent set of eigenvectors. 
The first case we consider here is the one in which the matrix \(\mathbf{B}\)  introduced above is diagonalizable. This can only happen if \(\mathbf{C}\) is also diagonalizable. 
Denoting by 
\(\mathbf{C}_D\) the diagonal matrix having the eigenvalues of \(\mathbf{C}\) as entries, an invertible matrix $\mathbf{U}$ exists such that
\(\mathbf{C}=\mathbf{U}^{-1}\mathbf{C}_D\mathbf{U}\) and therefore
\begin{equation}
\mathbf{C}_D = e^{\mathbf{B}_{D}T},
\label{CB-rel-diag}
\end{equation}
where \(\mathbf{B}_{D}\) is the  diagonal matrix with $\left[{\mathbf{B}_{D}}\right]_{ii}T =  \left[\ln{\mathbf{C}_D}\right]_{ii}$ which is the diagonal form of $\mathbf{B}$, as easily derived from Eq.~\eqref{CB-rel}:
\begin{align}
\mathbf{B}= \mathbf{U}^{-1}\mathbf{B}_{D}\mathbf{U}.
\end{align}

The second case we are interested in occurs when \(\mathbf{C}\) is not diagonalizable, i.e.,
when \(\mathbf{C}\) does not have \(n\) independent eigenvectors.
Then a matrix \(\mathbf{Q}\) exists such that one can write the matrix \(\mathbf{C}\) in the Jordan form, i.e.,
\begin{equation} \label{Jor}
\mathbf{C}=\mathbf{Q}^{-1}\mathbf{J}\mathbf{Q} \quad\mbox{with}\quad
\mathbf{J} = \mathbf{D}(\mathbf{I}+\mathbf{K}),
\end{equation}
where \(\mathbf{J}\) is the Jordan decomposition matrix of \(\mathbf{C}\) in terms of a diagonal matrix $\mathbf{D}$ and of the matrix 
\(\mathbf{D}\, \mathbf{K}\). The latter is a matrix 
whose entries right above the diagonal are the only non-vanishing ones.

Taking the logarithm of Eq.~\eqref{Jor}, one has \(\ln{\mathbf{C}}= \ln{\mathbf{J}}\), which yields  
\(\ln{\mathbf{J}}=\ln{\mathbf{D}(\mathbf{I}+\mathbf{K})} = \ln{\mathbf{D}} + \ln{(\mathbf{I}+\mathbf{K})}\).
Expanding,
\begin{equation}
\ln{(\mathbf{I}+\mathbf{K})} = \mathbf{K} - \frac{1}{2}\mathbf{K}^2 + \frac{1}{3} \mathbf{K}^3 + ....
\end{equation}
The above series actually terminates because $\mathbf{K}^n = 0$ for an \(n\) dimensional matrix with vanishing lower-diagonal elements.
We will encounter the above non-diagonalizable form 
in Section \ref{magdyn} when we study the magnetization dynamics along the critical line.

\subsection{The Mathieu equation}
\label{meapp}

We will now study Eq.~\eqref{mot}, but first we recast this second-order differential equation into two coupled first-order differential equations, taking a form similar to Eq.~\eqref{sys}:
\begin{equation}\label{mot1}
\frac{d}{dt}\left(
\begin{matrix}
f_k \\
\dot{f}_k
\end{matrix}
\right)=\left(
\begin{matrix}
0 & 1 \\
-r_k(t) & 0
\end{matrix}
\right)
\left(
\begin{matrix}
f_k \\
\dot{f}_k
\end{matrix}
\right),
\end{equation}
where
\begin{equation}
r_k(t) = r_c+k^2-r_1  \cos({\omega t}).
\label{eq-def-rkt}
\end{equation}

Using the Floquet-Bloch theorem, there are two real independent solutions of Eq.~\eqref{mot1}, which we denote by \(f_k^{(1)}(t)\) and \(f_k^{(2)}(t)\) and which can be arranged to form the matrix solution $\mathbf{\Phi}$ such that Eq.~\eqref{def-C} applies:
\begin{equation}
\left(
\begin{matrix}
f_k^{(1)}(t+T) & f_k^{(2)}(t+T) \\
\dot{f}_k^{(1)}(t+T) & \dot{f}_k^{(2)}(t+T)
\end{matrix}
\right)=\left(
\begin{matrix}
f_k^{(1)}(t) & f_k^{(2)}(t) \\
\dot{f}_k^{(1)}(t) & \dot{f}_k^{(2)}(t)
\end{matrix}
\right)\times\mathbf{C}.
\end{equation}

As linear and independent solutions $f_k^{(1,2)}(t)$ of the Mathieu equation we can consider the 
functions \(f_k^{(1)} \mapsto  M_{c,k}\) and \(f_k^{(2)} \mapsto M_{s,k}\) in Eqs.~\eqref{Mc} and \eqref{Ms}.
Defining \(\textbf{M}_k(t_2,t_1)\) as the matrix which generates the time evolution according to Eq.~\eqref{timeev},
one has
\begin{equation}
\frac{\partial \textbf{M}_k(t_2,t_1)}{\partial t_2} = \left(
\begin{tabular}{cc}
\(0\) & \(1\)  \\
\(-r_k(t_2)\) & \(0\)
\end{tabular}
\right)\textbf{M}_k(t_2,t_1),
\end{equation}
with \(\textbf{M}_k(t_1,t_1)=\textbf{I}_{2\times 2}\).
\(\textbf{M}_k(t_2,t_1)\) is a fundamental matrix for the Floquet system.
We consider a special case \(t_1=0\) which we write as
\begin{equation}\label{Mt0}
\textbf{M}_k(t_2,0)=\left(
\begin{tabular}{cc}
\(M_{c,k}(t_2)\) & \(M_{s,k}(t_2)\)  \\
\(\dot{M}_{c,k}(t_2)\) & \(\dot{M}_{s,k}(t_2)\)
\end{tabular}
\right).
\end{equation}
Note that \(\textbf{M}_k(t_2,0)\) is also a fundamental matrix for the Floquet system and therefore it can be expressed as a linear combination of $\textbf{M}_k(t_2,t_1)$ via a (possibly $t_1$-dependent) matrix $\mathbf{C}$ such that 
\begin{equation}
\textbf{M}_k(t_2,0)=\textbf{M}_k(t_2,t_1)\textbf{C}.
\label{def-C-app}
\end{equation}
For $t_2=0$ this relation yields \(\textbf{M}_k(0,0)=\textbf{M}_k(0,t_1)\textbf{C}\) and, given that $\textbf{M}_k(0,0)=\textbf{I}_{2\times 2}$, one finds $\textbf{C}=\textbf{M}_k(0,t_1)^{-1}$. Once inserted in Eq.~\eqref{def-C-app}, this expression implies 
$\textbf{M}_k(t_2,t_1) = \textbf{M}_k(t_2,0) \textbf{C}^{-1} = \textbf{M}_k(t_2,0) \textbf{M}_k(0,t_1)$.
Alternatively, by setting $t_2 = t_1$ in Eq.~\eqref{def-C-app} and by taking into account the initial condition for that equation, one finds $\textbf{C}=\textbf{M}_k(t_1,0)$ and therefore Eq.~\eqref{def-C-app} implies 
\begin{equation}
\textbf{M}_k(t_2,t_1) = \textbf{M}_k(t_2,0)\textbf{M}^{-1}_k(t_1,0).
\label{eq:app-Mt2t1-prod}
\end{equation}
The above manipulations will be helpful when we derive the
Floquet unitary in Sec.~\ref{microm} and Appendix \ref{microAp}.

Motivated by the analysis of the model we are interested in, we will now consider two cases. One where \(\mathbf{C}\) is diagonalizable, and the other where it is not. The latter occurs
in the analysis of the dynamics of the mode with \(k=0\) along the critical line given by Eq.~\eqref{critp}.

\subsubsection{\(\mathbf{C}\) is diagonalizable} \label{Cnond}

If \(\mathbf{C}\) is diagonalizable, then solutions \(f_k^{(1)}(t)\) and \(f_k^{(2)}(t)\)
arranged in the matrix $\mathbf{\Phi}$ satisfy Eq.~\eqref{app-FBT-2} as a consequence of the Floquet-Bloch theorem, with a diagonal $\mathbf{C}_D$ given in Eq.~\eqref{CB-rel-diag}, i.e., 
\begin{multline}\label{form}
\left(
\begin{matrix}
f_k^{(1)}(t) & f_k^{(2)}(t) \\[2mm]
\dot{f}_k^{(1)}(t) & \dot{f}_k^{(2)}(t)
\end{matrix}
\right)=\left(
\begin{matrix}
u_k^{(1)}(t) & u_k^{(2)}(t) \\[2mm]
w_k^{(1)}(t) & w_k^{(2)}(t)
\end{matrix}
\right)\\
\times
\left(
\begin{matrix}
\exp(i\epsilon_k^{(1)} t) & 0 \\[2mm]
0 & \exp(i\epsilon_k^{(2)} t)
\end{matrix}
\right),
\end{multline}
where
\(u_k^{(1)}\), \(u_k^{(2)}\), \(w_k^{(1)}\), and \(w_k^{(2)}\) are periodic functions with period \(T=2\pi/\omega\).
In solving the Mathieu equation \eqref{mot1} this instance occurs for \(k\neq0\), as we will be able to find two real and independent solutions.
However, \(\mathbf{C}\) turns out not to be diagonalizable for \(k=0\), a case which we consider in detail below.

\subsubsection{\(\mathbf{C}\) is non-diagonalizable} 
\label{keq0}

Using the Floquet-Bloch solution of the Mathieu equation derived in Sec.~\ref{FBth}, the functions \(M_{c,k=0}(t)\) and \(M_{s,k=0}(t)\) (c.f., Sec.~\ref{magdyn}) can be written as 
\begin{equation}
M_{c,k=0}(t)= \cos(\omega t/2),
\end{equation}
and
\begin{equation}
M_{s,k=0}(t)=\frac{2}{\omega}\frac{1}{1+q} \left[\sin\left(\frac{\omega t}{2}\right) 
+q\frac{\omega t}{2} \cos\left(\frac{\omega t}{2}\right)\right],
\end{equation}
where we keep,  
up to $O(q^2)$,
only the slowest oscillating terms, which for our parameters are characterized by the angular frequency \(\omega/2\).

The solutions \(M_{c,k=0}\) and \(M_{s,k=0}\) are two independent solutions of the Floquet equation \eqref{mot1} and therefore, according to the notation introduced after Eq.~\eqref{eq-def-rkt}, 
we can identify $f_k^{(1)}(t) \mapsto M_{c,k=0}(t)$ and $f_k^{(2)}(t) \mapsto M_{s,k=0}(t)$;
however, they do not have the form proposed in Eq.~\eqref{form}. Accordingly, the corresponding matrix 
\(\mathbf{C}\) is not diagonalizable but its logarithm can still be determined via the Jordan form Eq.~\eqref{Jor} with non-zero \(\mathbf{K}\).
Moreover, note that \(M_{c,k=0}\) is anti-periodic function as it changes sign for $t \mapsto t+T$, while \(M_{s,k=0}\) is not.
We now explicitly show that the above solutions \(M_{c,k=0}\) and \(M_{s,k=0}\) satisfy Floquet-Bloch theorem with a non-diagonalizable matrix  \(\mathbf{C}=e^{\pmb{B} T}\).

In fact, it is easy to check that
\begin{align}
&\left(
\begin{matrix}
M_{c,k=0}(t) & M_{s,k=0}(t) \nonumber\\
\dot{M}_{c,k=0}(t) & \dot{M}_{s,k=0}(t)
\end{matrix}
\right) \\ 
&\quad= \mathbf{P}_\mp(t) \times \exp\left\{\left(
\begin{matrix}
\pm i & \frac{q}{1+q}\frac{2}{\omega} \\
0 & \pm i
\end{matrix}
\right)\frac{\omega t}{2} \right\}\\
&\quad= \mathbf{P}_\mp(t)  \times e^{\pm i\frac{\omega t}{2}}\left[1+\frac{q}{1+q}t\left(
\begin{matrix}
0 & 1 \\
0 & 0
\end{matrix}
\right)\right],
\end{align}
where \(\mathbf{P}_\mp(t)\) is a \(2\times2\) periodic matrix with period \(T = 2\pi/\omega\) given by
\begin{equation}
\begin{split}
&\mathbf{P}_\mp(t) =\left(
\begin{matrix}
M_{c,k=0}(t) & M_{s,k=0}(t)-\frac{q}{1+q}t\ M_{c,k=0}(t)  \\
\dot{M}_{c,k=0}(t) & \dot{M}_{s,k=0}(t)-\frac{q}{1+q}t\ \dot{M}_{c,k=0}(t)
\end{matrix}
\right)e^{\mp i\frac{\omega t}{2}}.
\end{split}
\end{equation}
In this example \(\mathbf{C}\) is
\begin{align}
\mathbf{C}_{\pm} =  \exp{\left\{\left(
\begin{tabular}{cc}
\(\pm i\) & \(\frac{q}{1+q}\frac{2}{\omega}\) \\
0 & \(\pm i\)
\end{tabular}
\right)\pi\right\}},
\end{align}
and is non-diagonalizable.

\section{Approximate expressions for the quasienergy }
\label{disD}

In this section we provide some details concerning the derivation of the approximate expressions in Eq.~\eqref{dis1} for the quasienergy $\epsilon_k$ investigated in Sec.~\ref{FBth}.
Starting from Eq.~\eqref{dism}, which was derived by using the Floquet-Bloch theorem and under the assumption of a weak drive $q\ll 1$, we study the dispersion $\epsilon_k$ in the vicinity of the critical line defined in Eq.~\eqref{critp}. 
The condition for being at the critical line is equivalent to requiring
\begin{equation}
r=r_c=(\omega/2)^2+ r_1/2,
\end{equation}
where we neglect higher-order terms of the form $\omega^2 q^2$, with $q$ given in Eq.~\eqref{qdef}.
Substituting this expression in Eq.~\eqref{dism} and expressing the result in terms of the dimensionless drive amplitude $q$ defined in Eq.~\eqref{qdef}, we obtain
\begin{widetext}
\begin{equation}
\epsilon_k  = \frac{\omega}{2} + \frac{\omega}{2} \sqrt{2 + q + \left(\frac{2k}{\omega}\right)^2 - 2 \sqrt{1 + q + \left(\frac{q}{2}\right)^2 + \left(\frac{2k}{\omega}\right)^2}},
\label{disp-appen}
\end{equation}
\end{widetext}
which, for $k=0$, renders $\epsilon_{k=0} = \omega/2$ as it should do along the critical line. 
The expression above is characterized by the energy scale $\omega$ and by the two dimensionless ratios $q$ and $k/\omega$, which are associated with the drive amplitude and the external momentum, respectively, the former assumed to be much smaller than one, i.e., $q\ll 1$. 

The dependence on $k$ of $\epsilon_k$ in Eq.~\eqref{disp-appen} can be approximated by different expressions depending on the assumption on the ratio $k/\omega$. In particular, for $k \gg \omega$ one finds
\begin{equation}
\epsilon_k  = k + O(\omega/k),
\end{equation}
i.e., Eq.~\eqref{dis1c}. For $k \ll \omega$, instead, both ratios are much smaller than 1 and therefore one can expand up to the second order the innermost square root in Eq.~\eqref{disp-appen}, which eventually leads to
\begin{equation}
\epsilon_k  \simeq \frac{\omega}{2} + \frac{k}{2} \sqrt{ \left( \frac{2 k}{\omega}\right)^2 + 2q}. 
\end{equation}
This expression can be further approximated depending on the relationship between the two terms in the square root. In particular, if $k/\omega \gg \sqrt{q}$ one finds, up to order $q^0$,
\begin{equation}
\epsilon_k \simeq \frac{\omega}{2} + \frac{k^2}{\omega},
\end{equation}
i.e., Eq.~\eqref{dis1b}. If, instead,  $k/\omega \ll \sqrt{q}$, expanding the square root one finds
\begin{equation}
\epsilon_k \simeq \frac{\omega}{2} + k \sqrt{\frac{q}{2}},
\end{equation}
i.e., Eq.~\eqref{dis1a} taking into account Eq.~\eqref{kbar}.

\section{Critical quench in the undriven Gaussian model} 
\label{ON}

In order to compare in Sec.~\ref{corr} the predictions for correlation functions in the driven model with those in the absence of drive, we report here for completeness the expressions of the Keldysh and retarded Green's function for the latter, referring the reader to Refs.~\cite{Tavora2015,Chiocchetta2016} for additional details. 

The dynamics of \(\phi_k\) and \(\Pi_k\) for a quench from the thermal state
of the quadratic Hamiltonian with an initial value \(r_0\) of the parameter $r$ in Eq.~\eqref{Hg} with $r_1=0$, to the critical point with \(r=r_c=0\) is~\cite{Tavora2015,Chiocchetta2016}
\begin{align}
\phi_\textbf{k}(t)&= \cos\left(k t\right) \phi_\textbf{k}(0)
+\frac{\sin\left(k t\right)}{k}\Pi_\textbf{k}(0),\label{Phi1s}\\
\Pi_\textbf{k}(t)&= -k\sin\left(k t\right)\phi_\textbf{k}(0)
+\cos\left(k t\right)\Pi_\textbf{k}(0).
\end{align}
For a deep quench $r_0\gg \Lambda$ and at long wavelengths $k\ll\Lambda$,
\begin{align}
\langle \Pi_{i,\textbf{k}}(0) \Pi_{j,\textbf{q}}(0) \rangle &= \delta_{i,j}\delta_{\textbf{k},-\textbf{q}} \frac{\omega_{0k}}{2} \approx \delta_{i,j}\delta_{\textbf{k},-\textbf{q}} \frac{\sqrt{r_0}}{2},\\
\langle \phi_{i,\textbf{k}}(0) \phi_{j,\textbf{q}}(0) \rangle &= \delta_{i,j}\delta_{\textbf{k},-\textbf{q}} \frac{1}{2\omega_{0k}} \approx \delta_{i,j}\delta_{\textbf{k},-\textbf{q}} \frac{1}{2\sqrt{r_0}},\\
\langle \left\{\phi_{i,\textbf{k}}(0),\Pi_{j,\textbf{q}}(0) \right\}\rangle &= 0,
\end{align}
where we assumed the temperature $\beta^{-1}$ of the initial state to be such that $\beta r_0 \gg 1$.

The Keldysh Green's functions turn out to be
\begin{align}
iG_{K,u}^{\phi \phi}(k,t,t') &= \frac{\sqrt{r_0}}{2k^2} [\cos(k(t-t'))-\cos(k(t+t'))],\\
iG_{K,u}^{\Pi \Pi}(k,t,t') &= \frac{\sqrt{r_0}}{2}[\cos({k}(t-t'))+\cos({k}(t+t'))],\\
iG_{K,u}^{\phi \Pi}(k,t,t') &= \frac{\sqrt{r_0}}{2k} [\sin(k(t-t'))+\sin(k(t+t'))],
\end{align}
where we introduced above the subscript \(u\) in order to distinguish these quantities from the corresponding ones in the driven model. 

The retarded Green's functions, instead, are given by
\begin{align}
G_{R,u}^{\phi \phi}(k,t,t') &=-\theta(t-t')\ \frac{\sin(k (t-t'))}{k},\\
G_{R,u}^{\Pi \Pi}(k,t,t') &=-\theta(t-t')\ k \sin(k (t-t')),\\
G_{R,u}^{\phi \Pi}(k,t,t') &=\theta(t-t')\ \cos(k (t-t')).
\end{align}

At short times \(t,t'\ll k^{-1}\), the \(\langle \phi \phi\rangle \) correlators reduce to
\begin{align}
iG_{K,u}^{\phi \phi}(k,t,t')&=\sqrt{r_0} tt',\\
G_{R,u}^{\phi \phi}(k,t,t')&=-\theta(t-t') (t-t').
\end{align}

\section{Commutation Relations} 
\label{commr}

In this section we show that in order to satisfy the canonical commutation relations at all times  one needs to solve the Floquet problem exactly. In fact, in constructing our perturbative solution we introduce a deviation from the exact commutation relations which is controlled by the smallness of the drive amplitude \(q\), as we show below.

For simplicity, let us drop the momentum label from the various quantities which depend on them. 
The two independent solutions of the Mathieu equation \eqref{Mcseq} can be written as discussed in Sec.~\ref{FBth}, i.e., 
\begin{subequations}\label{Mcsf}
\begin{align}
  M_c(t) =2\alpha\, {\rm Re} f(t) = 2\alpha \sum_m c_m \cos\left((\epsilon+ m\omega)t\right), \\
  M_s(t) = 2\beta\, {\rm Im} f(t) = 2\beta \sum_m c_m \sin\left((\epsilon+ m\omega)t\right).
\end{align}
\end{subequations}
An exact solution should obey the canonical commutation
relations which is equivalent to 
obeying Eq.~\eqref{Mcan}
at all times.
Substituting Eq.~\eqref{Mcsf} in the latter condition gives the equivalent request that 
\begin{align}
 1=4\alpha \beta \sum_{m,n} c_m c_n(\epsilon + n\omega) \cos\left((m-n)\omega t\right).
 \label{app-sum-ccr}
\end{align}
By introducing the variable \(m-n=p\), and by
splitting the sum in Eq.~\eqref{app-sum-ccr} into a time-independent part corresponding to \(p=0\)
and a time-dependent part with \(p\neq 0\), we obtain
\begin{equation}
\begin{split}
  1&=4\alpha \beta \biggl\{\sum_n (\epsilon + n\omega) c_n^2  + \sum_{p\neq 0,n} (\epsilon + n\omega) c_n c_{n+p} \\
  &\quad + 
  \sum_{p\neq 0} \left[\cos(p \omega t)-1\right]\sum_n (\epsilon + n\omega) c_n c_{n+p}
  \biggr\}.
 \end{split}
\end{equation}
By requiring that the r.h.s.~of this equation is time-independent, we need the coefficient of the last term to vanish, i.e.,
\begin{align}
  0= \sum_n (\epsilon + n\omega) \left(c_n c_{n+p}+c_{n}c_{n-p}\right), \quad \text{for} \quad p\neq 0, \label{int1}
\end{align}
and the time-independent part needs to equal 1, i.e.,
\begin{align}
1 = 4\alpha \beta \sum_{m,n} c_mc_n(\epsilon + n\omega). \label{int2}
\end{align}
In our perturbative treatment, we kept only the two terms with coefficients \(c_0\) and \(c_{-1}\) (see Eq.~\eqref{c10gen}) and have argued that the smallness of the remaining coefficients is controlled by \(q\). 
Thus our truncated solution is
\begin{align}
  M_c = 2\alpha c_0 \biggl[\cos(\epsilon t) + \frac{c_{-1}}{c_0} \cos((\epsilon-\omega)t)\biggr],
  \label{app-ccr-Mc}\\
  M_s = 2\beta c_0 \biggl[\sin(\epsilon t) + \frac{c_{-1}}{c_0} \sin((\epsilon-\omega)t)\biggr],
  \label{app-ccr-Ms}
\end{align}
and we imposed
the validity of the commutation relation at the initial time 
$t=0$, corresponding to Eq.~\eqref{int2}, i.e.,
\begin{align}
1 = 4\alpha \beta\biggl[c_0^2 \epsilon + c_{-1}^2 (\epsilon-\omega) + c_0 c_{-1}(2\epsilon-\omega)\biggr]. \label{int3}
\end{align}

We can see from Eq.~\eqref{int1} that, in order to cancel the time-dependence with \(p=1\), we need to retain both \(c_1\) and \(c_{-2}\). In turn, keeping these terms requires keeping more terms in the expansion and therefore any truncation of the series will always result into residual oscillations.  
The magnitude of the associated error can be calculated by evaluating the r.h.s.~of Eq.~\eqref{Mcan} on the perturbative solutions \eqref{app-ccr-Mc} and \eqref{app-ccr-Ms} which, after imposing Eq.~\eqref{int3}, becomes
\begin{widetext}
\begin{align}
M_c \dot{M}_s - M_s \dot{M}_c  &=  1 -   
\frac{2(2\epsilon-\omega)  c_{-1}/c_0}{\epsilon + \left(\frac{c_{-1}}{c_0}\right)^2 (\epsilon-\omega) + \left(\frac{c_{-1}}{c_0}\right) (2\epsilon-\omega)} 
\sin^2(\omega t/2).
\label{eq:app-err-ccr}
\end{align}
\end{widetext}
By direct inspection of this equation one realizes that the largest magnitude of the error in the canonical commutation occurs at small momenta $k\ll \sqrt{q}\omega$. Accordingly, the corresponding coefficient of the time-dependent term in Eq.~\eqref{eq:app-err-ccr} can be determined by using Eqs.~\eqref{dis1a} and \eqref{ksm} with the conclusion that the error in the commutation relations is $q \sin^2(\omega t/2)$, i.e., of \(O(q)\) at small \(k\).
This error is further suppressed at intermediate and large \(k\), as discussed in Appendix~\ref{microAp} and explicitly shown in Eq.~\eqref{Verr1}.

\section{Micromotion Operator}\label{microAp}

Applying the Floquet-Bloch theorem, reviewed in Appendix \ref{appA}, the matrix which generates 
the time evolution (see Eq.~\eqref{timeev}) obeys
\begin{align}
\textbf{M}_k(t_2,0) &= \textbf{P}_k(t_2)e^{i\textbf{B}_kt_2},\nonumber \\
&= \textbf{P}_k(t_2)\textbf{U}_k^{-1} e^{i\textbf{B}_{D,k}t_2}\textbf{U}_k,\label{mkt0}
\end{align}
where \(\textbf{B}_k=\textbf{U}_k^{-1}\textbf{B}_{D,k}\textbf{U}_k\), and \(\textbf{B}_{D,k}\) is a diagonal matrix. Inserting this equality and its inverse evaluated for $t_2\mapsto t_1$ into 
Eq.~\eqref{eq:app-Mt2t1-prod} 
we obtain
\begin{align}
\textbf{M}_k(t_2,t_1) &= \textbf{P}_k(t_2)e^{i\textbf{B}_k(t_2-t_1)} \textbf{P}_k^{-1}(t_1), \nonumber \\
&= \textbf{P}_k(t_2)\textbf{U}_k^{-1}e^{i\textbf{B}_{D,k}(t_2-t_1)}\textbf{U}_k \textbf{P}_k^{-1}(t_1).\label{Mkdef}
\end{align}
Our goal here is to write $\textbf{M}_k$ above in terms of the two rotation matrices \(\mathbf{V}_k\) and \(\mathbf{F}_k\) introduced in Eq.~\eqref{eq:M-VFV}, which we have to determine.
The two matrices  \(\mathbf{V}_k(t_2)\) and \(\mathbf{V}^{-1}_k(t_1)\) in Eq.~\eqref{eq:M-VFV} capture micromotion, while the third matrix in the same equation performs the rotation due to the time evolution controlled by the Floquet Hamiltonian \(H_F\), see Eq.~\eqref{Fdef1}.

We will now use the fact that \(\textbf{M}_k(t,0)\)  can be written in terms of \(M_{c,k}\) and \(M_{s,k}\) as in Eq.~\eqref{Mt0} and that the latter can be related to the Floquet quasi modes $f_k$ and $f^*_k$ as in Eq.~\eqref{Mcsdef}.
We also find it convenient to define the phase \(\Omega_k(t)\) of the modes $u_k$ introduced in Eq.~\eqref{ukdef} as \begin{equation}
u_k(t) = |u_k(t)| e^{i \Omega_k(t)}\label{Omdef1}
\end{equation}
so that the latter equation implies
\begin{equation}
f_k(t) = u_k(t)e^{i\epsilon_k t} = |f_k(t)|e^{i\Omega_k(t)}e^{i\epsilon_k t}.
\label{Omdef}
\end{equation}
Using these expressions we can write
\begin{widetext}
\begin{align}\label{Mkt0}
\textbf{M}_k(t,0)&=\left(
\begin{tabular}{cc}
\(M_{c,k}(t)\) & \(M_{s,c}(t)\)  \\
\(\dot{M}_{c,k}(t)\) & \(\dot{M}_{s,k}(t)\)
\end{tabular}
\right)=
\left(
\begin{tabular}{cc}
\(f_k^*(t)\) & \(f_k(t)\)  \\
\(\dot{f}_k^*(t)\) & \(\dot{f}_k(t)\)
\end{tabular}
\right)
\left(
\begin{tabular}{cc}
\(\alpha_k\) & \(i\beta_k\)  \\
\(\alpha_k\) & \(-i\beta_k\)
\end{tabular}
\right)\nonumber \\
&=
\left(
\begin{tabular}{cc}
\(u_k^*(t)\) & \(u_k(t)\)  \\
\(\dot{u}_k^*(t)-i\epsilon_k u_k^*(t)\) & \(\dot{u}_k(t)+i\epsilon_k u_k(t)\)
\end{tabular}
\right)
\left(
\begin{tabular}{cc}
\(e^{-i\epsilon_k t}\) & \(0\)  \\
\(0\) & \(e^{i\epsilon_k t}\)
\end{tabular}
\right)
\left(
\begin{tabular}{cc}
\(\alpha_k\) & \(i\beta_k\)  \\
\(\alpha_k\) & \(-i\beta_k\)
\end{tabular}
\right) \nonumber \\
&=
\left(
\begin{tabular}{cc}
\(u_k^*(t)\) & \(u_k(t)\)  \\
\(\dot{u}_k^*(t)-i\epsilon_k u_k^*(t)\) & \(\dot{u}_k(t)+i\epsilon_k u_k(t)\)
\end{tabular}
\right)\left(
\begin{tabular}{cc}
\(\alpha_k\) & \(i\beta_k\)  \\
\(\alpha_k\) & \(-i\beta_k\)
\end{tabular}
\right)\frac{1}{-2i\alpha_k\beta_k}\left(
\begin{tabular}{cc}
\(-i\beta_k\) & \(-i\beta_k\)  \\
\(-\alpha_k\) & \(\alpha_k\)
\end{tabular}
\right)
\left(
\begin{tabular}{cc}
\(e^{-i\epsilon_k t}\) & \(0\)  \\
\(0\) & \(e^{i\epsilon_k t}\)
\end{tabular}
\right)
\left(
\begin{tabular}{cc}
\(\alpha_k\) & \(i\beta_k\)  \\
\(\alpha_k\) & \(-i\beta_k\)
\end{tabular}
\right),
\end{align}
\end{widetext}
where above, we have inserted the identity
\begin{align}
\mathbf{I}_{2\times2} = \left(\begin{tabular}{cc}
\(\alpha_k\) & \(i\beta_k\)  \\
\(\alpha_k\) & \(-i\beta_k\)
\end{tabular}
\right)\frac{1}{-2i\alpha_k\beta_k}\left(
\begin{tabular}{cc}
\(-i\beta_k\) & \(-i\beta_k\)  \\
\(-\alpha_k\) & \(\alpha_k\)
\end{tabular}\right). \nonumber
\end{align}
Comparing Eqs.~\eqref{mkt0} and \eqref{Mkt0}, we conclude that
\begin{equation}
\textbf{B}_{D,k} = \left(
\begin{matrix}
-\epsilon_k & 0  \\
0 & \epsilon_k
\end{matrix}
\right),
\end{equation}
\begin{equation}
\begin{split}
\textbf{U}^{-1}_ke^{i\textbf{B}_{D,k} t}\textbf{U}_k &= \frac{1}{-2i\alpha_k\beta_k}\left(
\begin{matrix}
-i\beta_k & -i\beta_k  \\
-\alpha_k & \alpha_k
\end{matrix}
\right)\\
&\quad \times
\left(
\begin{matrix}
e^{-i\epsilon_k t} & 0  \\
0 & e^{i\epsilon_k t}
\end{matrix}
\right)
\left(
\begin{matrix}
\alpha_k & i\beta_k  \\
\alpha_k & -i\beta_k
\end{matrix}
\right), \\
\end{split}
\end{equation}
and
\begin{equation}
\begin{split}
\textbf{P}_k(t) &= \left(
\begin{tabular}{cc}
\(u_k^*(t)\) & \(u_k(t)\)  \\
\(\dot{u}_k^*(t)-i\epsilon_k u_k^*(t)\) & \(\dot{u}_k(t)+i\epsilon_k u_k(t)\)
\end{tabular}
\right)\\
&\quad \times\left(
\begin{tabular}{cc}
\(\alpha_k\) & \(i\beta_k\)  \\
\(\alpha_k\) & \(-i\beta_k\)
\end{tabular}
\right).\label{Pk}
\end{split}
\end{equation}

The canonical commutation relation \([\phi_k(0),\Pi_k(0)]=1\) further imposes
\begin{equation}\label{dett2}
\begin{split}
\textrm{det}[\textbf{M}_k(0,0)] &=\textrm{det}[\textbf{P}_k(0)] \textrm{det}[\textbf{U}_k^{-1}]\textrm{det}[\textbf{U}_k] \\
 &= \textrm{det}[\textbf{P}_k(0)] = 1.
 \end{split}
\end{equation}
Using the explicit form of \(\textbf{P}_k(0)\) in Eq.~\eqref{Pk} we obtain
\begin{equation}
\textrm{det}[\textbf{P}_k(0)]= -2i\alpha_k \beta_k \left(2i\epsilon_k+2i\dot{\Omega}_k(0)\right)|f_k(0)|^2=1,
\end{equation}
which gives the condition
\begin{equation}\label{albe2}
\alpha_k \beta_k  =\frac{1}{4\left(\epsilon_k+\dot{\Omega}_k(0)\right)|f_k(0)|^2}.
\end{equation}

As shown in Appendix \ref{commr}, for the canonical commutation relation 
to hold at all times, an exact solution of the Mathieu equation
is needed. 
Since the solution in Eq.~\eqref{ftu} is truncated, it yields a solution with an \(O(q)\) error to the commutation relation at small momenta (the error is smaller at larger momenta, as we show below). 
In addition, if \(f_k^{(e)}(t)\) is an exact solution of the Mathieu equation,
then \(\textrm{det}[\textbf{P}^{(e)}_k(t)]\) is an integral of motion which is proportional to \(\text{Im}[{f_k^{(e)}}^* \dot{f}^{(e)}_k]\),
\begin{equation}
\textrm{det}[\textbf{P}^{(e)}_k(t)]= -2i\alpha_k \beta_k \left(2i\epsilon_k+2i\dot{\Omega}_k\right)|f_k^{(e)}(t)|^2=1.
\end{equation}

Let us define the matrix \(\mathbf{R}_k\) which performs the rotation from position-momentum fields to
creation-annihilation operators,
\begin{equation}\label{adef}
\left(
\begin{tabular}{c}
\(\phi_k\) \\
\(\Pi_k\)
\end{tabular}
\right) = \mathbf{R}_k \left(
\begin{tabular}{c}
\(\ a_\mathbf{k}\ \) \\
\(\ a^\dagger_\mathbf{-k}\)
\end{tabular}
\right),
\end{equation}
where,
\begin{equation}\label{Rk}
\mathbf{R}_k = \frac{1}{\sqrt{2\epsilon_k}}
\left(
\begin{tabular}{cc}
\(1\) & \(1\)  \\
\(-i\epsilon_k\) & \(i\epsilon_k\) \end{tabular}
\right), 
\end{equation}
with $\textrm{det}[\textbf{R}_k]=i$.  The creation and annihilation operators $a^\dagger_\mathbf{k}$ and $a_\mathbf{k}$ indicated here are those
which diagonalize \(H_F\) in Eq.~\eqref{Hf}.
Upon inserting the matrices \(\mathbf{R}_k\) and \(\mathbf{R}_k^{-1}\) in Eq.~\eqref{Mkdef} we obtain,
\begin{align}
 \textbf{M}_k(t_2,t_1) &= \textbf{P}_k(t_2)\textbf{U}_k^{-1}\textbf{R}_k^{-1} \nonumber\\
  &\quad\times\textbf{R}_ke^{i\textbf{B}_{D,k}(t_2-t_1)}\textbf{R}_k^{-1}\textbf{R}_k\textbf{U}_k \textbf{P}_k^{-1}(t_1) \nonumber \\
&= \mathbf{V}_k(t_2) \mathbf{F}_k(t_2-t_1) \mathbf{V}^{-1}_k(t_1).\label{VFder}
\end{align}
Accordingly, \(\mathbf{F}_k(t)\) can be obtained from above as 
\begin{equation}
\begin{split}
\textbf{F}_k(t)&=\textbf{R}_k e^{i\mathbf{B}_{D,k} t}\textbf{R}^{-1}_k=\textbf{R}_k
\left(
\begin{tabular}{cc}
\(e^{-i\epsilon_k t}\) & \(0\)  \\
\(0\) & \(e^{i\epsilon_k t}\)
\end{tabular}
\right)
\textbf{R}_k^{-1} \\
&=\left(
\begin{tabular}{cc}
\(\cos(\epsilon_k t)\) & \(\frac{1}{\epsilon_k}\sin(\epsilon_k t)\) \\
\(-\epsilon_k \sin(\epsilon_k t)\) & \(\cos(\epsilon_k t)\)
\end{tabular}
\right).
\end{split}
\end{equation}
Moreover, from Eq.~\eqref{VFder}, we identify \(\mathbf{V}_k(t)\) to be
\begin{equation}
\mathbf{V}_k(t) = \mathbf{P}_k (t) \mathbf{U}_k^{-1} \mathbf{R}_k^{-1}.
\end{equation}

Recall that in order for the commutation relation between \(\phi_k(0)\) and \(\Pi_k(0)\) to be equal to 1, \(\mathrm{det}[\mathbf{P}_k(0)]=1\) as shown in Eq.~\eqref{dett2}.
Moreover, preserving the commutation relation between the rotated fields obtained after the 
application of
 \(\mathbf{V}_k(0)\) requires \(\mathrm{det}[\mathbf{V}_k(0)]=\mathrm{det}[\mathbf{P}_k(0)]\mathrm{det}
 [\mathbf{U}_k^{-1}]\mathrm{det}[\mathbf{R}_k^{-1}]=1\),
with \(\mathrm{det}[\mathbf{R}_k]=i\) and \(\mathrm{det}[\mathbf{U}_k]=-i\). 
The matrix \(\mathbf{U}_k\) which satisfies this requirement is
\begin{equation}
\textbf{U}_k = \frac{1}{\sqrt{2\alpha_k\beta_k}}\left(
\begin{tabular}{cc}
\(\alpha_k\) & \(i\beta_k\)  \\
\(\alpha_k\) & \(-i\beta_k\)
\end{tabular}
\right).
\label{Uk}
\end{equation}

Using Eqs.~\eqref{Pk}, \eqref{Uk}, and \eqref{albe2} we can write
\begin{widetext}
\begin{align}
\textbf{P}_k (t) \mathbf{U}_k^{-1} &=\frac{1}{\sqrt{2\epsilon_k+2\dot{\Omega}_k(0)}|f_k(0)|}\left(
\begin{tabular}{cc}
\(u_k^*(t)\) & \(u_k(t)\)  \\
\(\dot{u}_k^*(t)-i\epsilon_k u_k^*(t)\) & \(\dot{u}_k(t)+i\epsilon_k u_k(t)\)
\end{tabular}
\right).
\end{align}
Thus the micromotion matrix is
\begin{equation}
\mathbf{V}_k(t) = \textbf{P}_k (t) \mathbf{U}_k^{-1}\textbf{R}_k^{-1} = \frac{1}{\sqrt{1+\frac{\dot{\Omega}_k(0)}{\epsilon_k}}|f_k(0)|}\left(
\begin{tabular}{cc}
\(\text{Re}[u_k(t)]\) & \(\frac{1}{\epsilon_k}\text{Im}[u_k(t)]\)  \\
\(\text{Re}\left[\dot{u}_k(t)+i\epsilon_k u_k(t)\right]\ \) & \(\ \frac{1}{\epsilon_k}\text{Im}\left[\dot{u}_k(t)+i\epsilon_k u_k(t)\right]\) \end{tabular}
\right).
\end{equation}
Using Eq.~\eqref{Omdef1}, the previous equation becomes
\begin{multline}
\mathbf{V}_k(t) = \frac{1}{\sqrt{1+\frac{\dot{\Omega}_k(0)}{\epsilon_k}}} \frac{|f_k(t)|}{|f_k(0)|}
\\ \times
\left(
\begin{tabular}{cc}
\(\cos(\Omega_k(t))\) & \(\ \ \ \frac{1}{\epsilon_k}\sin{(\Omega_k(t))}\) \\
\(-\epsilon_k \left(1+\frac{\dot{\Omega}_k(t)}{\epsilon_k}\right) \sin(\Omega_k(t))+\frac{d\ln(|f_k(t)|)}{dt}\cos(\Omega_k(t))\) & \(\ \ \ \left(1+\frac{\dot{\Omega}_k(t)}{\epsilon_k}\right)\cos{(\Omega_k(t))}+\frac{1}{\epsilon_k}\frac{d\ln(|f_k(t)|)}{dt}\sin(\Omega_k(t))\)
\end{tabular}
\right), \label{eq:app-expr-V}
\end{multline}
\end{widetext}
and
\begin{equation}
\det[\mathbf{V}_k(t)] = \frac{1+\frac{\dot{\Omega}_k(t)}{\epsilon_k}} {1+\frac{\dot{\Omega}_k(0)}{\epsilon_k}} \frac{|f_k(t)|^2}{|f_k(0)|^2}, \label{detVa}
\end{equation}
with
\begin{equation}
\begin{split}
&\det\left[\mathbf{M}_k(t_2,t_1)\right]\\
&=\det[\mathbf{V}_k(t_2)]\det[\mathbf{F}_k(t_2-t_1)]\det[\mathbf{V}^{-1}_k(t_1)]\\
&= \frac{1+\frac{\dot{\Omega}_k(t_2)}{\epsilon_k}} {1+\frac{\dot{\Omega}_k(t_1)}{\epsilon_k}} \frac{|f_k(t_2)|^2}{|f_k(t_1)|^2}. \label{detMa}
\end{split}
\end{equation}
For an exact solution, Eqs.~\eqref{detVa} and \eqref{detMa} would equal 1. Thus these
two equations provide a way to quantify the error in the commutation relations arising from
the truncation in Sambe space.

Near the critical line defined in Eq.~\eqref{critp} and for small drive amplitudes \(q\ll 1\), \( u_k(t)\) can be approximated by truncating the infinite series where all the coefficients except \(c_0\) and \(c_{-1}\) vanish. In addition, \(u_k(t)\) can be normalized such that \(c_0=1\).  We write
\begin{equation}\label{udef}
u_k(t) \approx 1 + \frac{c_{-1}}{c_0}e^{-i\omega t} = 1 + \sigma_k e^{-i\omega t},
\end{equation}
where, for later convenience, we introduce \(\sigma_k = c_{-1}/c_0\) having the following form at small and intermediate momenta (see Eq.~\eqref{c10gen}),
\begin{equation}\label{sigk}
\sigma_k 
\approx
\left\{
\begin{split}
&
1 - 4\frac{\overline{k}}{q\omega} \quad \mbox{for} 
& k\ll \sqrt{q} \omega \ll \omega,\\
&
\frac{1}{16}\frac{q^2 \omega^2}{\bar{k}^2} \quad \mbox{for} 
&
\sqrt{q} \omega \ll k\ll \omega.
\end{split}
\right.
\end{equation}

In the subsequent derivations, the following identities, derived on the basis of
Eqs.~\eqref{Omdef} and \eqref{udef} will be helpful, 
\begin{align}
|f_k(t)| &= \sqrt{1+\sigma_k^2+2\sigma_k \cos(\omega t)},\\
\frac{d \ln(|f_k(t)|)}{dt}&= \frac{-\omega\sigma_k \sin(\omega t)}{1+\sigma_k^2+2\sigma_k \cos(\omega t)},\\
\cos\Omega_k(t) &= \frac{1+\sigma_k \cos(\omega t)}{\sqrt{1+\sigma_k^2+2\sigma_k \cos(\omega t)}},\\
\sin\Omega_k(t) &= \frac{-\sigma_k \sin(\omega t)}{\sqrt{1+\sigma^2+2\sigma_k \cos(\omega t)}},\\
\dot{\Omega}_k &=-\omega \frac{\sigma_k^2+\sigma_k \cos(\omega t)}{1+\sigma_k^2+2\sigma_k \cos(\omega t)}.
\end{align}
In the two subsections below we investigate the micromotion operator in the two relevant limits we have identified in this work, i.e., the one of small momenta \(k \ll \sqrt{q} \omega \ll \omega\) and the other of intermediate momenta $k$ with \(\sqrt{q} \omega \ll k\ll \omega\).

\subsection{Micromotion operator for \(\sqrt{q} \omega \ll k\ll \omega\)}

In this case of intermediate momenta, Eq.~\eqref{sigk} implies
\(\sigma_k \approx q^2\omega^2/16\bar{k}^2 \ll 1 \). Keeping terms which are
linear in \(q^2\), we obtain
\begin{align}
|f_k(t)| &\approx 1 + \frac{1}{16} \frac{q^2\omega^2}{\overline{k}^2}\cos(\omega t),\\
\frac{d \ln(|f_k(t)|)}{dt}&\approx -\frac{1}{16} \frac{q^2\omega^2}{\overline{k}^2}\omega \sin(\omega t),\\
\sin\Omega_k(t) &\approx -\frac{1}{16} \frac{q^2\omega^2}{\overline{k}^2}\sin(\omega t),\\
\Omega_k(t) &\approx -\frac{1}{16} \frac{q^2\omega^2}{\overline{k}^2}\sin(\omega t),\\
\cos\Omega_k(t) &\approx 1.
\end{align}
These expressions, inserted in Eq.~\eqref{eq:app-expr-V} render Eq.~\eqref{V1}.
The error in the determinant of \(\mathbf{V}_k\) due to the truncation in Sambe space is of the form
\begin{equation}\label{Verr1}
\det[\mathbf{V}_k(t)] =1 + O\left(q^4\omega^4/\overline{k}^4\right),
\end{equation}
i.e., as anticipated, of higher-order in $q$ compared to Eq.~\eqref{eq:app-err-ccr}.

\subsection{Micromotion operator for \(k\ll \sqrt{q} \omega \ll \omega\)}

In this limit of small momenta, Eq.~\eqref{sigk} gives \(\sigma_k \approx 1-4\bar{k}/(q\omega)\).
Defining \(\delta_k = 1-\sigma_k \approx 4\bar{k}/q\omega \ll 1\), some helpful relations are
\begin{align}
|f_k(t)| &= \sqrt{2(1-\delta_k)(1+\cos(\omega t))+\delta_k^2},\\
\frac{d \ln(|f_k(t)|)}{dt}&= \frac{-\omega(1-\delta_k) \sin(\omega t)}{2(1-\delta_k)(1+\cos(\omega t))+\delta_k^2},\\
\cos\Omega_k(t) &= \frac{1+(1-\delta_k) \cos(\omega t)}{\sqrt{2(1-\delta_k)(1+\cos(\omega t))+\delta_k^2}},\\
\sin\Omega_k(t) &= \frac{-(1-\delta_k) \sin(\omega t)}{\sqrt{2(1-\delta_k)(1+\cos(\omega t))+\delta_k^2}},\\
\dot{\Omega}_k &=-\omega \frac{(1-\delta_k)^2+(1-\delta_k) \cos(\omega t)}{1+(1-\delta_k)^2+2(1-\delta_k) \cos(\omega t)}.
\end{align}
Expanding \(\mathbf{V}_k(0)\) from Eq.~\eqref{eq:app-expr-V} in powers of \(\bar{k}/(q\omega)\), we will keep the first two terms, as keeping only the first leading term will result in a singular matrix with zero determinant.
Accordingly, we have
\begin{align}
\cos\Omega_k(t) &\approx \frac{1}{\sqrt{2}} \sqrt{1+\cos(\omega t)}+\frac{\sqrt{2}\bar{k}}{q\omega}\frac{1-\cos(\omega t)}{\sqrt{1+\cos(\omega t)}}\nonumber\\
&=\left|\cos\left(\frac{\omega}{2}t\right)\right|+\frac{2\bar{k}}{q\omega}\frac{\sin^2(\omega t)}{\left|\cos\left(\frac{\omega}{2}t\right)\right|},
\end{align}
\begin{align}
\sin\Omega_k(t) &\approx -\left(1-\frac{2\bar{k}}{q \omega}\right)\frac{\sin(\omega t)}{\sqrt{2+2\cos(\omega t)}}
\nonumber\\
&=-\left(1-\frac{2\bar{k}}{q \omega}\right)\frac{\left|\cos\left(\frac{\omega}{2}t\right)\right|}{\cos\left(\frac{\omega}{2}t\right)}\sin\left(\frac{\omega}{2}t\right),
\end{align}
\begin{align}
1+\frac{\dot{\Omega}_k(t)}{\epsilon_k}&\approx \left(\frac{2\bar{k}}{q\omega} + \left(\frac{2\bar{k}}{q\omega} \right)^2 \right) \frac{1}{\cos^2\left(\frac{\omega t}{2}\right)} +\frac{2\bar{k}}{q\omega}  q,\\
|f_k(t)| &\approx \left(1+\frac{2\bar{k}}{q\omega}\right)2\left|\cos\left(\frac{\omega }{2}t\right)\right|,
\end{align}
\begin{align}
\frac{d \ln(|f_k(t)|)}{dt} &\approx -\left(1-\frac{2\bar{k}}{q\omega}\right)\frac{\omega}{2}\tan\left(\frac{\omega}{2}t\right).
\end{align}
These approximate expressions, once inserted into Eq.~\eqref{eq:app-expr-V}, give
\begin{widetext}
\begin{multline}\label{Vkf}
\mathbf{V}_k(t) =\frac{1}{2} \sqrt{\frac{q\omega}{2\overline{k}}}\left[\left(
\begin{tabular}{cc}
\(1+\cos\left(\omega t\right)\) & \(\ \ \ \ \ -\frac{2}{\omega} \sin\left(\omega t\right)\) \\
\(-\frac{\omega}{2} \sin\left(\omega t\right)\) & \(\ \ \ \ \ 1-\cos\left(\omega t\right)\)
\end{tabular}
\right)\right. \\ 
+\frac{2\bar{k}}{q \omega}\left(
\begin{tabular}{cc}
\( \frac{1}{2} \left[1-\cos\left(\omega t\right)-2\cos\left(2\omega t\right)\right]\) & \(\ \ \ \ \ \frac{3}{\omega} \sin\left(\omega t\right)\) \\
\(-\frac{\omega}{4} \left[\sin\left(\omega t\right)-2\sin\left(2\omega t\right) -4\tan\left(\omega t/2\right)\right]\) & \(\ \ \ \ \ \frac{1}{2} \left[5\cos\left(\omega t\right)-1\right]\)
\end{tabular}
\right)\\ 
+q \left(
\begin{tabular}{cc}
\(-\frac{1}{2}\left(1+\cos\left(\omega t\right)\right)\) & \(\ \ \ \ \ \frac{1}{\omega} \sin\left(\omega t\right)\) \\
\(\frac{\omega}{4} \sin\left(\omega t\right)\) & \(\ \ \ \ \  \frac{1}{2} \left(\cos\left(\omega t\right)-1\right)\)
\end{tabular}
\right)\\ + \left. q \frac{2\bar{k}}{q \omega} \left(
\begin{tabular}{cc}
\(\frac{1}{4}\left(1+3\cos\left(\omega t\right)+2\cos\left(2\omega t\right)\right)\) & \(\ \ \ \ \ -\frac{5}{2\omega}\sin(\omega t)\) \\
\(\frac{\omega}{8} \left( 3\sin(\omega t)-2\sin(2\omega t) -4\tan(\omega t/2)\right) \) & \(\ \ \ \ \  \frac{1}{4} \left(7-3\cos\left(\omega t\right)\right)\)
\end{tabular}
\right)
\right].
\end{multline}
\end{widetext}
At the leading order in the expansion for small momenta this expression renders Eq.~\eqref{V2}.

The error in the determinant of \(\mathbf{V}_k\) from the truncation in Sambe space is of the form
\begin{equation}\label{Verr2}
\det[\mathbf{V}_k(t)] = 1 - q \sin^2(\omega t/2) + O\left(q^2\right)+O\left(\frac{2\bar{k}}{q\omega}\right),
\end{equation}
i.e., of the same order as that found in Eq.~\eqref{eq:app-err-ccr}.

\section{Derivation of Eq.~\eqref{UF1}} 
\label{UFder}

In order to derive the expression reported in Eq.~\eqref{UF1} for $U_F$, we start from Eq.~\eqref{Vkf}, which obeys Eq.~\eqref{Vdef}.  Consider the matrix $\mathbf{R}_k$ in Eq.~\eqref{Rk} which transforms 
the rotation from position-momentum fields to creation-annhiliation operators. 
Combining Eqs.~\eqref{adef} and \eqref{Vdef}, we obtain,
\begin{equation}
U_F^{\dagger} \left(
\begin{tabular}{c}
\(\ a_\mathbf{k}\ \) \\
\(\ a^\dagger_\mathbf{-k}\)
\end{tabular}
\right) U_F= \mathbf{R}^{-1}_k  \mathbf{V}_k \mathbf{R}_k \left(
\begin{tabular}{c}
\(\ a_\mathbf{k}\ \) \\
\(\ a^\dagger_\mathbf{-k}\)
\end{tabular}
\right),
\label{eq:app-defUF}
\end{equation}
where we define \(u_k\) and \(v_k\) such that
\begin{equation}
\label{RVR}
\mathbf{R}^{-1}_k  \mathbf{V}_k \mathbf{R}_k = \left(
\begin{tabular}{cc}
\(u_k  \) & \(v_k \)  \\
\(v_k^*\) & \(u_k^*\) \end{tabular}
\right).
\end{equation}
From Eqs.~\eqref{Vkf} and \eqref{Rk} it follows that, for \(\bar{k}/q\omega \ll 1\),
\begin{widetext}
\begin{subequations}\label{uvdef}
\begin{align}
u_k &= \frac{1}{2}\sqrt{\frac{q\omega}{2\bar{k}}} \left\{ 1 +\frac{2\bar{k}}{q\omega}\left[ \cos(\omega t) -\frac{1}{2} \cos(2\omega t) + i \cos^2(\omega t) \tan(\omega t/2) \right] \right\},\\
v_k &= \frac{1}{2}\sqrt{\frac{q\omega}{2\bar{k}}} \left\{ e^{-i\omega t}   +\frac{2\bar{k}}{q\omega}\left[ \frac{1}{2}-\frac{3}{2}\cos(\omega t) -\frac{1}{2}\cos(2\omega t)+ i\left( \frac{1}{2}\sin(\omega t) + \frac{1}{2}\sin(2\omega t) + \tan(\omega t/2) \right) \right] \right\}.
\end{align}
\end{subequations}
\end{widetext}
The eigenvalues of \(U_F(t)\) in Eq.~\eqref{eq:app-defUF} are those of the matrix in Eq.~\eqref{RVR}, with the elements reported in Eq.~\eqref{uvdef}. At small $k$ we find these eigenvalues to be, at the leading order,
\begin{align}\label{eq:UF-eigenv}
\left(\frac{q\omega}{2 \bar{k}}\right)^{1/2}\quad\mbox{and}\quad \left(\frac{q\omega}{2 \bar{k}}\right)^{-1/2},
\end{align}
as anticipated in the text after Eq.~\eqref{UF1}. 
In the limit of small momenta and weak drive, these eigenvalues are time-independent.

Now that the action of $U_F$ on the creation and annihilation operators is known from Eqs.~\eqref{eq:app-defUF},  \eqref{RVR}, and \eqref{uvdef} we would like to determine the form of the operator \(U_F\).
Since the various momenta labeled by $k$ are independent, this is essentially a single-mode problem
and therefore we can simplify the notation by suppressing the momentum label. 
The form of the transformation induced by $U_F$ on the operators $a$ and $a^\dagger$ suggests that $U_F$ should take the generic form 
\begin{equation}
U(\beta,\sigma) = e^{O},
\label{eq:app-def-UO}
\end{equation}
parameterized by a real and a complex number $\beta$ and $\sigma$, respectively with
\begin{equation}
O = i \left(
\begin{tabular}{cc}
\(a^\dagger \) & \(a\)
\end{tabular}
\right) \left(
\begin{tabular}{cc}
\(\beta \) & \(\sigma\) \\
\(\sigma^* \) & \(\beta\)
\end{tabular}
\right) \left( \begin{tabular}{c}
\(a \) \\ \(a^\dagger\)
\end{tabular}
\right),
\label{eq:app-defO}
\end{equation}
such that $O^\dagger = - O$.
The action of the operator $U$ on $a$ can be easily determined by expanding the exponential:
\begin{equation}
U^\dagger a U = e^{-O} a e^{O} = 
\sum_{n=0}^{\infty} \frac{1}{n!}C^{(n)},
\label{eq:app-Uexp}
\end{equation}
where 
\begin{equation}
C^{(0)} = a \quad\mbox{and}\quad
C^{(n+1)} = \left[ C^{(n)} ,O\right], \quad n=0, 1, \ldots.
\end{equation}
Substituting in the above equations the operator $O$ defined in Eq.~\eqref{eq:app-defO} one obtains 
\begin{align}
C^{(2n+1)} &= \lambda^{2n} \left( 2i \sigma a^\dagger + 2i \beta a \right), \quad n=0, 1, \ldots\nonumber\\
C^{(2n)} &= \lambda^{2n} a, \quad n = 1, 2, \ldots ,\nonumber
\end{align}
where, for later convenience, we introduced
\begin{equation}
\lambda = \sqrt{4|\sigma|^2 - 4\beta^2}.
\end{equation}
Inserting these expressions in Eq.~\eqref{eq:app-Uexp}, one finds
\begin{equation}
U^\dagger a U 
= \sum_{n=0}^{\infty} \frac{\lambda^{2n+1}}{(2n+1)!} 
\frac{\left( 2i \sigma a^\dagger + 2i \beta a \right)}{\lambda} + \sum_{n=0}^{\infty} \frac{\lambda^{2n}}{(2n)!}  a ,
\end{equation}
in which the series can be resummed and yields 
\begin{equation}\label{Ue}
  U^\dagger a U 
  = \left[ \cosh \lambda  +2i \beta \frac{\sinh\lambda}{\lambda}  \right] a
  +2i\sigma\frac{\sinh \lambda}{\lambda}  a^\dagger.
\end{equation}
By comparing Eq.~\eqref{Ue} with Eqs.~\eqref{eq:app-defUF} and \eqref{RVR}
one can easily identify 
\begin{equation}
u =  \cosh \lambda  +2i \beta\frac{\sinh\lambda}{\lambda}\quad\mbox{and}\quad
v =  2i\sigma \frac{\sinh \lambda}{\lambda}.
\end{equation}
Solving for \(\beta\) and \(\sigma\) in terms of \(u\) and \(v\), we obtain
\begin{subequations}
\label{eq:app-beta-sigma}
\begin{align}
\beta &= \frac{\ln  \left(u_R + \sqrt{u^2_R-1} \right)}{2\sqrt{u^2_R-1}} u_I ,\\
\sigma &= -\frac{\ln  \left(u_R + \sqrt{u^2_R-1} \right)}{2\sqrt{u^2_R-1}} iv,
\end{align}
\end{subequations}
where $u_R = (u+u^*)/2$ and $u_I = (u - u^*)/(2 i)$ are the real and imaginary parts of $u$, respectively.
In the case we are actually interested in, \(u\rightarrow u_k\) and \(v\rightarrow v_k\). 
Using the explicit expressions of $u_k$ and $v_k$ in Eq.~\eqref{uvdef} in order to determine the corresponding $\beta_k$ and $\sigma_k$ from Eq.~\eqref{eq:app-beta-sigma}, 
one obtains the forms Eqs.~\eqref{eq:app-def-UO} and \eqref{eq:app-defO} for the operator \(U_F\) in Eq.~\eqref{UF1}, where the latter is written by
keeping only the dominant terms at small momenta.


\begin{thebibliography}{84}%
\makeatletter
\providecommand \@ifxundefined [1]{%
 \@ifx{#1\undefined}
}%
\providecommand \@ifnum [1]{%
 \ifnum #1\expandafter \@firstoftwo
 \else \expandafter \@secondoftwo
 \fi
}%
\providecommand \@ifx [1]{%
 \ifx #1\expandafter \@firstoftwo
 \else \expandafter \@secondoftwo
 \fi
}%
\providecommand \natexlab [1]{#1}%
\providecommand \enquote  [1]{``#1''}%
\providecommand \bibnamefont  [1]{#1}%
\providecommand \bibfnamefont [1]{#1}%
\providecommand \citenamefont [1]{#1}%
\providecommand \href@noop [0]{\@secondoftwo}%
\providecommand \href [0]{\begingroup \@sanitize@url \@href}%
\providecommand \@href[1]{\@@startlink{#1}\@@href}%
\providecommand \@@href[1]{\endgroup#1\@@endlink}%
\providecommand \@sanitize@url [0]{\catcode `\\12\catcode `\$12\catcode
  `\&12\catcode `\#12\catcode `\^12\catcode `\_12\catcode `\%12\relax}%
\providecommand \@@startlink[1]{}%
\providecommand \@@endlink[0]{}%
\providecommand \url  [0]{\begingroup\@sanitize@url \@url }%
\providecommand \@url [1]{\endgroup\@href {#1}{\urlprefix }}%
\providecommand \urlprefix  [0]{URL }%
\providecommand \Eprint [0]{\href }%
\providecommand \doibase [0]{http://dx.doi.org/}%
\providecommand \selectlanguage [0]{\@gobble}%
\providecommand \bibinfo  [0]{\@secondoftwo}%
\providecommand \bibfield  [0]{\@secondoftwo}%
\providecommand \translation [1]{[#1]}%
\providecommand \BibitemOpen [0]{}%
\providecommand \bibitemStop [0]{}%
\providecommand \bibitemNoStop [0]{.\EOS\space}%
\providecommand \EOS [0]{\spacefactor3000\relax}%
\providecommand \BibitemShut  [1]{\csname bibitem#1\endcsname}%
\let\auto@bib@innerbib\@empty
\bibitem [{\citenamefont {Wilczek}(2012)}]{Wilczek2012_Q}%
  \BibitemOpen
  \bibfield  {author} {\bibinfo {author} {\bibfnamefont {F.}~\bibnamefont
  {Wilczek}},\ }\href {\doibase 10.1103/PhysRevLett.109.160401} {\bibfield
  {journal} {\bibinfo  {journal} {Phys. Rev. Lett.}\ }\textbf {\bibinfo
  {volume} {109}},\ \bibinfo {pages} {160401} (\bibinfo {year}
  {2012})}\BibitemShut {NoStop}%
\bibitem [{\citenamefont {Shapere}\ and\ \citenamefont
  {Wilczek}(2012)}]{Wilczek2012_C}%
  \BibitemOpen
  \bibfield  {author} {\bibinfo {author} {\bibfnamefont {A.}~\bibnamefont
  {Shapere}}\ and\ \bibinfo {author} {\bibfnamefont {F.}~\bibnamefont
  {Wilczek}},\ }\href {\doibase 10.1103/PhysRevLett.109.160402} {\bibfield
  {journal} {\bibinfo  {journal} {Phys. Rev. Lett.}\ }\textbf {\bibinfo
  {volume} {109}},\ \bibinfo {pages} {160402} (\bibinfo {year}
  {2012})}\BibitemShut {NoStop}%
\bibitem [{\citenamefont {Li}\ \emph {et~al.}(2012)\citenamefont {Li},
  \citenamefont {Gong}, \citenamefont {Yin}, \citenamefont {Quan},
  \citenamefont {Yin}, \citenamefont {Zhang}, \citenamefont {Duan},\ and\
  \citenamefont {Zhang}}]{Li2012}%
  \BibitemOpen
  \bibfield  {author} {\bibinfo {author} {\bibfnamefont {T.}~\bibnamefont
  {Li}}, \bibinfo {author} {\bibfnamefont {Z.-X.}\ \bibnamefont {Gong}},
  \bibinfo {author} {\bibfnamefont {Z.-Q.}\ \bibnamefont {Yin}}, \bibinfo
  {author} {\bibfnamefont {H.~T.}\ \bibnamefont {Quan}}, \bibinfo {author}
  {\bibfnamefont {X.}~\bibnamefont {Yin}}, \bibinfo {author} {\bibfnamefont
  {P.}~\bibnamefont {Zhang}}, \bibinfo {author} {\bibfnamefont {L.-M.}\
  \bibnamefont {Duan}}, \ and\ \bibinfo {author} {\bibfnamefont
  {X.}~\bibnamefont {Zhang}},\ }\href {\doibase 10.1103/PhysRevLett.109.163001}
  {\bibfield  {journal} {\bibinfo  {journal} {Phys. Rev. Lett.}\ }\textbf
  {\bibinfo {volume} {109}},\ \bibinfo {pages} {163001} (\bibinfo {year}
  {2012})}\BibitemShut {NoStop}%
\bibitem [{\citenamefont {Bruno}(2013{\natexlab{a}})}]{Bruno2013_1}%
  \BibitemOpen
  \bibfield  {author} {\bibinfo {author} {\bibfnamefont {P.}~\bibnamefont
  {Bruno}},\ }\href {\doibase 10.1103/PhysRevLett.110.118901} {\bibfield
  {journal} {\bibinfo  {journal} {Phys. Rev. Lett.}\ }\textbf {\bibinfo
  {volume} {110}},\ \bibinfo {pages} {118901} (\bibinfo {year}
  {2013}{\natexlab{a}})}\BibitemShut {NoStop}%
\bibitem [{\citenamefont {Bruno}(2013{\natexlab{b}})}]{Bruno2013_2}%
  \BibitemOpen
  \bibfield  {author} {\bibinfo {author} {\bibfnamefont {P.}~\bibnamefont
  {Bruno}},\ }\href {\doibase 10.1103/PhysRevLett.111.029301} {\bibfield
  {journal} {\bibinfo  {journal} {Phys. Rev. Lett.}\ }\textbf {\bibinfo
  {volume} {111}},\ \bibinfo {pages} {029301} (\bibinfo {year}
  {2013}{\natexlab{b}})}\BibitemShut {NoStop}%
\bibitem [{\citenamefont {Bruno}(2013{\natexlab{c}})}]{Bruno2013_3}%
  \BibitemOpen
  \bibfield  {author} {\bibinfo {author} {\bibfnamefont {P.}~\bibnamefont
  {Bruno}},\ }\href {\doibase 10.1103/PhysRevLett.111.070402} {\bibfield
  {journal} {\bibinfo  {journal} {Phys. Rev. Lett.}\ }\textbf {\bibinfo
  {volume} {111}},\ \bibinfo {pages} {070402} (\bibinfo {year}
  {2013}{\natexlab{c}})}\BibitemShut {NoStop}%
\bibitem [{\citenamefont {Watanabe}\ and\ \citenamefont
  {Oshikawa}(2015)}]{Oshikawa2015}%
  \BibitemOpen
  \bibfield  {author} {\bibinfo {author} {\bibfnamefont {H.}~\bibnamefont
  {Watanabe}}\ and\ \bibinfo {author} {\bibfnamefont {M.}~\bibnamefont
  {Oshikawa}},\ }\href {\doibase 10.1103/PhysRevLett.114.251603} {\bibfield
  {journal} {\bibinfo  {journal} {Phys. Rev. Lett.}\ }\textbf {\bibinfo
  {volume} {114}},\ \bibinfo {pages} {251603} (\bibinfo {year}
  {2015})}\BibitemShut {NoStop}%
\bibitem [{\citenamefont {Prokof'ev}\ and\ \citenamefont
  {Svistunov}(2020)}]{Svistunov2020}%
  \BibitemOpen
  \bibfield  {author} {\bibinfo {author} {\bibfnamefont {N.}~\bibnamefont
  {Prokof'ev}}\ and\ \bibinfo {author} {\bibfnamefont {B.}~\bibnamefont
  {Svistunov}},\ }\href {\doibase 10.1103/PhysRevB.101.020505} {\bibfield
  {journal} {\bibinfo  {journal} {Phys. Rev. B}\ }\textbf {\bibinfo {volume}
  {101}},\ \bibinfo {pages} {020505} (\bibinfo {year} {2020})}\BibitemShut
  {NoStop}%
\bibitem [{\citenamefont {Else}\ \emph {et~al.}(2017)\citenamefont {Else},
  \citenamefont {Bauer},\ and\ \citenamefont {Nayak}}]{Else2017}%
  \BibitemOpen
  \bibfield  {author} {\bibinfo {author} {\bibfnamefont {D.~V.}\ \bibnamefont
  {Else}}, \bibinfo {author} {\bibfnamefont {B.}~\bibnamefont {Bauer}}, \ and\
  \bibinfo {author} {\bibfnamefont {C.}~\bibnamefont {Nayak}},\ }\href
  {\doibase 10.1103/PhysRevX.7.011026} {\bibfield  {journal} {\bibinfo
  {journal} {Phys. Rev. X}\ }\textbf {\bibinfo {volume} {7}},\ \bibinfo {pages}
  {011026} (\bibinfo {year} {2017})}\BibitemShut {NoStop}%
\bibitem [{\citenamefont {Else}\ \emph {et~al.}(2020)\citenamefont {Else},
  \citenamefont {Monroe}, \citenamefont {Nayak},\ and\ \citenamefont
  {Yao}}]{Else2019}%
  \BibitemOpen
  \bibfield  {author} {\bibinfo {author} {\bibfnamefont {D.~V.}\ \bibnamefont
  {Else}}, \bibinfo {author} {\bibfnamefont {C.}~\bibnamefont {Monroe}},
  \bibinfo {author} {\bibfnamefont {C.}~\bibnamefont {Nayak}}, \ and\ \bibinfo
  {author} {\bibfnamefont {N.~Y.}\ \bibnamefont {Yao}},\ }\href {\doibase
  10.1146/annurev-conmatphys-031119-050658} {\bibfield  {journal} {\bibinfo
  {journal} {Annual Review of Condensed Matter Physics}\ }\textbf {\bibinfo
  {volume} {11}},\ \bibinfo {pages} {467} (\bibinfo {year} {2020})}\BibitemShut
  {NoStop}%
\bibitem [{\citenamefont {Khemani}\ \emph {et~al.}(2019)\citenamefont
  {Khemani}, \citenamefont {Moessner},\ and\ \citenamefont
  {Sondhi}}]{Khemani2019}%
  \BibitemOpen
  \bibfield  {author} {\bibinfo {author} {\bibfnamefont {V.}~\bibnamefont
  {Khemani}}, \bibinfo {author} {\bibfnamefont {R.}~\bibnamefont {Moessner}}, \
  and\ \bibinfo {author} {\bibfnamefont {S.}~\bibnamefont {Sondhi}},\
  }\href@noop {} {\bibfield  {journal} {\bibinfo  {journal} {arXiv:1910.10745}\
  } (\bibinfo {year} {2019})}\BibitemShut {NoStop}%
\bibitem [{\citenamefont {Kozin}\ and\ \citenamefont
  {Kyriienko}(2019)}]{Kozin2019}%
  \BibitemOpen
  \bibfield  {author} {\bibinfo {author} {\bibfnamefont {V.~K.}\ \bibnamefont
  {Kozin}}\ and\ \bibinfo {author} {\bibfnamefont {O.}~\bibnamefont
  {Kyriienko}},\ }\href {\doibase 10.1103/PhysRevLett.123.210602} {\bibfield
  {journal} {\bibinfo  {journal} {Phys. Rev. Lett.}\ }\textbf {\bibinfo
  {volume} {123}},\ \bibinfo {pages} {210602} (\bibinfo {year}
  {2019})}\BibitemShut {NoStop}%
\bibitem [{\citenamefont {Khemani}\ \emph {et~al.}(2020)\citenamefont
  {Khemani}, \citenamefont {Moessner},\ and\ \citenamefont
  {Sondhi}}]{Sondhi2020}%
  \BibitemOpen
  \bibfield  {author} {\bibinfo {author} {\bibfnamefont {V.}~\bibnamefont
  {Khemani}}, \bibinfo {author} {\bibfnamefont {R.}~\bibnamefont {Moessner}}, \
  and\ \bibinfo {author} {\bibfnamefont {S.~L.}\ \bibnamefont {Sondhi}},\
  }\href@noop {} {\bibfield  {journal} {\bibinfo  {journal} {arXiv:2001.11037}\
  } (\bibinfo {year} {2020})}\BibitemShut {NoStop}%
\bibitem [{\citenamefont {Kozin}\ and\ \citenamefont
  {Kyriienko}(2020)}]{Kozin2020}%
  \BibitemOpen
  \bibfield  {author} {\bibinfo {author} {\bibfnamefont {V.~L.}\ \bibnamefont
  {Kozin}}\ and\ \bibinfo {author} {\bibfnamefont {O.}~\bibnamefont
  {Kyriienko}},\ }\href@noop {} {\bibfield  {journal} {\bibinfo  {journal}
  {arXiv:2005.06321}\ } (\bibinfo {year} {2020})}\BibitemShut {NoStop}%
\bibitem [{\citenamefont {\"Ohberg}\ and\ \citenamefont
  {Wright}(2019)}]{Wright19}%
  \BibitemOpen
  \bibfield  {author} {\bibinfo {author} {\bibfnamefont {P.}~\bibnamefont
  {\"Ohberg}}\ and\ \bibinfo {author} {\bibfnamefont {E.~M.}\ \bibnamefont
  {Wright}},\ }\href {\doibase 10.1103/PhysRevLett.123.250402} {\bibfield
  {journal} {\bibinfo  {journal} {Phys. Rev. Lett.}\ }\textbf {\bibinfo
  {volume} {123}},\ \bibinfo {pages} {250402} (\bibinfo {year}
  {2019})}\BibitemShut {NoStop}%
\bibitem [{\citenamefont {Syrwid}\ \emph {et~al.}(2020)\citenamefont {Syrwid},
  \citenamefont {Kosior},\ and\ \citenamefont {Sacha}}]{Sacha20}%
  \BibitemOpen
  \bibfield  {author} {\bibinfo {author} {\bibfnamefont {A.}~\bibnamefont
  {Syrwid}}, \bibinfo {author} {\bibfnamefont {A.}~\bibnamefont {Kosior}}, \
  and\ \bibinfo {author} {\bibfnamefont {K.}~\bibnamefont {Sacha}},\ }\href
  {\doibase 10.1103/PhysRevLett.124.178901} {\bibfield  {journal} {\bibinfo
  {journal} {Phys. Rev. Lett.}\ }\textbf {\bibinfo {volume} {124}},\ \bibinfo
  {pages} {178901} (\bibinfo {year} {2020})}\BibitemShut {NoStop}%
\bibitem [{\citenamefont {\"Ohberg}\ and\ \citenamefont
  {Wright}(2020)}]{Wright20}%
  \BibitemOpen
  \bibfield  {author} {\bibinfo {author} {\bibfnamefont {P.}~\bibnamefont
  {\"Ohberg}}\ and\ \bibinfo {author} {\bibfnamefont {E.~M.}\ \bibnamefont
  {Wright}},\ }\href {\doibase 10.1103/PhysRevLett.124.178902} {\bibfield
  {journal} {\bibinfo  {journal} {Phys. Rev. Lett.}\ }\textbf {\bibinfo
  {volume} {124}},\ \bibinfo {pages} {178902} (\bibinfo {year}
  {2020})}\BibitemShut {NoStop}%
\bibitem [{\citenamefont {Sacha}\ and\ \citenamefont
  {Zakrzewski}(2017)}]{Sacha2018}%
  \BibitemOpen
  \bibfield  {author} {\bibinfo {author} {\bibfnamefont {K.}~\bibnamefont
  {Sacha}}\ and\ \bibinfo {author} {\bibfnamefont {J.}~\bibnamefont
  {Zakrzewski}},\ }\href {\doibase 10.1088/1361-6633/aa8b38} {\bibfield
  {journal} {\bibinfo  {journal} {Reports on Progress in Physics}\ }\textbf
  {\bibinfo {volume} {81}},\ \bibinfo {pages} {016401} (\bibinfo {year}
  {2017})}\BibitemShut {NoStop}%
\bibitem [{\citenamefont {Sacha}(2015)}]{Sacha15}%
  \BibitemOpen
  \bibfield  {author} {\bibinfo {author} {\bibfnamefont {K.}~\bibnamefont
  {Sacha}},\ }\href {\doibase 10.1103/PhysRevA.91.033617} {\bibfield  {journal}
  {\bibinfo  {journal} {Phys. Rev. A}\ }\textbf {\bibinfo {volume} {91}},\
  \bibinfo {pages} {033617} (\bibinfo {year} {2015})}\BibitemShut {NoStop}%
\bibitem [{\citenamefont {Else}\ \emph {et~al.}(2016)\citenamefont {Else},
  \citenamefont {Bauer},\ and\ \citenamefont {Nayak}}]{Else2016}%
  \BibitemOpen
  \bibfield  {author} {\bibinfo {author} {\bibfnamefont {D.~V.}\ \bibnamefont
  {Else}}, \bibinfo {author} {\bibfnamefont {B.}~\bibnamefont {Bauer}}, \ and\
  \bibinfo {author} {\bibfnamefont {C.}~\bibnamefont {Nayak}},\ }\href
  {\doibase 10.1103/PhysRevLett.117.090402} {\bibfield  {journal} {\bibinfo
  {journal} {Phys. Rev. Lett.}\ }\textbf {\bibinfo {volume} {117}},\ \bibinfo
  {pages} {090402} (\bibinfo {year} {2016})}\BibitemShut {NoStop}%
\bibitem [{\citenamefont {Else}\ and\ \citenamefont {Nayak}(2016)}]{Else2016a}%
  \BibitemOpen
  \bibfield  {author} {\bibinfo {author} {\bibfnamefont {D.~V.}\ \bibnamefont
  {Else}}\ and\ \bibinfo {author} {\bibfnamefont {C.}~\bibnamefont {Nayak}},\
  }\href {\doibase 10.1103/PhysRevB.93.201103} {\bibfield  {journal} {\bibinfo
  {journal} {Phys. Rev. B}\ }\textbf {\bibinfo {volume} {93}},\ \bibinfo
  {pages} {201103} (\bibinfo {year} {2016})}\BibitemShut {NoStop}%
\bibitem [{\citenamefont {Chandran}\ and\ \citenamefont
  {Sondhi}(2016)}]{Chandran2016}%
  \BibitemOpen
  \bibfield  {author} {\bibinfo {author} {\bibfnamefont {A.}~\bibnamefont
  {Chandran}}\ and\ \bibinfo {author} {\bibfnamefont {S.~L.}\ \bibnamefont
  {Sondhi}},\ }\href {\doibase 10.1103/PhysRevB.93.174305} {\bibfield
  {journal} {\bibinfo  {journal} {Phys. Rev. B}\ }\textbf {\bibinfo {volume}
  {93}},\ \bibinfo {pages} {174305} (\bibinfo {year} {2016})}\BibitemShut
  {NoStop}%
\bibitem [{\citenamefont {Khemani}\ \emph {et~al.}(2016)\citenamefont
  {Khemani}, \citenamefont {Lazarides}, \citenamefont {Moessner},\ and\
  \citenamefont {Sondhi}}]{Khemani2016}%
  \BibitemOpen
  \bibfield  {author} {\bibinfo {author} {\bibfnamefont {V.}~\bibnamefont
  {Khemani}}, \bibinfo {author} {\bibfnamefont {A.}~\bibnamefont {Lazarides}},
  \bibinfo {author} {\bibfnamefont {R.}~\bibnamefont {Moessner}}, \ and\
  \bibinfo {author} {\bibfnamefont {S.~L.}\ \bibnamefont {Sondhi}},\ }\href
  {\doibase 10.1103/PhysRevLett.116.250401} {\bibfield  {journal} {\bibinfo
  {journal} {Phys. Rev. Lett.}\ }\textbf {\bibinfo {volume} {116}},\ \bibinfo
  {pages} {250401} (\bibinfo {year} {2016})}\BibitemShut {NoStop}%
\bibitem [{\citenamefont {von Keyserlingk}\ \emph {et~al.}(2016)\citenamefont
  {von Keyserlingk}, \citenamefont {Khemani},\ and\ \citenamefont
  {Sondhi}}]{Keyserlingk2016}%
  \BibitemOpen
  \bibfield  {author} {\bibinfo {author} {\bibfnamefont {C.~W.}\ \bibnamefont
  {von Keyserlingk}}, \bibinfo {author} {\bibfnamefont {V.}~\bibnamefont
  {Khemani}}, \ and\ \bibinfo {author} {\bibfnamefont {S.~L.}\ \bibnamefont
  {Sondhi}},\ }\href {\doibase 10.1103/PhysRevB.94.085112} {\bibfield
  {journal} {\bibinfo  {journal} {Phys. Rev. B}\ }\textbf {\bibinfo {volume}
  {94}},\ \bibinfo {pages} {085112} (\bibinfo {year} {2016})}\BibitemShut
  {NoStop}%
\bibitem [{\citenamefont {von Keyserlingk}\ and\ \citenamefont
  {Sondhi}(2016{\natexlab{a}})}]{Keyserlingk2016_2}%
  \BibitemOpen
  \bibfield  {author} {\bibinfo {author} {\bibfnamefont {C.~W.}\ \bibnamefont
  {von Keyserlingk}}\ and\ \bibinfo {author} {\bibfnamefont {S.~L.}\
  \bibnamefont {Sondhi}},\ }\href {\doibase 10.1103/PhysRevB.93.245145}
  {\bibfield  {journal} {\bibinfo  {journal} {Phys. Rev. B}\ }\textbf {\bibinfo
  {volume} {93}},\ \bibinfo {pages} {245145} (\bibinfo {year}
  {2016}{\natexlab{a}})}\BibitemShut {NoStop}%
\bibitem [{\citenamefont {von Keyserlingk}\ and\ \citenamefont
  {Sondhi}(2016{\natexlab{b}})}]{Keyserlingk2016_3}%
  \BibitemOpen
  \bibfield  {author} {\bibinfo {author} {\bibfnamefont {C.~W.}\ \bibnamefont
  {von Keyserlingk}}\ and\ \bibinfo {author} {\bibfnamefont {S.~L.}\
  \bibnamefont {Sondhi}},\ }\href {\doibase 10.1103/PhysRevB.93.245146}
  {\bibfield  {journal} {\bibinfo  {journal} {Phys. Rev. B}\ }\textbf {\bibinfo
  {volume} {93}},\ \bibinfo {pages} {245146} (\bibinfo {year}
  {2016}{\natexlab{b}})}\BibitemShut {NoStop}%
\bibitem [{\citenamefont {Yao}\ \emph {et~al.}(2017)\citenamefont {Yao},
  \citenamefont {Potter}, \citenamefont {Potirniche},\ and\ \citenamefont
  {Vishwanath}}]{Yao2017}%
  \BibitemOpen
  \bibfield  {author} {\bibinfo {author} {\bibfnamefont {N.~Y.}\ \bibnamefont
  {Yao}}, \bibinfo {author} {\bibfnamefont {A.~C.}\ \bibnamefont {Potter}},
  \bibinfo {author} {\bibfnamefont {I.-D.}\ \bibnamefont {Potirniche}}, \ and\
  \bibinfo {author} {\bibfnamefont {A.}~\bibnamefont {Vishwanath}},\ }\href
  {\doibase 10.1103/PhysRevLett.118.030401} {\bibfield  {journal} {\bibinfo
  {journal} {Phys. Rev. Lett.}\ }\textbf {\bibinfo {volume} {118}},\ \bibinfo
  {pages} {030401} (\bibinfo {year} {2017})}\BibitemShut {NoStop}%
\bibitem [{\citenamefont {Ho}\ \emph {et~al.}(2017)\citenamefont {Ho},
  \citenamefont {Choi}, \citenamefont {Lukin},\ and\ \citenamefont
  {Abanin}}]{Abanin_Ho17}%
  \BibitemOpen
  \bibfield  {author} {\bibinfo {author} {\bibfnamefont {W.~W.}\ \bibnamefont
  {Ho}}, \bibinfo {author} {\bibfnamefont {S.}~\bibnamefont {Choi}}, \bibinfo
  {author} {\bibfnamefont {M.~D.}\ \bibnamefont {Lukin}}, \ and\ \bibinfo
  {author} {\bibfnamefont {D.~A.}\ \bibnamefont {Abanin}},\ }\href {\doibase
  10.1103/PhysRevLett.119.010602} {\bibfield  {journal} {\bibinfo  {journal}
  {Phys. Rev. Lett.}\ }\textbf {\bibinfo {volume} {119}},\ \bibinfo {pages}
  {010602} (\bibinfo {year} {2017})}\BibitemShut {NoStop}%
\bibitem [{\citenamefont {Moessner}\ and\ \citenamefont
  {Sondhi}(2017)}]{Moessner2017}%
  \BibitemOpen
  \bibfield  {author} {\bibinfo {author} {\bibfnamefont {R.}~\bibnamefont
  {Moessner}}\ and\ \bibinfo {author} {\bibfnamefont {S.~L.}\ \bibnamefont
  {Sondhi}},\ }\href@noop {} {\bibfield  {journal} {\bibinfo  {journal} {Nature
  Physics}\ }\textbf {\bibinfo {volume} {13}},\ \bibinfo {pages} {424}
  (\bibinfo {year} {2017})}\BibitemShut {NoStop}%
\bibitem [{\citenamefont {Russomanno}\ \emph {et~al.}(2017)\citenamefont
  {Russomanno}, \citenamefont {Iemini}, \citenamefont {Dalmonte},\ and\
  \citenamefont {Fazio}}]{Russomanno2017}%
  \BibitemOpen
  \bibfield  {author} {\bibinfo {author} {\bibfnamefont {A.}~\bibnamefont
  {Russomanno}}, \bibinfo {author} {\bibfnamefont {F.}~\bibnamefont {Iemini}},
  \bibinfo {author} {\bibfnamefont {M.}~\bibnamefont {Dalmonte}}, \ and\
  \bibinfo {author} {\bibfnamefont {R.}~\bibnamefont {Fazio}},\ }\href
  {\doibase 10.1103/PhysRevB.95.214307} {\bibfield  {journal} {\bibinfo
  {journal} {Phys. Rev. B}\ }\textbf {\bibinfo {volume} {95}},\ \bibinfo
  {pages} {214307} (\bibinfo {year} {2017})}\BibitemShut {NoStop}%
\bibitem [{\citenamefont {Zeng}\ and\ \citenamefont {Sheng}(2017)}]{Zeng2017}%
  \BibitemOpen
  \bibfield  {author} {\bibinfo {author} {\bibfnamefont {T.-S.}\ \bibnamefont
  {Zeng}}\ and\ \bibinfo {author} {\bibfnamefont {D.~N.}\ \bibnamefont
  {Sheng}},\ }\href {\doibase 10.1103/PhysRevB.96.094202} {\bibfield  {journal}
  {\bibinfo  {journal} {Phys. Rev. B}\ }\textbf {\bibinfo {volume} {96}},\
  \bibinfo {pages} {094202} (\bibinfo {year} {2017})}\BibitemShut {NoStop}%
\bibitem [{\citenamefont {Huang}\ \emph {et~al.}(2018)\citenamefont {Huang},
  \citenamefont {Wu},\ and\ \citenamefont {Liu}}]{Huang2018}%
  \BibitemOpen
  \bibfield  {author} {\bibinfo {author} {\bibfnamefont {B.}~\bibnamefont
  {Huang}}, \bibinfo {author} {\bibfnamefont {Y.-H.}\ \bibnamefont {Wu}}, \
  and\ \bibinfo {author} {\bibfnamefont {W.~V.}\ \bibnamefont {Liu}},\ }\href
  {\doibase 10.1103/PhysRevLett.120.110603} {\bibfield  {journal} {\bibinfo
  {journal} {Phys. Rev. Lett.}\ }\textbf {\bibinfo {volume} {120}},\ \bibinfo
  {pages} {110603} (\bibinfo {year} {2018})}\BibitemShut {NoStop}%
\bibitem [{\citenamefont {Gong}\ \emph {et~al.}(2018)\citenamefont {Gong},
  \citenamefont {Hamazaki},\ and\ \citenamefont {Ueda}}]{Gong2018}%
  \BibitemOpen
  \bibfield  {author} {\bibinfo {author} {\bibfnamefont {Z.}~\bibnamefont
  {Gong}}, \bibinfo {author} {\bibfnamefont {R.}~\bibnamefont {Hamazaki}}, \
  and\ \bibinfo {author} {\bibfnamefont {M.}~\bibnamefont {Ueda}},\ }\href
  {https://link.aps.org/doi/10.1103/PhysRevLett.120.040404} {\bibfield
  {journal} {\bibinfo  {journal} {Phys. Rev. Lett.}\ }\textbf {\bibinfo
  {volume} {120}},\ \bibinfo {pages} {040404} (\bibinfo {year}
  {2018})}\BibitemShut {NoStop}%
\bibitem [{\citenamefont {Kosior}\ and\ \citenamefont
  {Sacha}(2018)}]{Kosior2018}%
  \BibitemOpen
  \bibfield  {author} {\bibinfo {author} {\bibfnamefont {A.}~\bibnamefont
  {Kosior}}\ and\ \bibinfo {author} {\bibfnamefont {K.}~\bibnamefont {Sacha}},\
  }\href {\doibase 10.1103/PhysRevA.97.053621} {\bibfield  {journal} {\bibinfo
  {journal} {Phys. Rev. A}\ }\textbf {\bibinfo {volume} {97}},\ \bibinfo
  {pages} {053621} (\bibinfo {year} {2018})}\BibitemShut {NoStop}%
\bibitem [{\citenamefont {Wang}\ \emph {et~al.}(2018)\citenamefont {Wang},
  \citenamefont {Xing}, \citenamefont {Carlo},\ and\ \citenamefont
  {Poletti}}]{Wang2018}%
  \BibitemOpen
  \bibfield  {author} {\bibinfo {author} {\bibfnamefont {R.~R.~W.}\
  \bibnamefont {Wang}}, \bibinfo {author} {\bibfnamefont {B.}~\bibnamefont
  {Xing}}, \bibinfo {author} {\bibfnamefont {G.~G.}\ \bibnamefont {Carlo}}, \
  and\ \bibinfo {author} {\bibfnamefont {D.}~\bibnamefont {Poletti}},\ }\href
  {\doibase 10.1103/PhysRevE.97.020202} {\bibfield  {journal} {\bibinfo
  {journal} {Phys. Rev. E}\ }\textbf {\bibinfo {volume} {97}},\ \bibinfo
  {pages} {020202} (\bibinfo {year} {2018})}\BibitemShut {NoStop}%
\bibitem [{\citenamefont {Yao}\ \emph {et~al.}(2020)\citenamefont {Yao},
  \citenamefont {Nayak}, \citenamefont {Balents},\ and\ \citenamefont
  {Zaletel}}]{Yao2018}%
  \BibitemOpen
  \bibfield  {author} {\bibinfo {author} {\bibfnamefont {N.~Y.}\ \bibnamefont
  {Yao}}, \bibinfo {author} {\bibfnamefont {C.}~\bibnamefont {Nayak}}, \bibinfo
  {author} {\bibfnamefont {L.}~\bibnamefont {Balents}}, \ and\ \bibinfo
  {author} {\bibfnamefont {M.~P.}\ \bibnamefont {Zaletel}},\ }\href@noop {}
  {\bibfield  {journal} {\bibinfo  {journal} {Nature Physics}\ }\textbf
  {\bibinfo {volume} {16}},\ \bibinfo {pages} {438} (\bibinfo {year}
  {2020})}\BibitemShut {NoStop}%
\bibitem [{\citenamefont {Heugel}\ \emph {et~al.}(2019)\citenamefont {Heugel},
  \citenamefont {Oscity}, \citenamefont {Eichler}, \citenamefont {Zilberberg},\
  and\ \citenamefont {Chitra}}]{Heugel2019}%
  \BibitemOpen
  \bibfield  {author} {\bibinfo {author} {\bibfnamefont {T.~L.}\ \bibnamefont
  {Heugel}}, \bibinfo {author} {\bibfnamefont {M.}~\bibnamefont {Oscity}},
  \bibinfo {author} {\bibfnamefont {A.}~\bibnamefont {Eichler}}, \bibinfo
  {author} {\bibfnamefont {O.}~\bibnamefont {Zilberberg}}, \ and\ \bibinfo
  {author} {\bibfnamefont {R.}~\bibnamefont {Chitra}},\ }\href {\doibase
  10.1103/PhysRevLett.123.124301} {\bibfield  {journal} {\bibinfo  {journal}
  {Phys. Rev. Lett.}\ }\textbf {\bibinfo {volume} {123}},\ \bibinfo {pages}
  {124301} (\bibinfo {year} {2019})}\BibitemShut {NoStop}%
\bibitem [{\citenamefont {Gambetta}\ \emph {et~al.}(2019)\citenamefont
  {Gambetta}, \citenamefont {Carollo}, \citenamefont {Lazarides}, \citenamefont
  {Lesanovsky},\ and\ \citenamefont {Garrahan}}]{Garrahan19}%
  \BibitemOpen
  \bibfield  {author} {\bibinfo {author} {\bibfnamefont {F.~M.}\ \bibnamefont
  {Gambetta}}, \bibinfo {author} {\bibfnamefont {F.}~\bibnamefont {Carollo}},
  \bibinfo {author} {\bibfnamefont {A.}~\bibnamefont {Lazarides}}, \bibinfo
  {author} {\bibfnamefont {I.}~\bibnamefont {Lesanovsky}}, \ and\ \bibinfo
  {author} {\bibfnamefont {J.~P.}\ \bibnamefont {Garrahan}},\ }\href {\doibase
  10.1103/PhysRevE.100.060105} {\bibfield  {journal} {\bibinfo  {journal}
  {Phys. Rev. E}\ }\textbf {\bibinfo {volume} {100}},\ \bibinfo {pages}
  {060105} (\bibinfo {year} {2019})}\BibitemShut {NoStop}%
\bibitem [{\citenamefont {Zhang}\ \emph {et~al.}(2017)\citenamefont {Zhang},
  \citenamefont {Hess}, \citenamefont {Kyprianidis}, \citenamefont {Becker},
  \citenamefont {Lee}, \citenamefont {Smith}, \citenamefont {Pagano},
  \citenamefont {Potirniche}, \citenamefont {Potter}, \citenamefont
  {Vishwanath}, \citenamefont {Yao},\ and\ \citenamefont {Monroe}}]{Zhang2017}%
  \BibitemOpen
  \bibfield  {author} {\bibinfo {author} {\bibfnamefont {J.}~\bibnamefont
  {Zhang}}, \bibinfo {author} {\bibfnamefont {P.~W.}\ \bibnamefont {Hess}},
  \bibinfo {author} {\bibfnamefont {A.-C.}\ \bibnamefont {Kyprianidis}},
  \bibinfo {author} {\bibfnamefont {P.}~\bibnamefont {Becker}}, \bibinfo
  {author} {\bibfnamefont {A.}~\bibnamefont {Lee}}, \bibinfo {author}
  {\bibfnamefont {J.~K.}\ \bibnamefont {Smith}}, \bibinfo {author}
  {\bibfnamefont {G.}~\bibnamefont {Pagano}}, \bibinfo {author} {\bibfnamefont
  {I.-D.}\ \bibnamefont {Potirniche}}, \bibinfo {author} {\bibfnamefont
  {A.~C.}\ \bibnamefont {Potter}}, \bibinfo {author} {\bibfnamefont
  {A.}~\bibnamefont {Vishwanath}}, \bibinfo {author} {\bibfnamefont {N.~Y.}\
  \bibnamefont {Yao}}, \ and\ \bibinfo {author} {\bibfnamefont
  {C.}~\bibnamefont {Monroe}},\ }\href@noop {} {\bibfield  {journal} {\bibinfo
  {journal} {Nature}\ }\textbf {\bibinfo {volume} {543}},\ \bibinfo {pages}
  {217} (\bibinfo {year} {2017})}\BibitemShut {NoStop}%
\bibitem [{\citenamefont {Choi}\ \emph {et~al.}(2017)\citenamefont {Choi},
  \citenamefont {Choi}, \citenamefont {Landig}, \citenamefont {Kucsko},
  \citenamefont {Zhou}, \citenamefont {Isoya}, \citenamefont {Jelezko},
  \citenamefont {Onoda}, \citenamefont {Sumiya}, \citenamefont {Khemani},
  \citenamefont {von Keyserlingk}, \citenamefont {Yao}, \citenamefont
  {Demler},\ and\ \citenamefont {Lukin}}]{Choi2017}%
  \BibitemOpen
  \bibfield  {author} {\bibinfo {author} {\bibfnamefont {S.}~\bibnamefont
  {Choi}}, \bibinfo {author} {\bibfnamefont {J.}~\bibnamefont {Choi}}, \bibinfo
  {author} {\bibfnamefont {R.}~\bibnamefont {Landig}}, \bibinfo {author}
  {\bibfnamefont {G.}~\bibnamefont {Kucsko}}, \bibinfo {author} {\bibfnamefont
  {H.}~\bibnamefont {Zhou}}, \bibinfo {author} {\bibfnamefont {J.}~\bibnamefont
  {Isoya}}, \bibinfo {author} {\bibfnamefont {F.}~\bibnamefont {Jelezko}},
  \bibinfo {author} {\bibfnamefont {S.}~\bibnamefont {Onoda}}, \bibinfo
  {author} {\bibfnamefont {H.}~\bibnamefont {Sumiya}}, \bibinfo {author}
  {\bibfnamefont {V.}~\bibnamefont {Khemani}}, \bibinfo {author} {\bibfnamefont
  {C.~W.}\ \bibnamefont {von Keyserlingk}}, \bibinfo {author} {\bibfnamefont
  {N.~Y.}\ \bibnamefont {Yao}}, \bibinfo {author} {\bibfnamefont {E.~A.}\
  \bibnamefont {Demler}}, \ and\ \bibinfo {author} {\bibfnamefont {M.~D.}\
  \bibnamefont {Lukin}},\ }\href@noop {} {\bibfield  {journal} {\bibinfo
  {journal} {Nature}\ }\textbf {\bibinfo {volume} {543}},\ \bibinfo {pages}
  {221} (\bibinfo {year} {2017})}\BibitemShut {NoStop}%
\bibitem [{\citenamefont {Rovny}\ \emph {et~al.}(2018)\citenamefont {Rovny},
  \citenamefont {Blum},\ and\ \citenamefont {Barrett}}]{Barrett2018}%
  \BibitemOpen
  \bibfield  {author} {\bibinfo {author} {\bibfnamefont {J.}~\bibnamefont
  {Rovny}}, \bibinfo {author} {\bibfnamefont {R.~L.}\ \bibnamefont {Blum}}, \
  and\ \bibinfo {author} {\bibfnamefont {S.~E.}\ \bibnamefont {Barrett}},\
  }\href {\doibase 10.1103/PhysRevLett.120.180603} {\bibfield  {journal}
  {\bibinfo  {journal} {Phys. Rev. Lett.}\ }\textbf {\bibinfo {volume} {120}},\
  \bibinfo {pages} {180603} (\bibinfo {year} {2018})}\BibitemShut {NoStop}%
\bibitem [{\citenamefont {Autti}\ \emph {et~al.}(2018)\citenamefont {Autti},
  \citenamefont {Eltsov},\ and\ \citenamefont {Volovik}}]{Volovik2018}%
  \BibitemOpen
  \bibfield  {author} {\bibinfo {author} {\bibfnamefont {S.}~\bibnamefont
  {Autti}}, \bibinfo {author} {\bibfnamefont {V.~B.}\ \bibnamefont {Eltsov}}, \
  and\ \bibinfo {author} {\bibfnamefont {G.~E.}\ \bibnamefont {Volovik}},\
  }\href {\doibase 10.1103/PhysRevLett.120.215301} {\bibfield  {journal}
  {\bibinfo  {journal} {Phys. Rev. Lett.}\ }\textbf {\bibinfo {volume} {120}},\
  \bibinfo {pages} {215301} (\bibinfo {year} {2018})}\BibitemShut {NoStop}%
\bibitem [{\citenamefont {Pal}\ \emph {et~al.}(2018)\citenamefont {Pal},
  \citenamefont {Nishad}, \citenamefont {Mahesh},\ and\ \citenamefont
  {Sreejith}}]{Sohan18}%
  \BibitemOpen
  \bibfield  {author} {\bibinfo {author} {\bibfnamefont {S.}~\bibnamefont
  {Pal}}, \bibinfo {author} {\bibfnamefont {N.}~\bibnamefont {Nishad}},
  \bibinfo {author} {\bibfnamefont {T.~S.}\ \bibnamefont {Mahesh}}, \ and\
  \bibinfo {author} {\bibfnamefont {G.~J.}\ \bibnamefont {Sreejith}},\ }\href
  {\doibase 10.1103/PhysRevLett.120.180602} {\bibfield  {journal} {\bibinfo
  {journal} {Phys. Rev. Lett.}\ }\textbf {\bibinfo {volume} {120}},\ \bibinfo
  {pages} {180602} (\bibinfo {year} {2018})}\BibitemShut {NoStop}%
\bibitem [{\citenamefont {Smits}\ \emph {et~al.}(2018)\citenamefont {Smits},
  \citenamefont {Liao}, \citenamefont {Stoof},\ and\ \citenamefont {van~der
  Straten}}]{Smits18}%
  \BibitemOpen
  \bibfield  {author} {\bibinfo {author} {\bibfnamefont {J.}~\bibnamefont
  {Smits}}, \bibinfo {author} {\bibfnamefont {L.}~\bibnamefont {Liao}},
  \bibinfo {author} {\bibfnamefont {H.~T.~C.}\ \bibnamefont {Stoof}}, \ and\
  \bibinfo {author} {\bibfnamefont {P.}~\bibnamefont {van~der Straten}},\
  }\href {\doibase 10.1103/PhysRevLett.121.185301} {\bibfield  {journal}
  {\bibinfo  {journal} {Phys. Rev. Lett.}\ }\textbf {\bibinfo {volume} {121}},\
  \bibinfo {pages} {185301} (\bibinfo {year} {2018})}\BibitemShut {NoStop}%
\bibitem [{\citenamefont {Giergiel}\ \emph {et~al.}(2018)\citenamefont
  {Giergiel}, \citenamefont {Kosior}, \citenamefont {Hannaford},\ and\
  \citenamefont {Sacha}}]{Sacha18}%
  \BibitemOpen
  \bibfield  {author} {\bibinfo {author} {\bibfnamefont {K.}~\bibnamefont
  {Giergiel}}, \bibinfo {author} {\bibfnamefont {A.}~\bibnamefont {Kosior}},
  \bibinfo {author} {\bibfnamefont {P.}~\bibnamefont {Hannaford}}, \ and\
  \bibinfo {author} {\bibfnamefont {K.}~\bibnamefont {Sacha}},\ }\href
  {\doibase 10.1103/PhysRevA.98.013613} {\bibfield  {journal} {\bibinfo
  {journal} {Phys. Rev. A}\ }\textbf {\bibinfo {volume} {98}},\ \bibinfo
  {pages} {013613} (\bibinfo {year} {2018})}\BibitemShut {NoStop}%
\bibitem [{\citenamefont {Surace}\ \emph {et~al.}(2019)\citenamefont {Surace},
  \citenamefont {Russomanno}, \citenamefont {Dalmonte}, \citenamefont {Silva},
  \citenamefont {Fazio},\ and\ \citenamefont {Iemini}}]{Fazio19}%
  \BibitemOpen
  \bibfield  {author} {\bibinfo {author} {\bibfnamefont {F.~M.}\ \bibnamefont
  {Surace}}, \bibinfo {author} {\bibfnamefont {A.}~\bibnamefont {Russomanno}},
  \bibinfo {author} {\bibfnamefont {M.}~\bibnamefont {Dalmonte}}, \bibinfo
  {author} {\bibfnamefont {A.}~\bibnamefont {Silva}}, \bibinfo {author}
  {\bibfnamefont {R.}~\bibnamefont {Fazio}}, \ and\ \bibinfo {author}
  {\bibfnamefont {F.}~\bibnamefont {Iemini}},\ }\href {\doibase
  10.1103/PhysRevB.99.104303} {\bibfield  {journal} {\bibinfo  {journal} {Phys.
  Rev. B}\ }\textbf {\bibinfo {volume} {99}},\ \bibinfo {pages} {104303}
  (\bibinfo {year} {2019})}\BibitemShut {NoStop}%
\bibitem [{\citenamefont {Eyal}\ \emph {et~al.}(1996)\citenamefont {Eyal},
  \citenamefont {Moshe}, \citenamefont {Nishigaki},\ and\ \citenamefont
  {Zinn-Justin}}]{Eyal1996}%
  \BibitemOpen
  \bibfield  {author} {\bibinfo {author} {\bibfnamefont {G.}~\bibnamefont
  {Eyal}}, \bibinfo {author} {\bibfnamefont {M.}~\bibnamefont {Moshe}},
  \bibinfo {author} {\bibfnamefont {S.}~\bibnamefont {Nishigaki}}, \ and\
  \bibinfo {author} {\bibfnamefont {J.}~\bibnamefont {Zinn-Justin}},\ }\href
  {\doibase 10.1016/0550-3213(96)00168-x} {\bibfield  {journal} {\bibinfo
  {journal} {Nuclear Physics B}\ }\textbf {\bibinfo {volume} {470}},\ \bibinfo
  {pages} {369} (\bibinfo {year} {1996})}\BibitemShut {NoStop}%
\bibitem [{\citenamefont {Moshe}\ and\ \citenamefont
  {Zinn-Justin}(2003)}]{Moshe2003}%
  \BibitemOpen
  \bibfield  {author} {\bibinfo {author} {\bibfnamefont {M.}~\bibnamefont
  {Moshe}}\ and\ \bibinfo {author} {\bibfnamefont {J.}~\bibnamefont
  {Zinn-Justin}},\ }\href {\doibase 10.1016/s0370-1573(03)00263-1} {\bibfield
  {journal} {\bibinfo  {journal} {Physics Reports}\ }\textbf {\bibinfo {volume}
  {385}},\ \bibinfo {pages} {69} (\bibinfo {year} {2003})}\BibitemShut
  {NoStop}%
\bibitem [{\citenamefont {Sotiriadis}\ \emph {et~al.}(2009)\citenamefont
  {Sotiriadis}, \citenamefont {Calabrese},\ and\ \citenamefont
  {Cardy}}]{Sotiriadis2009}%
  \BibitemOpen
  \bibfield  {author} {\bibinfo {author} {\bibfnamefont {S.}~\bibnamefont
  {Sotiriadis}}, \bibinfo {author} {\bibfnamefont {P.}~\bibnamefont
  {Calabrese}}, \ and\ \bibinfo {author} {\bibfnamefont {J.}~\bibnamefont
  {Cardy}},\ }\href {\doibase 10.1209/0295-5075/87/20002} {\bibfield  {journal}
  {\bibinfo  {journal} {{EPL} (Europhysics Letters)}\ }\textbf {\bibinfo
  {volume} {87}},\ \bibinfo {pages} {20002} (\bibinfo {year}
  {2009})}\BibitemShut {NoStop}%
\bibitem [{\citenamefont {Sotiriadis}\ and\ \citenamefont
  {Cardy}(2010)}]{Sotiriadis2010}%
  \BibitemOpen
  \bibfield  {author} {\bibinfo {author} {\bibfnamefont {S.}~\bibnamefont
  {Sotiriadis}}\ and\ \bibinfo {author} {\bibfnamefont {J.}~\bibnamefont
  {Cardy}},\ }\href {\doibase 10.1103/PhysRevB.81.134305} {\bibfield  {journal}
  {\bibinfo  {journal} {Phys. Rev. B}\ }\textbf {\bibinfo {volume} {81}},\
  \bibinfo {pages} {134305} (\bibinfo {year} {2010})}\BibitemShut {NoStop}%
\bibitem [{\citenamefont {Sciolla}\ and\ \citenamefont
  {Biroli}(2011)}]{Sciolla2011}%
  \BibitemOpen
  \bibfield  {author} {\bibinfo {author} {\bibfnamefont {B.}~\bibnamefont
  {Sciolla}}\ and\ \bibinfo {author} {\bibfnamefont {G.}~\bibnamefont
  {Biroli}},\ }\href {\doibase 10.1088/1742-5468/2011/11/p11003} {\bibfield
  {journal} {\bibinfo  {journal} {Journal of Statistical Mechanics: Theory and
  Experiment}\ }\textbf {\bibinfo {volume} {2011}},\ \bibinfo {pages} {P11003}
  (\bibinfo {year} {2011})}\BibitemShut {NoStop}%
\bibitem [{\citenamefont {Sciolla}\ and\ \citenamefont
  {Biroli}(2013)}]{Sciolla2013}%
  \BibitemOpen
  \bibfield  {author} {\bibinfo {author} {\bibfnamefont {B.}~\bibnamefont
  {Sciolla}}\ and\ \bibinfo {author} {\bibfnamefont {G.}~\bibnamefont
  {Biroli}},\ }\href {\doibase 10.1103/PhysRevB.88.201110} {\bibfield
  {journal} {\bibinfo  {journal} {Phys. Rev. B}\ }\textbf {\bibinfo {volume}
  {88}},\ \bibinfo {pages} {201110} (\bibinfo {year} {2013})}\BibitemShut
  {NoStop}%
\bibitem [{\citenamefont {Chandran}\ \emph {et~al.}(2013)\citenamefont
  {Chandran}, \citenamefont {Nanduri}, \citenamefont {Gubser},\ and\
  \citenamefont {Sondhi}}]{Chandran2013}%
  \BibitemOpen
  \bibfield  {author} {\bibinfo {author} {\bibfnamefont {A.}~\bibnamefont
  {Chandran}}, \bibinfo {author} {\bibfnamefont {A.}~\bibnamefont {Nanduri}},
  \bibinfo {author} {\bibfnamefont {S.~S.}\ \bibnamefont {Gubser}}, \ and\
  \bibinfo {author} {\bibfnamefont {S.~L.}\ \bibnamefont {Sondhi}},\ }\href
  {\doibase 10.1103/PhysRevB.88.024306} {\bibfield  {journal} {\bibinfo
  {journal} {Phys. Rev. B}\ }\textbf {\bibinfo {volume} {88}},\ \bibinfo
  {pages} {024306} (\bibinfo {year} {2013})}\BibitemShut {NoStop}%
\bibitem [{\citenamefont {Gagel}\ \emph {et~al.}(2014)\citenamefont {Gagel},
  \citenamefont {Orth},\ and\ \citenamefont {Schmalian}}]{Schmalian2014}%
  \BibitemOpen
  \bibfield  {author} {\bibinfo {author} {\bibfnamefont {P.}~\bibnamefont
  {Gagel}}, \bibinfo {author} {\bibfnamefont {P.~P.}\ \bibnamefont {Orth}}, \
  and\ \bibinfo {author} {\bibfnamefont {J.}~\bibnamefont {Schmalian}},\ }\href
  {\doibase 10.1103/PhysRevLett.113.220401} {\bibfield  {journal} {\bibinfo
  {journal} {Phys. Rev. Lett.}\ }\textbf {\bibinfo {volume} {113}},\ \bibinfo
  {pages} {220401} (\bibinfo {year} {2014})}\BibitemShut {NoStop}%
\bibitem [{\citenamefont {Gagel}\ \emph {et~al.}(2015)\citenamefont {Gagel},
  \citenamefont {Orth},\ and\ \citenamefont {Schmalian}}]{Schmalian2015}%
  \BibitemOpen
  \bibfield  {author} {\bibinfo {author} {\bibfnamefont {P.}~\bibnamefont
  {Gagel}}, \bibinfo {author} {\bibfnamefont {P.~P.}\ \bibnamefont {Orth}}, \
  and\ \bibinfo {author} {\bibfnamefont {J.}~\bibnamefont {Schmalian}},\ }\href
  {\doibase 10.1103/PhysRevB.92.115121} {\bibfield  {journal} {\bibinfo
  {journal} {Phys. Rev. B}\ }\textbf {\bibinfo {volume} {92}},\ \bibinfo
  {pages} {115121} (\bibinfo {year} {2015})}\BibitemShut {NoStop}%
\bibitem [{\citenamefont {Chiocchetta}\ \emph {et~al.}(2015)\citenamefont
  {Chiocchetta}, \citenamefont {Tavora}, \citenamefont {Gambassi},\ and\
  \citenamefont {Mitra}}]{Tavora2015}%
  \BibitemOpen
  \bibfield  {author} {\bibinfo {author} {\bibfnamefont {A.}~\bibnamefont
  {Chiocchetta}}, \bibinfo {author} {\bibfnamefont {M.}~\bibnamefont {Tavora}},
  \bibinfo {author} {\bibfnamefont {A.}~\bibnamefont {Gambassi}}, \ and\
  \bibinfo {author} {\bibfnamefont {A.}~\bibnamefont {Mitra}},\ }\href
  {\doibase 10.1103/PhysRevB.91.220302} {\bibfield  {journal} {\bibinfo
  {journal} {Phys. Rev. B}\ }\textbf {\bibinfo {volume} {91}},\ \bibinfo
  {pages} {220302} (\bibinfo {year} {2015})}\BibitemShut {NoStop}%
\bibitem [{\citenamefont {Maraga}\ \emph {et~al.}(2015)\citenamefont {Maraga},
  \citenamefont {Chiocchetta}, \citenamefont {Mitra},\ and\ \citenamefont
  {Gambassi}}]{Maraga2015}%
  \BibitemOpen
  \bibfield  {author} {\bibinfo {author} {\bibfnamefont {A.}~\bibnamefont
  {Maraga}}, \bibinfo {author} {\bibfnamefont {A.}~\bibnamefont {Chiocchetta}},
  \bibinfo {author} {\bibfnamefont {A.}~\bibnamefont {Mitra}}, \ and\ \bibinfo
  {author} {\bibfnamefont {A.}~\bibnamefont {Gambassi}},\ }\href {\doibase
  10.1103/PhysRevE.92.042151} {\bibfield  {journal} {\bibinfo  {journal} {Phys.
  Rev. E}\ }\textbf {\bibinfo {volume} {92}},\ \bibinfo {pages} {042151}
  (\bibinfo {year} {2015})}\BibitemShut {NoStop}%
\bibitem [{\citenamefont {Smacchia}\ \emph {et~al.}(2015)\citenamefont
  {Smacchia}, \citenamefont {Knap}, \citenamefont {Demler},\ and\ \citenamefont
  {Silva}}]{Smacchia2015}%
  \BibitemOpen
  \bibfield  {author} {\bibinfo {author} {\bibfnamefont {P.}~\bibnamefont
  {Smacchia}}, \bibinfo {author} {\bibfnamefont {M.}~\bibnamefont {Knap}},
  \bibinfo {author} {\bibfnamefont {E.}~\bibnamefont {Demler}}, \ and\ \bibinfo
  {author} {\bibfnamefont {A.}~\bibnamefont {Silva}},\ }\href {\doibase
  10.1103/PhysRevB.91.205136} {\bibfield  {journal} {\bibinfo  {journal} {Phys.
  Rev. B}\ }\textbf {\bibinfo {volume} {91}},\ \bibinfo {pages} {205136}
  (\bibinfo {year} {2015})}\BibitemShut {NoStop}%
\bibitem [{\citenamefont {Maraga}\ \emph {et~al.}(2016)\citenamefont {Maraga},
  \citenamefont {Smacchia},\ and\ \citenamefont {Silva}}]{Maraga2016}%
  \BibitemOpen
  \bibfield  {author} {\bibinfo {author} {\bibfnamefont {A.}~\bibnamefont
  {Maraga}}, \bibinfo {author} {\bibfnamefont {P.}~\bibnamefont {Smacchia}}, \
  and\ \bibinfo {author} {\bibfnamefont {A.}~\bibnamefont {Silva}},\ }\href
  {https://link.aps.org/doi/10.1103/PhysRevB.94.245122} {\bibfield  {journal}
  {\bibinfo  {journal} {Phys. Rev. B}\ }\textbf {\bibinfo {volume} {94}},\
  \bibinfo {pages} {245122} (\bibinfo {year} {2016})}\BibitemShut {NoStop}%
\bibitem [{\citenamefont {Chiocchetta}\ \emph {et~al.}(2016)\citenamefont
  {Chiocchetta}, \citenamefont {Tavora}, \citenamefont {Gambassi},\ and\
  \citenamefont {Mitra}}]{Chiocchetta2016}%
  \BibitemOpen
  \bibfield  {author} {\bibinfo {author} {\bibfnamefont {A.}~\bibnamefont
  {Chiocchetta}}, \bibinfo {author} {\bibfnamefont {M.}~\bibnamefont {Tavora}},
  \bibinfo {author} {\bibfnamefont {A.}~\bibnamefont {Gambassi}}, \ and\
  \bibinfo {author} {\bibfnamefont {A.}~\bibnamefont {Mitra}},\ }\href
  {\doibase 10.1103/PhysRevB.94.134311} {\bibfield  {journal} {\bibinfo
  {journal} {Phys. Rev. B}\ }\textbf {\bibinfo {volume} {94}},\ \bibinfo
  {pages} {134311} (\bibinfo {year} {2016})}\BibitemShut {NoStop}%
\bibitem [{\citenamefont {Lemonik}\ and\ \citenamefont
  {Mitra}(2016)}]{Lemonik2016}%
  \BibitemOpen
  \bibfield  {author} {\bibinfo {author} {\bibfnamefont {Y.}~\bibnamefont
  {Lemonik}}\ and\ \bibinfo {author} {\bibfnamefont {A.}~\bibnamefont
  {Mitra}},\ }\href {\doibase 10.1103/PhysRevB.94.024306} {\bibfield  {journal}
  {\bibinfo  {journal} {Phys. Rev. B}\ }\textbf {\bibinfo {volume} {94}},\
  \bibinfo {pages} {024306} (\bibinfo {year} {2016})}\BibitemShut {NoStop}%
\bibitem [{\citenamefont {Chiocchetta}\ \emph {et~al.}(2017)\citenamefont
  {Chiocchetta}, \citenamefont {Gambassi}, \citenamefont {Diehl},\ and\
  \citenamefont {Marino}}]{Chiocchetta2017}%
  \BibitemOpen
  \bibfield  {author} {\bibinfo {author} {\bibfnamefont {A.}~\bibnamefont
  {Chiocchetta}}, \bibinfo {author} {\bibfnamefont {A.}~\bibnamefont
  {Gambassi}}, \bibinfo {author} {\bibfnamefont {S.}~\bibnamefont {Diehl}}, \
  and\ \bibinfo {author} {\bibfnamefont {J.}~\bibnamefont {Marino}},\ }\href
  {\doibase 10.1103/PhysRevLett.118.135701} {\bibfield  {journal} {\bibinfo
  {journal} {Phys. Rev. Lett.}\ }\textbf {\bibinfo {volume} {118}},\ \bibinfo
  {pages} {135701} (\bibinfo {year} {2017})}\BibitemShut {NoStop}%
\bibitem [{\citenamefont {D'Alessio}\ and\ \citenamefont
  {Rigol}(2014)}]{Rigol14}%
  \BibitemOpen
  \bibfield  {author} {\bibinfo {author} {\bibfnamefont {L.}~\bibnamefont
  {D'Alessio}}\ and\ \bibinfo {author} {\bibfnamefont {M.}~\bibnamefont
  {Rigol}},\ }\href {\doibase 10.1103/PhysRevX.4.041048} {\bibfield  {journal}
  {\bibinfo  {journal} {Phys. Rev. X}\ }\textbf {\bibinfo {volume} {4}},\
  \bibinfo {pages} {041048} (\bibinfo {year} {2014})}\BibitemShut {NoStop}%
\bibitem [{\citenamefont {Lazarides}\ \emph {et~al.}(2014)\citenamefont
  {Lazarides}, \citenamefont {Das},\ and\ \citenamefont
  {Moessner}}]{Lazarides14}%
  \BibitemOpen
  \bibfield  {author} {\bibinfo {author} {\bibfnamefont {A.}~\bibnamefont
  {Lazarides}}, \bibinfo {author} {\bibfnamefont {A.}~\bibnamefont {Das}}, \
  and\ \bibinfo {author} {\bibfnamefont {R.}~\bibnamefont {Moessner}},\ }\href
  {\doibase 10.1103/PhysRevE.90.012110} {\bibfield  {journal} {\bibinfo
  {journal} {Phys. Rev. E}\ }\textbf {\bibinfo {volume} {90}},\ \bibinfo
  {pages} {012110} (\bibinfo {year} {2014})}\BibitemShut {NoStop}%
\bibitem [{\citenamefont {Ponte}\ \emph {et~al.}(2015)\citenamefont {Ponte},
  \citenamefont {Chandran}, \citenamefont {Papic},\ and\ \citenamefont
  {Abanin}}]{Ponte15}%
  \BibitemOpen
  \bibfield  {author} {\bibinfo {author} {\bibfnamefont {P.}~\bibnamefont
  {Ponte}}, \bibinfo {author} {\bibfnamefont {A.}~\bibnamefont {Chandran}},
  \bibinfo {author} {\bibfnamefont {Z.}~\bibnamefont {Papic}}, \ and\ \bibinfo
  {author} {\bibfnamefont {D.~A.}\ \bibnamefont {Abanin}},\ }\href {\doibase
  https://doi.org/10.1016/j.aop.2014.11.008} {\bibfield  {journal} {\bibinfo
  {journal} {Annals of Physics}\ }\textbf {\bibinfo {volume} {353}},\ \bibinfo
  {pages} {196 } (\bibinfo {year} {2015})}\BibitemShut {NoStop}%
\bibitem [{\citenamefont {Natsheh}\ \emph {et~al.}()\citenamefont {Natsheh},
  \citenamefont {Gambassi},\ and\ \citenamefont {Mitra}}]{Natsheh2020b}%
  \BibitemOpen
  \bibfield  {author} {\bibinfo {author} {\bibfnamefont {M.}~\bibnamefont
  {Natsheh}}, \bibinfo {author} {\bibfnamefont {A.}~\bibnamefont {Gambassi}}, \
  and\ \bibinfo {author} {\bibfnamefont {A.}~\bibnamefont {Mitra}},\
  }\href@noop {} {\bibinfo  {journal} {in preparation}\ }\BibitemShut {NoStop}%
\bibitem [{\citenamefont {Calabrese}\ and\ \citenamefont
  {Cardy}(2007)}]{calabrese2007}%
  \BibitemOpen
\bibfield  {journal} {  }\bibfield  {author} {\bibinfo {author} {\bibfnamefont
  {P.}~\bibnamefont {Calabrese}}\ and\ \bibinfo {author} {\bibfnamefont
  {J.}~\bibnamefont {Cardy}},\ }\href@noop {} {\bibfield  {journal} {\bibinfo
  {journal} {Journal of Statistical Mechanics: Theory and Experiment}\ }\textbf
  {\bibinfo {volume} {2007}},\ \bibinfo {pages} {P06008} (\bibinfo {year}
  {2007})}\BibitemShut {NoStop}%
\bibitem [{\citenamefont {Mitra}(2018)}]{Mitra2018}%
  \BibitemOpen
  \bibfield  {author} {\bibinfo {author} {\bibfnamefont {A.}~\bibnamefont
  {Mitra}},\ }\href {\doibase 10.1146/annurev-conmatphys-031016-025451}
  {\bibfield  {journal} {\bibinfo  {journal} {Annual Review of Condensed Matter
  Physics}\ }\textbf {\bibinfo {volume} {9}},\ \bibinfo {pages} {245} (\bibinfo
  {year} {2018})}\BibitemShut {NoStop}%
\bibitem [{\citenamefont {Aarts}\ \emph {et~al.}(2002)\citenamefont {Aarts},
  \citenamefont {Ahrensmeier}, \citenamefont {Baier}, \citenamefont {Berges},\
  and\ \citenamefont {Serreau}}]{Berges02}%
  \BibitemOpen
  \bibfield  {author} {\bibinfo {author} {\bibfnamefont {G.}~\bibnamefont
  {Aarts}}, \bibinfo {author} {\bibfnamefont {D.}~\bibnamefont {Ahrensmeier}},
  \bibinfo {author} {\bibfnamefont {R.}~\bibnamefont {Baier}}, \bibinfo
  {author} {\bibfnamefont {J.}~\bibnamefont {Berges}}, \ and\ \bibinfo {author}
  {\bibfnamefont {J.}~\bibnamefont {Serreau}},\ }\href {\doibase
  10.1103/PhysRevD.66.045008} {\bibfield  {journal} {\bibinfo  {journal} {Phys.
  Rev. D}\ }\textbf {\bibinfo {volume} {66}},\ \bibinfo {pages} {045008}
  (\bibinfo {year} {2002})}\BibitemShut {NoStop}%
\bibitem [{\citenamefont {Berges}\ and\ \citenamefont
  {Gasenzer}(2007)}]{Gasenzer07}%
  \BibitemOpen
  \bibfield  {author} {\bibinfo {author} {\bibfnamefont {J.}~\bibnamefont
  {Berges}}\ and\ \bibinfo {author} {\bibfnamefont {T.}~\bibnamefont
  {Gasenzer}},\ }\href {\doibase 10.1103/PhysRevA.76.033604} {\bibfield
  {journal} {\bibinfo  {journal} {Phys. Rev. A}\ }\textbf {\bibinfo {volume}
  {76}},\ \bibinfo {pages} {033604} (\bibinfo {year} {2007})}\BibitemShut
  {NoStop}%
\bibitem [{\citenamefont {Oliver}\ \emph {et~al.}(2010)\citenamefont {Oliver},
  \citenamefont {Lozier}, \citenamefont {Boisvert},\ and\ \citenamefont
  {Clark}}]{Nist}%
  \BibitemOpen
  \bibfield  {author} {\bibinfo {author} {\bibfnamefont {F.}~\bibnamefont
  {Oliver}}, \bibinfo {author} {\bibfnamefont {D.}~\bibnamefont {Lozier}},
  \bibinfo {author} {\bibfnamefont {R.}~\bibnamefont {Boisvert}}, \ and\
  \bibinfo {author} {\bibfnamefont {C.}~\bibnamefont {Clark}},\ }\href@noop {}
  {\emph {\bibinfo {title} {NIST Handbook of Mathematical Functions}}}\
  (\bibinfo  {publisher} {Cambridge university press},\ \bibinfo {year}
  {2010})\BibitemShut {NoStop}%
\bibitem [{\citenamefont {McLachlan}(1947)}]{McLachlin1947}%
  \BibitemOpen
  \bibfield  {author} {\bibinfo {author} {\bibfnamefont {N.}~\bibnamefont
  {McLachlan}},\ }\href@noop {} {\emph {\bibinfo {title} {Theory and
  Application of Mathieu Functions}}}\ (\bibinfo  {publisher} {Oxford
  university press},\ \bibinfo {year} {1947})\BibitemShut {NoStop}%
\bibitem [{\citenamefont {Richards}(1983)}]{Richards1983}%
  \BibitemOpen
  \bibfield  {author} {\bibinfo {author} {\bibfnamefont {J.~A.}\ \bibnamefont
  {Richards}},\ }\href@noop {} {\emph {\bibinfo {title} {Analysis of
  Periodically Time-Varying Systems}}}\ (\bibinfo  {publisher} {Springer},\
  \bibinfo {year} {1983})\BibitemShut {NoStop}%
\bibitem [{\citenamefont {Guo}\ \emph {et~al.}(2013)\citenamefont {Guo},
  \citenamefont {Marthaler},\ and\ \citenamefont {Sch\"on}}]{Guo13}%
  \BibitemOpen
  \bibfield  {author} {\bibinfo {author} {\bibfnamefont {L.}~\bibnamefont
  {Guo}}, \bibinfo {author} {\bibfnamefont {M.}~\bibnamefont {Marthaler}}, \
  and\ \bibinfo {author} {\bibfnamefont {G.}~\bibnamefont {Sch\"on}},\ }\href
  {\doibase 10.1103/PhysRevLett.111.205303} {\bibfield  {journal} {\bibinfo
  {journal} {Phys. Rev. Lett.}\ }\textbf {\bibinfo {volume} {111}},\ \bibinfo
  {pages} {205303} (\bibinfo {year} {2013})}\BibitemShut {NoStop}%
\bibitem [{\citenamefont {Shirley}(1965)}]{Shirley65}%
  \BibitemOpen
  \bibfield  {author} {\bibinfo {author} {\bibfnamefont {J.~H.}\ \bibnamefont
  {Shirley}},\ }\href {\doibase 10.1103/PhysRev.138.B979} {\bibfield  {journal}
  {\bibinfo  {journal} {Phys. Rev.}\ }\textbf {\bibinfo {volume} {138}},\
  \bibinfo {pages} {B979} (\bibinfo {year} {1965})}\BibitemShut {NoStop}%
\bibitem [{\citenamefont {Sambe}(1973)}]{Sambe73}%
  \BibitemOpen
  \bibfield  {author} {\bibinfo {author} {\bibfnamefont {H.}~\bibnamefont
  {Sambe}},\ }\href {\doibase 10.1103/PhysRevA.7.2203} {\bibfield  {journal}
  {\bibinfo  {journal} {Phys. Rev. A}\ }\textbf {\bibinfo {volume} {7}},\
  \bibinfo {pages} {2203} (\bibinfo {year} {1973})}\BibitemShut {NoStop}%
\bibitem [{\citenamefont {Kamenev}(2011)}]{Kamenevbook}%
  \BibitemOpen
  \bibfield  {author} {\bibinfo {author} {\bibfnamefont {A.}~\bibnamefont
  {Kamenev}},\ }\href@noop {} {\bibfield  {journal} {\bibinfo  {journal} {{\sl
  Field Theory of Non-Equilibrium Systems}, Cambridge University Press,
  Cambridge}\ } (\bibinfo {year} {2011})}\BibitemShut {NoStop}%
\bibitem [{\citenamefont {Cugliandolo}(2011)}]{Cugliandolo2011}%
  \BibitemOpen
  \bibfield  {author} {\bibinfo {author} {\bibfnamefont {L.~F.}\ \bibnamefont
  {Cugliandolo}},\ }\href {\doibase 10.1088/1751-8113/44/48/483001} {\bibfield
  {journal} {\bibinfo  {journal} {Journal of Physics A: Mathematical and
  Theoretical}\ }\textbf {\bibinfo {volume} {44}},\ \bibinfo {pages} {483001}
  (\bibinfo {year} {2011})}\BibitemShut {NoStop}%
\bibitem [{\citenamefont {Foini}\ \emph {et~al.}(2011)\citenamefont {Foini},
  \citenamefont {Cugliandolo},\ and\ \citenamefont {Gambassi}}]{Foini2011b}%
  \BibitemOpen
  \bibfield  {author} {\bibinfo {author} {\bibfnamefont {L.}~\bibnamefont
  {Foini}}, \bibinfo {author} {\bibfnamefont {L.~F.}\ \bibnamefont
  {Cugliandolo}}, \ and\ \bibinfo {author} {\bibfnamefont {A.}~\bibnamefont
  {Gambassi}},\ }\href {\doibase 10.1103/PhysRevB.84.212404} {\bibfield
  {journal} {\bibinfo  {journal} {Phys. Rev. B}\ }\textbf {\bibinfo {volume}
  {84}},\ \bibinfo {pages} {212404} (\bibinfo {year} {2011})}\BibitemShut
  {NoStop}%
\bibitem [{\citenamefont {Foini}\ \emph {et~al.}(2012)\citenamefont {Foini},
  \citenamefont {Cugliandolo},\ and\ \citenamefont {Gambassi}}]{Foini2012}%
  \BibitemOpen
  \bibfield  {author} {\bibinfo {author} {\bibfnamefont {L.}~\bibnamefont
  {Foini}}, \bibinfo {author} {\bibfnamefont {L.~F.}\ \bibnamefont
  {Cugliandolo}}, \ and\ \bibinfo {author} {\bibfnamefont {A.}~\bibnamefont
  {Gambassi}},\ }\href {\doibase 10.1088/1742-5468/2012/09/P09011} {\bibfield
  {journal} {\bibinfo  {journal} {Journal of Statistical Mechanics: Theory and
  Experiment}\ }\textbf {\bibinfo {volume} {2012}},\ \bibinfo {pages} {P09011}
  (\bibinfo {year} {2012})}\BibitemShut {NoStop}%
\bibitem [{\citenamefont {Dehghani}\ and\ \citenamefont
  {Mitra}(2016)}]{Dehghani16}%
  \BibitemOpen
  \bibfield  {author} {\bibinfo {author} {\bibfnamefont {H.}~\bibnamefont
  {Dehghani}}\ and\ \bibinfo {author} {\bibfnamefont {A.}~\bibnamefont
  {Mitra}},\ }\href@noop {} {\bibfield  {journal} {\bibinfo  {journal} {Phys.
  Rev. B}\ }\textbf {\bibinfo {volume} {93}},\ \bibinfo {pages} {245416}
  (\bibinfo {year} {2016})}\BibitemShut {NoStop}%
\bibitem [{\citenamefont {Calabrese}\ and\ \citenamefont
  {Cardy}(2005)}]{Calabrese2005}%
  \BibitemOpen
  \bibfield  {author} {\bibinfo {author} {\bibfnamefont {P.}~\bibnamefont
  {Calabrese}}\ and\ \bibinfo {author} {\bibfnamefont {J.}~\bibnamefont
  {Cardy}},\ }\href@noop {} {\bibfield  {journal} {\bibinfo  {journal} {Journal
  of Statistical Mechanics: Theory and Experiment}\ }\textbf {\bibinfo {volume}
  {2005}},\ \bibinfo {pages} {P04010} (\bibinfo {year} {2005})}\BibitemShut
  {NoStop}%
\bibitem [{\citenamefont {Marcuzzi}\ and\ \citenamefont
  {Gambassi}(2014)}]{Marcuzzi2014}%
  \BibitemOpen
  \bibfield  {author} {\bibinfo {author} {\bibfnamefont {M.}~\bibnamefont
  {Marcuzzi}}\ and\ \bibinfo {author} {\bibfnamefont {A.}~\bibnamefont
  {Gambassi}},\ }\href {\doibase 10.1103/PhysRevB.89.134307} {\bibfield
  {journal} {\bibinfo  {journal} {Phys. Rev. B}\ }\textbf {\bibinfo {volume}
  {89}},\ \bibinfo {pages} {134307} (\bibinfo {year} {2014})}\BibitemShut
  {NoStop}%
\bibitem [{\citenamefont {Eckardt}\ and\ \citenamefont
  {Anisimovas}(2015)}]{Eckardt2015}%
  \BibitemOpen
  \bibfield  {author} {\bibinfo {author} {\bibfnamefont {A.}~\bibnamefont
  {Eckardt}}\ and\ \bibinfo {author} {\bibfnamefont {E.}~\bibnamefont
  {Anisimovas}},\ }\href {\doibase 10.1088/1367-2630/17/9/093039} {\bibfield
  {journal} {\bibinfo  {journal} {New Journal of Physics}\ }\textbf {\bibinfo
  {volume} {17}},\ \bibinfo {pages} {093039} (\bibinfo {year}
  {2015})}\BibitemShut {NoStop}%
\end{thebibliography}
%

\end{document}